\documentclass[longauth]{aa}  
\usepackage[colorlinks=true,citecolor=blue]{hyperref}
\usepackage{txfonts}
\usepackage{longtable,lscape}
\usepackage{amsmath}
\usepackage{graphicx}
\usepackage{xcolor}
\usepackage{subfigure}
\usepackage{xspace}
\usepackage{gensymb}

\newcommand{\snr}{$S/N$\xspace}
\newcommand{\perpix}{pixel$^{-1}$\xspace}
\usepackage[normalem]{ulem}
\usepackage{geometry}

\newcommand{\stkout}[1]{\ifmmode\text{\sout{\ensuremath{#1}}}\else\sout{#1}\fi}

\begin{document} 
\newcommand*\tess{\textit{TESS }}
\newcommand{\ktwo}{\emph{K2}} 
\newcommand{\gaia}{\emph{Gaia}}
\newcommand{\kepler}{\emph{Kepler}}
\newcommand{\corot}{\emph{CoRoT}}
\newcommand{\plato}{\emph{PLATO}}
\newcommand{\sname}{TOI-942}
\newcommand{\planetb}{TOI-942\,b}
\newcommand{\planetc}{TOI-942\,c}
\newcommand{\logrhk}{$\rm log\,R^{\prime}_\mathrm{HK}$}
\newcommand{\kms}{\,km\,s$^{-1}$} % kilometres per second
\newcommand{\ms}{\,m\,s$^{-1}$} % metres per second

\newcommand{\mstar}{M$_{\star}$}
\newcommand{\rstar}{R$_{\star}$}
\newcommand{\lstar}{L$_{\star}$}

\newcommand{\msun}{$M_{\odot}$}
\newcommand{\rsun}{$R_{\odot}$}
\newcommand{\lsun}{$L_{\odot}$}

\newcommand{\vsini}{$v$\,sin\,$i_\star$}   
\newcommand{\sini}{sin\,$i_\star$}   
\newcommand{\vrad}{$v_{\rm rad}$} 
\newcommand{\vmic}{$v_{\rm mic}$}
\newcommand{\vmac}{$v_{\rm mac}$}
\newcommand{\dex}{$\rm dex$}
\newcommand{\teff}{$T_{\rm eff}$}
\newcommand{\logg}{log\,{\it g$_\star$}}
\newcommand{\feh}{[Fe/H]}
\newcommand{\tih}{[Ti/H]}
\newcommand{\mearth}{$M_{\oplus}$}
\newcommand{\rearth}{$R_{\oplus}$}
\newcommand{\prot}{$P_{\rm rot}$}

   \title{The GAPS Programme at TNG}

%   \subtitle{MCDXXX. Something about TOI-1430}
   \subtitle{LXV. Precise density measurement of TOI-1430~b, a young planet with an evaporating atmosphere}

   \author{D.\ Nardiello\inst{1,2,3}
          \and
          J.\ M.\ Akana Murphy\inst{4}
          \and
          R.\ Spinelli\inst{5,6}
          \and
          M.\ Baratella\inst{7}
          \and
          S.\ Desidera\inst{2}
          \and
          V.\ Nascimbeni\inst{2}
          \and
          L.\ Malavolta\inst{1,2}
          \and
          K.\ Biazzo\inst{8}
          \and
          A.\ Maggio\inst{5}
          \and
          D.\ Locci\inst{5}
          \and
          S.\ Benatti\inst{5}
          \and
          N.\ M.\ Batalha\inst{4}
          \and
          V.\ D'Orazi\inst{9}
          \and
          L.\ Borsato\inst{2}
          \and
          G.\ Piotto\inst{1,2,3}
          \and
          R.\ J.\ Oelkers\inst{10}
          \and
          M.\ Mallonn\inst{11}
          \and
          A.\ Sozzetti\inst{12}
          \and
          L.\ R.\ Bedin\inst{2}
          \and
          G.\ Mantovan\inst{1,2}
          \and
          T.\ Zingales\inst{1,2}
    %%%%%%%%%%%%%%%%%%%%%%%%%%%%%%%%%%%%%%%%%%%%%%%%%%%%%%%%%%%%%%%%%%%%%%%%%%%%%%
          \and
          L.\ Affer\inst{5}
          \and
          A.\ Bignamini\inst{13}
          \and
          A.\ S.\ Bonomo\inst{12}
          \and
          L.\ Cabona\inst{14}
          \and
          K.\ A.\ Collins\inst{15}
          \and
          M.\ Damasso\inst{12} 
          \and
          S.\ Filomeno\inst{8,9,16}
          \and
          A.\ Ghedina\inst{17}
          \and
          A.\ Harutyunyan\inst{17}
          \and
          A.\ F.\ Lanza\inst{18}
          \and
          L.\ Mancini\inst{9,12,19}
          \and
          M.\ Rainer\inst{14}
          \and
          G.\ Scandariato\inst{18}
          \and
          R.\ P.\ Schwarz\inst{15}
          \and
          R.\ Sefako\inst{20}
          \and
          G.\ Srdoc\inst{21}
    %      al.\inst{1} \\
          %LCO co-authors
          }

   \institute{Dipartimento di Fisica e Astronomia "Galileo Galilei" -- Universit\`a degli Studi di Padova, Vicolo dell'Osservatorio 3, I-35122 Padova, Italy
              \email{domenico.nardiello@unipd.it}
             \and
             INAF -- Osservatorio Astronomico di Padova, Vicolo dell'Osservatorio 5, Padova I-35122, Italy
             \and 
             Centro di Ateneo di Studi e Attivit\`a Spaziali "G. Colombo" -- Universit\`a degli Studi di Padova, Via Venezia 15, I-35131, Padova, Italy
             \and
             Department of Astronomy and Astrophysics, University of California, Santa Cruz, CA 95060, USA 
             \and
             INAF -- Osservatorio Astronomico di Palermo, Piazza del Parlamento 1, I-90134 Palermo (PA), Italy
             \and
             Dipartimento di Scienza e Alta Tecnologia, Universit\`a dell’Insubria, Via Valleggio 11, I-22100 Como, Italy 
             \and
             ESO-European Southern Observatory, Alonso de Cordova 3107, Vitacura, Santiago, Chile
             \and
             INAF -- Osservatorio Astronomico di Roma, Via Frascati 33, I-00078 Monte Porzio Catone (Roma), Italy
             \and
             Dipartimento di Fisica, Universit\`a degli Studi di Roma Tor Vergata, via della Ricerca Scientifica 1, I-00133 Roma, Italy
             \and
             Department of Physics and Astronomy, The University of Texas, Rio Grande Valley, Brownsville, TX 78520, USA 
             \and
             Leibniz-Institut f\"ur Astrophysik Potsdam (AIP), An der Sternwarte 16, D-14482 Potsdam, Germany
             \and
             INAF -- Osservatorio Astrofisico di Torino, Via Osservatorio 20, I-10025, Pino Torinese, Italy
             \and
             INAF -- Osservatorio Astronomico di Trieste, via Tiepolo 11, I-34143, Trieste, Italy
             \and             
             INAF -- Osservatorio Astronomico di Brera, Via E. Bianchi 46, I-23807, Merate (LC), Italy
             \and
             Center for Astrophysics \textbar \ Harvard \& Smithsonian, 60 Garden Street, Cambridge, MA 02138, USA  
             \and
             Dipartimento di Fisica, Sapienza Università di Roma, Piazzale Aldo Moro 5, I-00185 Roma, Italy 
             \and
             Fundaci\'on Galileo Galilei – INAF, Rambla J.A. Fernandez P., 7, E-38712 S.C.Tenerife, Spain 
             \and
             INAF -- Osservatorio Astrofisico di Catania, Via S. Sofia 78, I-95123, Catania, Italy
             \and
             Max Planck Institute for Astronomy, Königstuhl 17, D-69117, Heidelberg, Germany
             \and
             South African Astronomical Observatory, P.O. Box 9, Observatory, Cape Town 7935, South Africa
             \and
             Kotizarovci Observatory, Sarsoni 90, 51216 Viskovo, Croatia
             }

   \date{Received 13 September 2024; accepted 16 November 2024}

% \abstract{}{}{}{}{} 
% 5 {} token are mandatory
 
  \abstract
   {Small-sized ($<4$~$R_{\oplus}$) exoplanets in tight orbits around young stars (10--1000~Myr) give us the opportunity to investigate the mechanisms that led to their formation, the evolution of their physical and orbital properties and, especially, of their atmospheres. Thanks to the all-sky survey carried out by the \tess spacecraft, many of these exoplanets have been discovered and have subsequently been characterized with dedicated follow-up observations. }
   {In the context of a collaboration among the Global Architecture of Planetary Systems (GAPS) team, the TESS-Keck Survey (TKS) team and, the California Planet Search (CPS) team, we measured with a high level of precision the mass and the radius of TOI-1430~b, a young ($\sim 700$~Myr) exoplanet with an escaping He atmosphere orbiting the K-dwarf star HD~235088 (TOI-1430).}
   {By adopting appropriate stellar parameters, which were measured in this work, we were able to simultaneously model the signals due to strong stellar activity and the transiting planet TOI-1430~b in both photometric and spectroscopic series. This allowed us to measure both the radius and mass (and consequently the density) of the planet with high precision, and  reconstruct the evolution of its atmosphere.}
   {TOI-1430 is an active K-dwarf star born $700 \pm 150$~Myr ago and  rotates in $P_{\rm rot} \sim 12$~days. This star hosts a mini-Neptune whose orbital period is $P_{\rm b} = 7.434133 \pm 0.000004$~days. Thanks to long-term photometric and spectroscopic monitoring of this target performed with {\it TESS}, HARPS-N, HIRES, and APF, we estimated a radius $R_{\rm P,b} = 1.98 \pm 0.07~R_{\oplus}$, a mass $M_{\rm P,b} = 4.2 \pm 0.8~M_{\oplus}$, and thus a planetary density $\rho_{\rm b} = 0.5 \pm 0.1~\rho_{\oplus}$. TOI-1430~b is hence a low-density mini-Neptune with an extended atmosphere, at the edge of the radius gap. Because this planet is known to have an evaporating atmosphere of He, we reconstructed its atmospheric history. Our analysis supports the scenario in which, shortly after its birth, TOI-1430~b may have been super-puffy, with a radius $5\times$--$13\times$ and a mass $1.5\times$--$2\times$ that of today; in $\sim 200$~Myr from now, TOI-1430~b should lose its envelope, showing its Earth-size core. We also looked for signals from a second planet in the spectroscopic/photometric series, without detecting any. }
   {}

   \maketitle
%
%-------------------------------------------------------------------
%
\section{Introduction}
%
%   \cite{2019MNRAS.490.3806N}
%
%
%The candidate was classified as a "likely planet" by \cite{giacalone2021}
%\cite{zhang2023} detected evaporating helium .......
%

Young exoplanets ($\lesssim 1$~Gyr) offer the unique opportunity to understand the mechanisms of formation and evolution of planetary systems. The Transiting Exoplanet Survey Satellite ({\it TESS}, \citealt{2015JATIS...1a4003R}) has allowed us to increase the number of planets belonging to this scarcely populated category. In fact, in the last five years, many projects started to search for young planets with \tess (e.g., \citealt{2019ApJS..245...13B,2019MNRAS.490.3806N,2019ApJ...880L..17N,2022MNRAS.511.4285B}). Despite the difficulties due to large variations of the light curves of active stars, many candidate exoplanets have been found both around members of young stellar clusters, associations, and moving groups (e.g., \citealt{2020AJ....160..239B,2020AJ....160...33R,2021AJ....161...65N,2020AJ....160..179M,2020MNRAS.498.5972N,2021MNRAS.505.3767N,2021AJ....161..171T, 2022AJ....163..156M,2023AJ....165...85W, 2024AJ....168...41T}) and orbiting young single stars (e.g., \citealt{2022ApJ...935L..10G,2022AJ....163...99G,2022AJ....164...71V,2022MNRAS.516.4432M}). The intensive spectroscopic follow-ups of these objects have allowed not only to put constraints on the planetary masses (e.g., \citealt{2019A&A...630A..81B,benatti2021,2022MNRAS.514.1606B,2022MNRAS.513.5955K,2023A&A...675A.158D,2023A&A...672A.126D,2024MNRAS.531.4275B,2024A&A...688A..15D,2023arXiv231016888M,2024A&A...682A.135C}) but also to identify extended (escaping) atmospheres (e.g., \citealt{orellmiquel2023,zhang2023,2023MNRAS.518.3777G,2024AJ....167..214P,2024arXiv240609225M}).
The discovery and the full characterization of young exoplanets help us constrain the timescales on which mechanisms like photo-evaporation (\citealt{2013ApJ...775..105O,LopFor14}) and core-powered mass loss (\citealt{2018MNRAS.476..759G,2019MNRAS.487...24G}) act.

The Global Architecture of Planetary Systems (GAPS) collaboration (\citealt{2013A&A...554A..28C}) is aimed at spectroscopic follow-up  of transiting and non-transiting exoplanets with the goal of  determining planetary masses, as well as measuring orbital properties (\citealt{2015A&A...581L...6D}), detecting planetary atmospheres (e.g., \citealt{2022A&A...668A.176P,2024A&A...687A.143S,2024A&A...686A..83G}) and characterizing the host stars (\citealt{2022A&A...663A.142M,2022A&A...664A.161B,2024A&A...682A.136C}). In particular, the Young Objects program (\citealt{2020A&A...638A...5C}) provides a spectroscopic follow-up of  planets younger than 750--1000 Myr. Under this program, we carried out a three-year observational campaign of the young K-dwarf star HD~235088 (a.k.a. TOI-1430, $V \sim 9.2$) that hosts a  potential transiting exoplanet (TOI-1430~b), as reported in \citet[False Positive Probability equal to 0.03]{2021AJ....161...24G}.
 Detection of an atmosphere for this planet were obtained by \citet{zhang2023} using Keck/NIRSPEC and by \citet{orellmiquel2023} adopting CARMENES data. They both detected the presence of an evaporating helium atmosphere.
\citet{2024ApJS..272...32P} detected an hint ($<2 \sigma$) of TOI-1430~b's signal using radial velocities (RVs) measurements.

In this work we detect and characterize TOI-1430~b making use of a combination of photometric data from 9 \tess sectors and spectroscopic data collected with the High Accuracy Radial velocity Planet Searcher for the Northern emisphere (HARPS-N, \citealt{2012SPIE.8446E..1VC}) on the Telescopio Nazionale Galileo (TNG) for the GAPS program and with the High Resolution Echelle Spectrometer (HIRES) on the 10~m Keck I telescope for the TESS-Keck Survey (\citealt{2020AJ....159..241D}) and California Planet Search programs. On the basis of our results, we characterize the evolutionary history of TOI-1430~b's atmosphere.

\section{Observations and data reduction}
\label{sec:obs}
\begin{figure}
  \centering
  \includegraphics[width=0.39\textwidth]{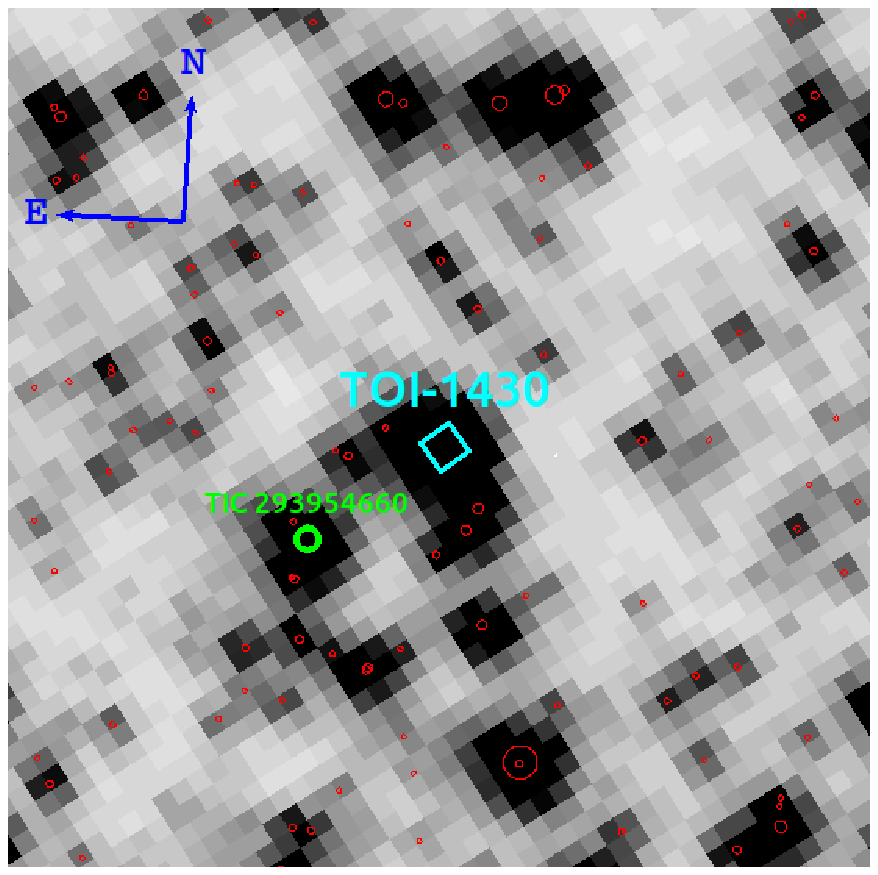} 
  \caption{Finding chart ($15\times 15$~arcmin$^2$) of TOI-1430 (cyan diamond) from a \tess Full Frame Image:  neighbor stars with $G\leq 15$ are identified with red circles whose radii are inversely proportional to the difference between the magnitude of TOI-1430 and that of its neighbors. The green circle indicates the eclipsing binary that generates the spurious signal in the SAP light curves. \label{fig:0}}
\end{figure}

\begin{figure*}
  \centering
  \includegraphics[bb=22 195 583 712, width=0.9\textwidth]{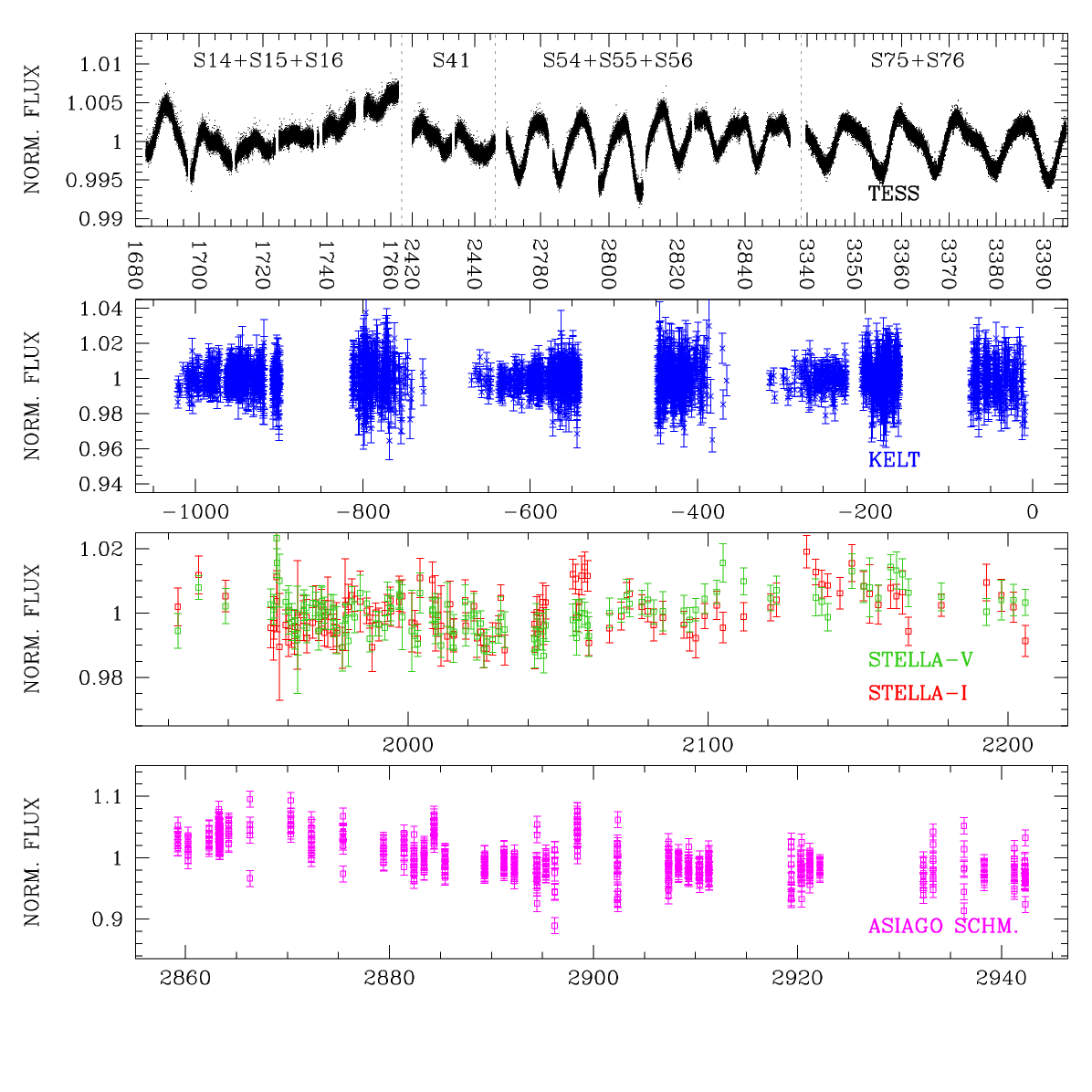} 
  \includegraphics[bb=22 558 583 712, width=0.9\textwidth]{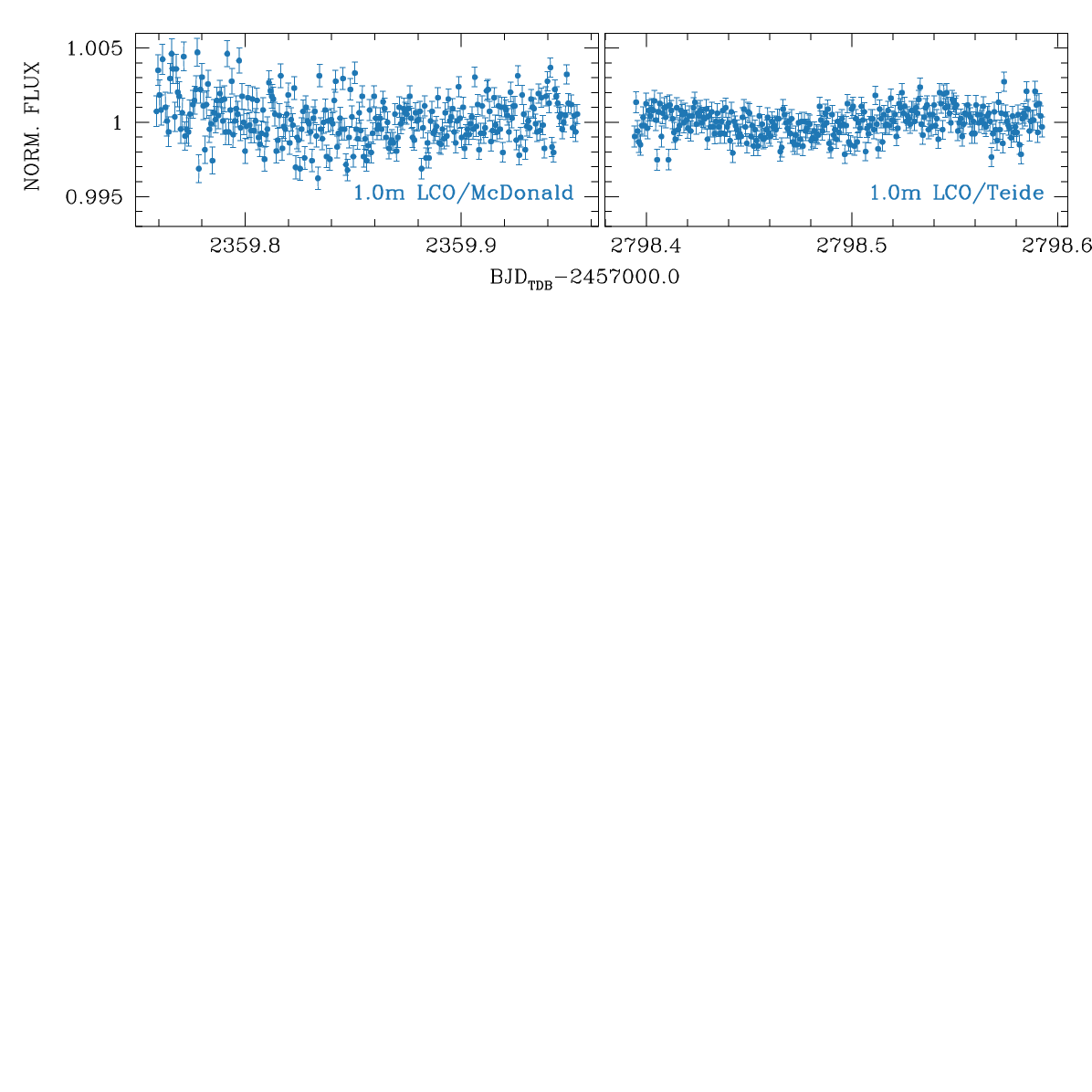} 
  \caption{Light curves of TOI-1430 obtained with different instruments. From top to the bottom: \tess light curve (black dots), obtained from short cadence data of sectors 14, 15, 16, 41, 54, 55, 56, 75, and 76, KELT light curve (blue crosses), STELLA light curves (green and red squares for $V$ and $I$ bands, respectively), Asiago Schmidt 67/92~cm $u$-sloan light curve (magenta squares), and LCO light curves in zs filter (azure circles). Vertical dashed lines in the upper panel indicate the gaps between the different years of the \tess mission.\label{fig:1}}
\end{figure*}

%%%%
\subsection{\tess data}
The dataset of TOI-1430 used in this work has been collected by \tess in short-cadence mode in nine different sectors (see Fig.~\ref{fig:0}): during the 2nd Cycle of \tess operations in Sectors 14, 15, and 16 (18 July 2019 -- 7 October 2019), during Sector 41 (4th Cycle of \tess mission, 23 July 2021 -- 20 August 2021), in Sectors 54, 55, and 56 between cycle 4 and cycle 5 (1 September 2022 -- 30 September 2022), and finally in cycle 6 during sectors 75 and 76 (30 January 2024 -- 26 March 2024).

In this work we did not make use of the Pre-search Data Conditioning Simple Aperture Photometry (PDCSAP) light curves (\citealt{2012PASP..124.1000S,2012PASP..124..985S,2014PASP..126..100S}) because of systematic effects introduced by the \tess official correction pipeline in the cases of variable stars (see Fig.~A.1 in \citealt{2022A&A...664A.163N} and Fig.~\ref{fig:A1} of this work). Moreover, we found that  the SAP light curves also suffer from a systematic effect introduced by contaminated local background (see panels (b) and (c) of Fig.~\ref{fig:A1}): in fact, we found a periodic signal ($\sim 2.81$~days) due to a close-by eclipsing binary (TIC~293954660, \citealt{2022ApJS..259...50S}) that clearly affected the local background that is subtracted to the pixel values inside the photometric aperture adopted to extract the light curve of TOI-1430.

We extracted the light curves of TOI-1430 (and of all the stars within the same Camera/CCD) and we corrected them for systematic effects adopting the PATHOS pipeline presented by \citet{2019MNRAS.490.3806N} (for a detailed description of the pipeline, see also \citealt{2020MNRAS.495.4924N, 2020MNRAS.498.5972N, 2021MNRAS.505.3767N}), based on the PSF-based approach developed by  \citet{2015MNRAS.447.3536N, 2016MNRAS.455.2337N} and also used for {\it Kepler/K2} data (see, e.g., \citealt{2016MNRAS.456.1137L,2016MNRAS.463.1831N}). During the extraction of stellar fluxes, for the calculation of the local background, we took care to exclude all the pixels contaminated by the flux of the neighbor stars by using masks obtained from the PSF models and the Gaia~DR3 catalog.

For the analysis described in the next sections, we removed all the points flagged with the quality parameter \texttt{DQUALITY>0}. The analysed light curve is shown in Fig.~\ref{fig:1}.

\subsection{KELT}
In this work, we used the light curves obtained during the Kilodegree Extremely Little Telescope (KELT) survey (\citealt{2007PASP..119..923P}).  The photometric series contains 2847 points (see Fig.~\ref{fig:1}). Observations were carried out with KELT-North (Winer Observatory, Sonoita, Arizona, USA) between 21 February 2012 and 30 November 2014. 
The same light curves were also analysed by \cite{2018AJ....155...39O}.

%%%
\subsection{STELLA}
TOI-1430 was observed with the WiFSIP imager mounted at the
robotic STELLA telescope (Izana Observatory, Tenerife, Spain; \citealt{2004AN....325..527S}) between March and
December 2020. Observations were carried out in nightly observing blocks
of five exposures in $V$-Johnson ($t_{\rm exp}=4$~s) and five exposures in $I$-Cousin
bands ($t_{\rm exp}=3$~s). The pipeline described by \citet{2015A&A...580A..60M,2018A&A...614A..35M} was used to obtain the differential light curves. The individual exposures per observing
block and filter were averaged. The final $V$ and $I$ light curves contain
126 and 129 points, respectively (see Fig.~\ref{fig:1}).

%%%
\subsection{Asiago Schmidt 67/92 cm Telescope}
We observed TOI-1430 with the Asiago Schmidt 67/92~cm telescope between September and December 2022. Observations were carried out in $u$-Sloan band with an exposure time of 6~s. Light curve was extracted by using the routines described in \citet{2015MNRAS.447.3536N,2016MNRAS.455.2337N}. The light curve contains 646 points (Fig.~\ref{fig:1}).

%%%
\begin{figure*}
  \centering
  \includegraphics[width=0.9\textwidth]{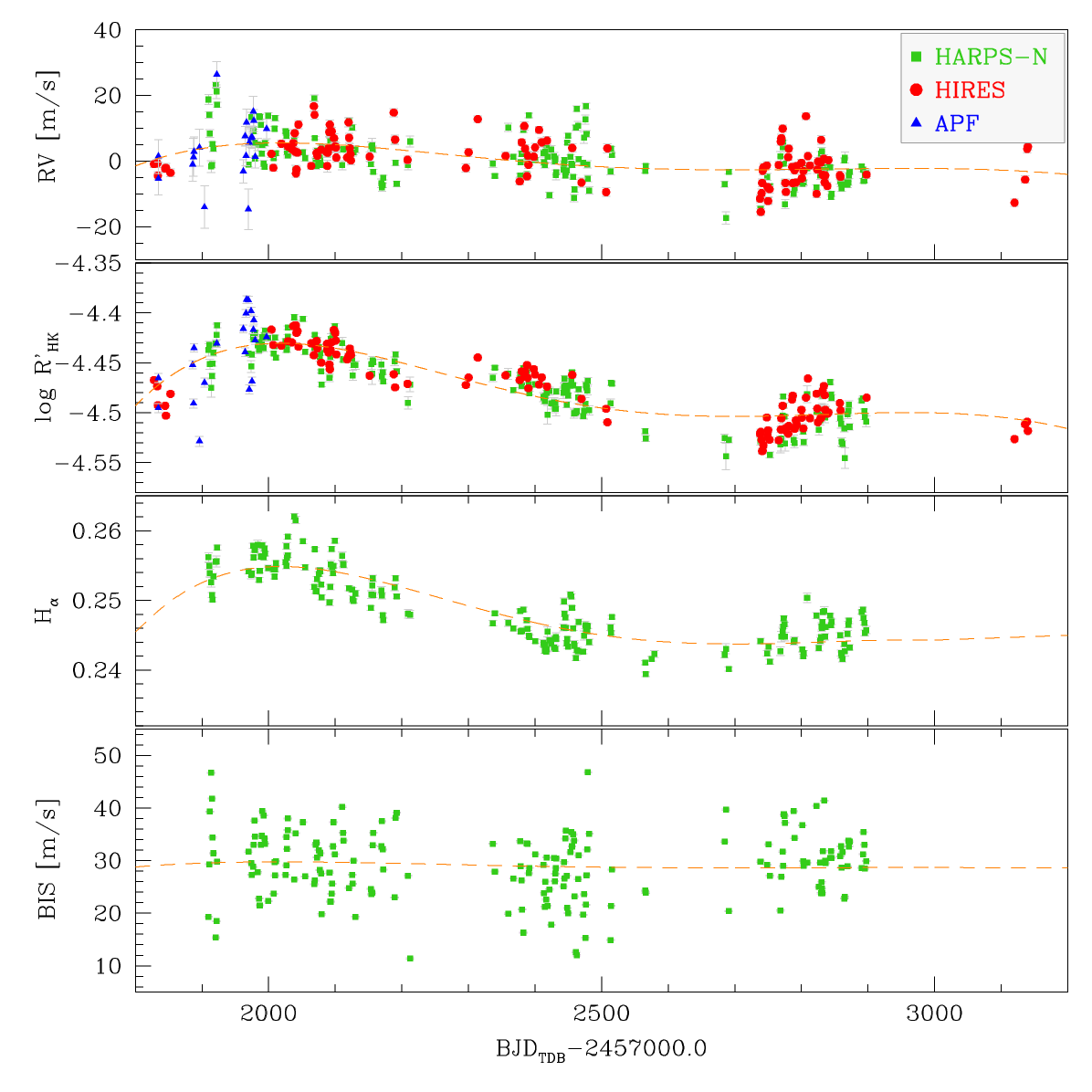} 
  \caption{Spectroscopic time series obtained with HARPS-N (green squares), HIRES (red circles) and APF (blue triangles) and used in this work. From the top to bottom panels: the RV, the \logrhk, the H$\alpha$, and the BIS time series. Dashed orange lines represent the 5th-degree polynomial models used for the modeling of the long trend in the spectroscopic series. The RVs have been reported in the same scale applying to them an offset; the \logrhk series are calibrated in the same Mt. Wilson scale (\citealt{1995ApJ...438..269B}). \label{fig:2}}
\end{figure*}

\subsection{LCO}
In this work, we used public archive images collected at the telescopes of the Las Cumbres Observatory (LCO). In particular, we used 284 frames collected during the night of May 24, 2021 with the 1.0m Telescope at the McDonald Observatory, and the 292 images obtained during the night of August 6, 2022 at the 1.0m Telescope at the Teide Observatory. Both the time series were obtained with the Sinistro imager in PanStarrs $zs$ filter and with an exposure time of 30s. In this work, we used the already pre-reduced images by the BANZAI pipeline (\citealt{2018zndo...1257560M,2021arXiv210614922X}).

We extracted the light curves from the LCO data with the \texttt{STARSKY} code (\citealt{2013A&A...549A..30N}), a software pipeline specifically designed for the TASTE project (\citealt{2011A&A...527A..85N}) to perform differential photometry on defocused images. The size of the circular apertures and the weights assigned to each reference star were automatically set by \texttt{STARSKY} in order to minimise the photometric scatter of the target. More details about the \texttt{STARSKY} pipeline are reported in \citet{2011A&A...532A..24N,2014AN....335..797G,2024A&A...686A..84L}. The time stamps were consistently converted to the BJD$_{\rm TDB}$ standard following \citet{2010PASP..122..935E}.

\subsection{HARPS-N data}

TOI-1430 is one of the stars included in the GAPS Young Objects program (\citealt{2020A&A...638A...5C}) to be monitored with HARPS-N at TNG. Observations of this star were obtained during three different seasons: the first one covered the period 1 March 2020 -- 28 December 2020; observations during the second season were carried out between 2 May 2021 and 17 December 2021; finally, in the third season TOI-1430 was observed between 15 April 2022 and 13 November 2022. We collected a total of 191 spectra with an average \snr$\sim 85.7$ and a dispersion $\sigma$(\snr)$\sim 21.1$. The average airmass and seeing conditions of the observations are $1.3\pm 0.3$ and $1.2 \pm 0.3$~arcsec, respectively. In this work, we excluded the observation obtained during the night of 18 April 2022, because the \snr associated with the spectrum was too low (\snr$\sim 13$).

%{\bf SERENA: Data reduction of HARPS-N data}
We reduced the HARPS-N spectra by using the standard Data Reduction Software (DRS) pipeline running through the YABI workflow interface implemented at the INAF Trieste Observatory\footnote{\url{https://www.ia2.inaf.it/}}.YABI allows the HARPS-N users to run the DRS in offline mode, with the possibility to custom the reduction by changing the binary mask or the RV range to evaluate the CCF (e.g. when the target is a moderate/fast rotator). It also allows to evaluate the \logrhk activity index from the spectra collected for the RV monitoring. The RVs are extracted by using the Cross-Correlation Function (CCF) approach as specified in \cite{2002A&A...388..632P} and references therein. With this method, the observed spectrum is cross-correlated with a binary mask depicting the typical features of a specific spectral type to obtain a normalized weighted mean of the line profiles of the spectrum. In the case of TOI-1430, we used a K5 mask\footnote{Although the K2 mask would be the best choice for this star, it is currently not available in the YABI interface. However, we do not expect this choice to affect the RV precision, which is limited by stellar activity.}. The typical value of the RV uncertainty is 0.8 \ms. From the resulting CCF, we can measure asymmetry indices such as the Bisector Span (BIS), which is useful for tracking line profile changes due to the stellar activity.
We also obtained two additional indices of the chromospheric activity, i.e., \logrhk and H${\alpha}$. The former is directly provided by the DRS according to the methodology reported in  \cite{2011arXiv1107.5325L} and references therein, the latter is extracted by using the ACTIN2 code  \citep{2018JOSS....3..667G,2021A&A...646A..77G}.
The HARPS-N RV, \logrhk, H${\alpha}$, and BIS time series are shown in Fig.~\ref{fig:2} (green squares). % and reported in Table~\ref{tab:E1}.

\subsection{HIRES data}
TOI-1430 was observed with  HIRES \citep[][]{vogt94} on the 10 m Keck I telescope at the W. M. Keck Observatory on Maunakea.HIRES spectral range covers from 0.3 to 1.1 $\mu$m. \cite{2024ApJS..272...32P} published 66 HIRES spectra of TOI-1430 obtained between 10 December 2019 and 21 October 2021 as part of the TESS-Keck Survey \citep[TKS;][]{chontos22}. We subsequently obtained 47 spectra of TOI-1430 between 7 June 2022 and 14 July 2023 under the University of California and Keck observing program 2022B-U084 (PI: N. Batalha), in collaboration with the California Planet Search (CPS). All HIRES observations were obtained with a warm (50\degree C) cell of molecular iodine at the entrance slit \citep{butler96}. The superposition of the iodine absorption lines on the stellar spectrum provides both a fiducial wavelength solution and a precise, observation-specific characterization of the instrument's PSF. The mean airmass of the observations is $1.5 \pm 0.4$. Across all 113 HIRES iodine-in spectra, the observations have a median exposure length of 305 s and a median \snr of 220 per pixel at 5500 \AA. The majority of the observations (99 out of 113) were taken using the B5 decker (3\farcs5 $\times$ 0\farcs861, $R=$ 45,000), while the remaining exposures used the C2 decker (14\arcsec $\times$ 0\farcs861, $R=$ 45,000). The only difference between the two is that the length of the C2 decker enables better sky subtraction during data reduction, though this is not critical for a bright star like TOI-1430 (choosing B5 versus C2 typically makes a difference for targets with $V > 10$ mag). The exposures were taken with a median seeing of 1\farcs2 and all observations were collected with a moon separation of $>30$\degree.

To measure precise RVs using the iodine method, a high-resolution, high-\snr, iodine-free ``template'' spectrum must also be obtained to create a deconvolved stellar spectral template (DSST) of the host star. To compute RVs, we used the iodine-free template of TOI-1430 taken by TKS on 10 December 2019. The template was produced by averaging together two iodine-free exposures taken back-to-back. Each spectrum was collected using the B3 decker (14\arcsec $\times$ 0\farcs574, $R=$ 60,000), which provides slightly higher resolution compared to the B5 or C2 deckers at the cost of throughput. The spectra were obtained at an airmass of 2.6, in seeing of 1\farcs1, and at a moon separation of 91\degree. The two template observations had exposure times of 291 s and 293 s, resulting in \snr of 215  per pixel at 5500 \AA\,for both. Triple-shot exposures of rapidly rotating B stars were taken with the iodine cell in the light path immediately before and after the template was collected to precisely constrain the instrumental PSF. The data collection and reduction followed the methods of the California Planet Search (CPS) as described in \cite{howard10}. 

We note here that while the HIRES template observations were obtained at a relatively high airmass, we do not find any evidence that this affects our RV measurements. Upon visual inspection, the DSST shows no signs of poor-performing deconvolution, which would typically manifest itself in the form of ripple-like features in the spectrum. Furthermore, none of the PFS fitting parameters are correlated with the measured RVs, indicating that the DSST itself is not introducing systematics into the RV measurements. Finally, the median RV precision for the HIRES data is 1.2 \ms, which is typical for CPS observations of stars of similar spectral type.

RVs were determined following the procedures of \cite{howard10}. As part of a forward model, the stellar spectrum was divided into about 700 pieces between $\sim$5000--6000 \AA, with each piece being 2 \AA\ in width. For each piece, the product of the DSST and the Fourier Transform Spectrograph iodine spectrum was convolved with the PSF to match the iodine-in observation. As one of the free parameters, an RV for each piece of spectrum was produced. The pieces were weighted using all observations of the star to produce a single RV for each iodine-in spectrum. Pointwise measurement uncertainties were estimated by taking the weighted standard deviation of the mean of the velocity measured from each of the 2 \AA-wide spectral chunks across all observations.

In addition to computing RVs from each spectrum, we also measured S-value activity indicators, which track Ca II H and K emission strength. The S-values were computed following the methods of \citet{2024ApJ...961...85I}. The HIRES RV and \logrhk\,time series are shown in Fig.~\ref{fig:2} (red circles). % and reported in Table~\ref{tab:E2}.

\subsection{APF literature data}
The TKS obtained 20 iodine-in spectra of TOI-1430 with the Levy spectrograph mounted on the 2.4 m Automated Planet Finder telescope \citep[APF;][]{vogt14} at Lick Observatory between 17 December 2019 and 
%15 August 2022 
27 May 2020
\citep{2024ApJS..272...32P}. The observations had a median exposure time of 1800 s and a median \snr of 54 \perpix at 5500 \AA. The observations were taken with the W decker (1\arcsec $\times$ 3\arcsec, $R =$ 95,000). 

The reduction pipeline used to compute RVs from the APF spectra mirrors the methods of \cite{howard10}. As with the HIRES observations, spectra were obtained with a warm cell of molecular iodine in the light path. We computed the APF RVs using the iodine-free HIRES template. HIRES templates have been shown to serve as effective replacements for APF templates in the CPS Doppler reduction pipeline \citep[e.g.,][]{dai20, macdougall21, lange24}, and provide an efficient alternative to the long exposure that would otherwise be required to achieve similar \snr on an iodine-free APF observation. The APF RV and \logrhk\ time series are shown in Fig.~\ref{fig:2} (blue triangles). % and reported in Table~\ref{tab:E3}.

\subsection{SOPHIE literature data}
Four SOPHIE RVs of TOI-1430 were published by \citet{soubiran2018}, spanning from August 2008 to September 2015. The rms dispersion of these RVs is 5.6 m/s and their mean value is $-$27.331 \kms, close to the 
mean value of HARPS-N RVs.
This supports the lack of large ($\sim$ few tens  m/s) variations over timescales of 15 years.

\begin{figure*}
  \centering
  \includegraphics[width=0.9\textwidth, bb=18 228 585 707]{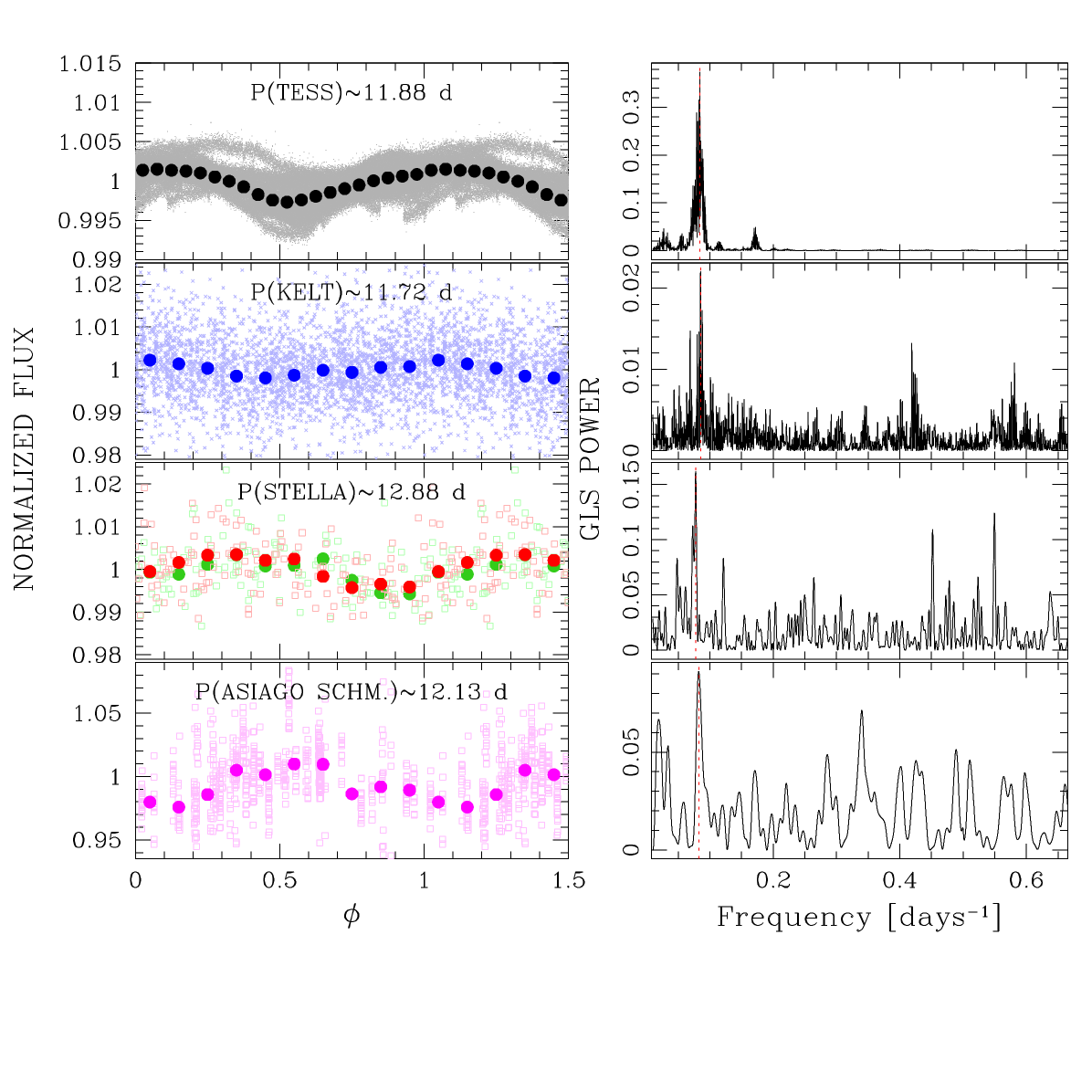} 
  \caption{Analysis of the rotation period of TOI-1430 performed with different photometric series. The left panels show the light curves folded with the period associated with the peak of the corresponding GLS periodogram (right panels). $\Phi$ is the rotational phase of the star.From the upper panel to the bottom, we reported the analysis carried on {\it TESS}, KELT, STELLA, and Asiago Schmidt light curves, respectively.  \label{fig:3}}
\end{figure*}

\section{Stellar parameters}
\label{sec:star}

\subsection{Kinematics and multiplicity}
\label{sec:kin}

TOI-1430 is not associated with any known moving groups and our search,  
based on Gaia DR3 (see \citealt[Appendix B]{2022A&A...664A.163N} for a detailed description) 
did not identify any convincing comoving objects within 5 deg, a separation much larger than any plausible physically associated companion. \citet{oh2017} flagged the K6 star HIP 98381 as a comoving object at 673 arcmin
but updated astrometry from Gaia DR3 and the highly discrepant RVs between the two stars \citep{soubiran2018} rule out a physical association.

TOI-1430's position on the U,V plane is similar to that of the Hyades and well within the kinematic space of nearby young stars proposed by \citet{montes2001}. The W velocity differs by about 12 km/s with respect to that of the Hyades (see Table~\ref{tab:1}). The a posteriori probability distribution function for the kinematical age as computed from the $U, V, W$ velocity components with the method of \citet{Almeida-FernandesRochaPinto18} has a rather sharp maximum centred around $\sim 1.3$~Gyr, supporting the youthness of TOI-1430.

The similar mean RV$\sim -27.3$~km\,s$^{-1}$  derived from SOPHIE (2008--2015) and HARPS-N (2020--2022) argues against the presence of unrecognized massive companions at a separation of a few astronomical units. 
This is also supported by the Renormalised Unit Weight Error (RUWE=0.966) in Gaia DR3 and the lack of significant Gaia-Hipparcos proper motion anomaly \citep[\snr $\sim 1.52$, ][]{kervella2022}.

\subsection{Stellar atmospheric parameters}
\label{sec:stellar_parameters}

For the determination of stellar parameters, we considered the coadded spectrum of the target, which has an \snr$\sim$250 at $\sim$6000\,\AA. Following the same strategy as in \cite{2022A&A...664A.163N}, \cite{2023A&A...672A.126D}, and \cite{2023arXiv231016888M}, we inferred the atmospheric stellar parameters using the approach presented in \cite{2020baratella} and based on equivalent widths (EW) of iron (Fe) and titanium (Ti) lines. The method was designed specifically to analyse young and intermediate-age stars which show high levels of stellar activity. Briefly, it uses a combination of Fe and Ti lines to impose the excitation equilibrium for deriving $T_{\mathrm{eff}}$, and only Ti for the measurement of surface gravity $\log g$ and microturbulence velocity $\xi$ through ionization equilibrium and the zeroing of the trend between \ion{Ti}{i} abundance and $EW/\lambda$, respectively. This way, we overcome possible systematic effects observed on the stellar spectra that affect the derivation of the $\xi$ parameter, and ultimately the iron abundance (see \citealt{2020baratella} for details). 

We adopted the same line list and codes as done in the  works mentioned above. The EWs are measured with ARES v2 \citep{ares}, excluding those lines with EW uncertainties larger than 10$\%$ and with EW$>120$ m$\AA$. We then adopted the 1D-LTE ATLAS9 model atmospheres with  the ODFNEW opacity treatment (\citealt{castellikurucz2003}) and we used the 2019 version of the MOOG code \citep{sneden1973} to derive the stellar parameters. Our final spectroscopic analysis indicates a $T_{\mathrm{eff}}$ of $5075\pm75$ K, a $\log g$ of $4.55\pm0.05$ dex, a microturbulence $\xi$ equal to $0.79\pm0.07$ km/s, and an iron abundance of [Fe/H]$=-0.02 \pm 0.03$\,dex and a titanium abundance[Ti/H]$=0.05 \pm 0.06$\,dex (see Table\,\ref{tab:1}). The uncertainties on iron and titanium abundances consider both the EWs scatter and the stellar parameters error contributions.

The spectroscopic results confirm with excellent agreement the values obtained using various relations exploiting $Gaia$ DR3 photometry and parallax \citep{2023A&A...674A...1G} and 2MASS photometry \citep{2mass}. Using the code \texttt{colte} \citep{2021casagrande}, we derived the photometric values of effective temperature in 21 different color indexes, which varies from a minimum of $5048\pm61$\,K (in $G_{RP} - K$) up to a maximum of $5175\pm 103$\,K (in $G_{RP}-J$), with a mean photometric $T_{\mathrm{eff}}$ of $5104\pm60$ K. Given the small distance of $\sim$42\,pc from the Sun, we assumed no reddening to estimate a surface gravity of $4.58 \pm 0.04$\,dex from the $Gaia$ parallax and a microturbulence $\xi$ of $0.79 \pm 0.04$ km/s from the relation by \cite{2016dutra}, again in excellent agreement with our spectroscopic values. Very close stellar parameters were also recently derived by \cite{orellmiquel2023} through CARMENES spectra. 

\subsection{Lithium detection}
\label{sec:lithium}

We looked for the lithium line at 6707.8\,\AA\,in the coadded spectrum and, unlike \cite{orellmiquel2023}, we detected a small feature, with a mean equivalent width of $EW_{\rm Li}=0.6\pm0.2$\,m\AA. The probable disagreement with this recent work could be indeed due to the lower resolution and $S/N$ of the coadded spectrum used by those authors, which set an upper limit of 3\,m\AA\, for the Li EW. 

However, from our stellar parameters and EW measurement, we could only obtain an upper limit of $<0.1$\,dex on the lithium abundance with non-LTE corrections by \cite{Lindetal2009}. This upper limit seems to be consistent with clusters not younger than the Hyades or Praesepe ($\sim$650\,Myr), as also claimed by \cite{orellmiquel2023}. Our age estimate will be discussed in Sect.\,\ref{sec:age}.

\subsection{Projected rotational velocity}
\label{sec:vsini}

As done in other previous works (see, e.g., \citealt{2023A&A...672A.126D, 2023arXiv231016888M}), we synthesized the absorption lines around three spectral regions (namely, 5400, 6200, and 6700\,\AA) to obtain an estimate of the projected rotational velocity ($v\sin i$), after assuming the stellar parameters derived in Sect.\,\ref{sec:stellar_parameters} and fixing the macroturbulence velocity to 1.9~\kms\, from \cite{Breweretal2016}, the limb-darkening coefficient, and the instrumental resolution. We considered the same MOOG code and ATLAS9 grids of model atmospheres as done above and obtained a $v\sin i=1.9\pm0.6$\,km/s (see Table\,\ref{tab:1}), which is slightly below the resolution of HARPS-N, suggesting a very slow stellar rotation unless the star is observed nearly pole-on. Similarly, \cite{orellmiquel2023} provided a projected rotational velocity of $<2.9$\,km/s, which is at the limit of the resolution of their spectra.

\subsection{Stellar rotation}
\label{sec:rotation}

\subsubsection{Stellar rotation from photometric series}
\label{sec:prot}

We used the light curves described in Sect.~\ref{sec:obs} to estimate the rotation period of TOI-1430. For each photometric series, we extracted the Generalized Lomb-Scargle (GLS) periodogram (\citealt{2009A&A...496..577Z}), and we identified the period associated with the most powerful peak. Results are shown in Fig.~\ref{fig:3}: the GLS periodogram of the entire \tess light curve shows a very important peak at $P_{\rm rot} =11.88 \pm 0.05$~d  (first row from the top of Fig.~\ref{fig:3}). Error on the rotation period is calculated by locally fitting a Gaussian function to the peak in the periodogram, and considering as period's error the standard deviation. We also measured the rotation period in each of the four \tess seasons, to test any sign of differential rotation: we found  $P_{\rm rot} =11.7 \pm 0.6$~d, $P_{\rm rot} =12.9 \pm 1.8$~d,  $P_{\rm rot} =11.8 \pm 0.7$~d, and $P_{\rm rot} =11.9 \pm 0.9$~d for observations obtained in the years 2019, 2021, 2023, and 2024, respectively. Within the errors, all the measured rotation periods agree and no clear evidence of differential rotation is detectable from \tess data.

The second row of Fig.~\ref{fig:3} shows the phased KELT light curve and the GLS periodogram that peaks at $P_{\rm rot}=11.72\pm 0.06$~d with an analytical False Alarm Probability FAP$\sim 10^{-11}$. The sampling and the number of points in the STELLA light curves are not ideal for identifying the rotation period of TOI-1430; indeed, in both the photometric bands, we identified a strong peak in the STELLA GLS periodograms at $P_{\rm rot} = 12.9 \pm 0.3$~d, with a FAP$\sim 0.17$. The Asiago Schmidt light curve was not ideal for our purpose either, because of its low photometric precision; however, we identified a peak in the GLS periodogram at $P_{\rm rot}=12.13 \pm 0.7$~d, associated with a FAP$\sim 0.09$.

\begin{figure}
  \centering
  \includegraphics[width=0.45\textwidth, bb=39 204 409 710]{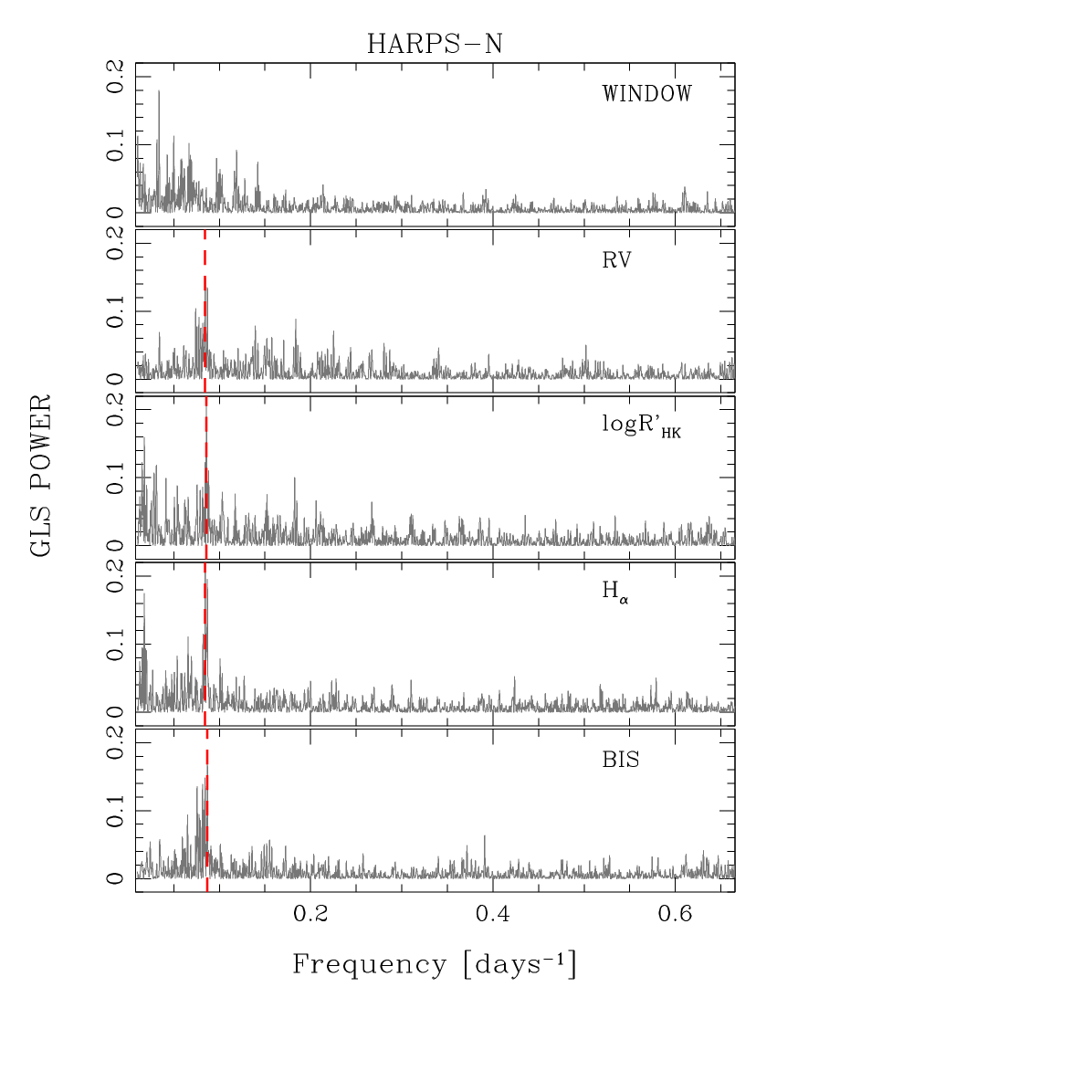} \\
  \includegraphics[width=0.45\textwidth, bb=39 371 409 710]{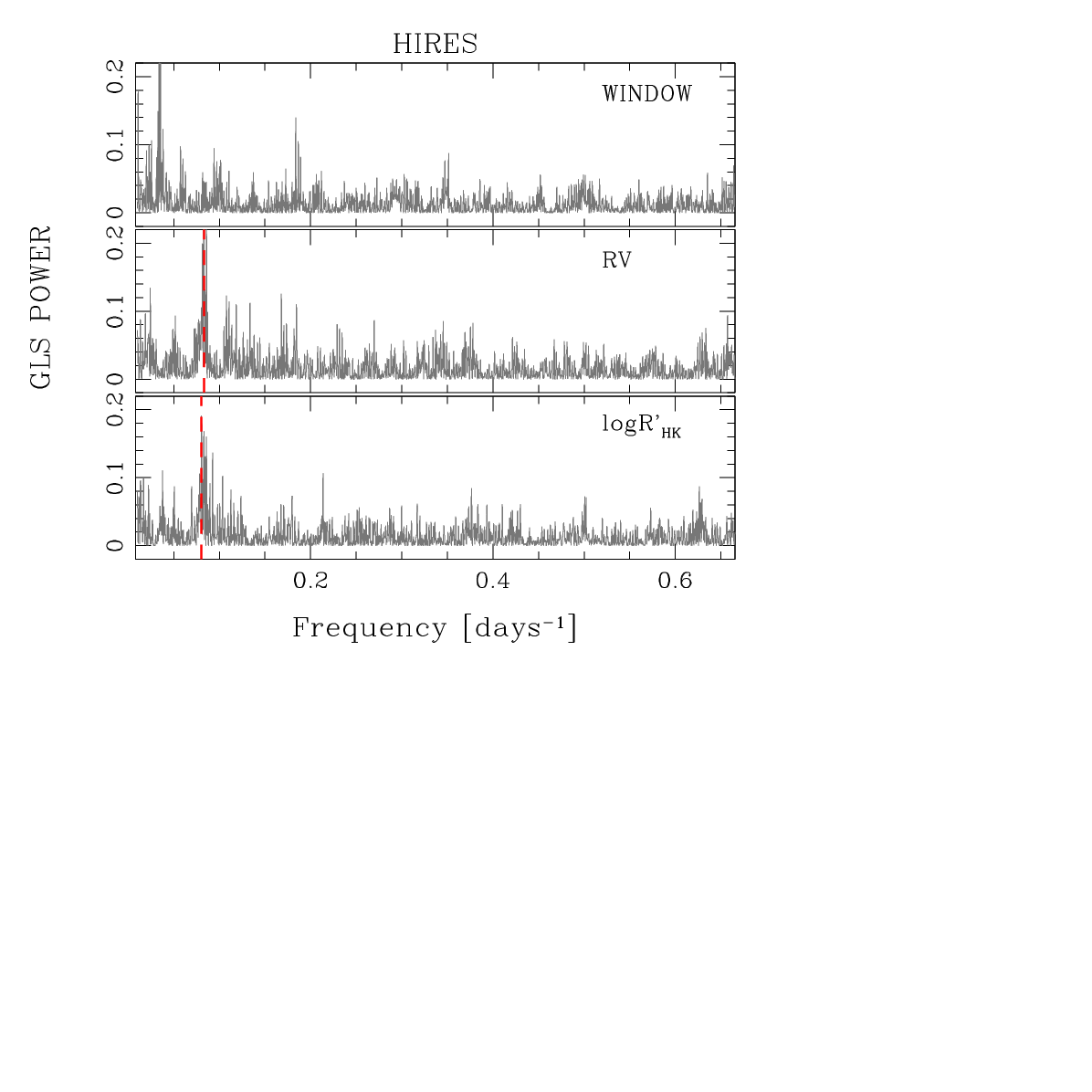} 
  \caption{Analysis of the rotation period of TOI-1430 carried on the HARPS-N (five upper panels) and HIRES (three bottom panels) spectroscopic series shown in Fig.~\ref{fig:2} (after the subtraction of the long trend). Panels show the GLS periodograms extracted from the RV, \logrhk, H${\alpha}$, and BIS time series. Red dashed lines indicate the frequency of the peak. The window function periodograms are also reported. \label{fig:4}}
\end{figure}

\subsubsection{Stellar rotation from spectroscopic series}
\label{sec:rvfreq}
We calculated the rotation period of the star from the HARPS-N and HIRES spectroscopic time series. As shown in Fig.~\ref{fig:2}, the \logrhk, H$\alpha$, and BIS series show a long, sinusoidal-trend likely due to a long-period cycle of activity ($P>1000$~d). To estimate the rotation period of the star we first removed this long trend by fitting a 5th-degree polynomial function to the time series characterized by coefficients  shared to all the datasets and a multiplying factor associated with each time series (see, e.g., \citealt{2021MNRAS.505..830C}). The 5th-degree polynomial models are shown in Fig.~\ref{fig:2} (orange dashed line).

We extracted the GLS periodograms of all the time series (after the subtraction of the long trend) illustrated in Fig.~\ref{fig:2}, first by individually analyzing the time series obtained with the individual instruments, and then by considering all the points at once. Periodograms are shown in Fig.~\ref{fig:4}: from the HARPS-N RV, \logrhk, H$\alpha$, and BIS series we observed a peak in the periodogram at  $P_{\rm rot}=11.85 \pm 0.12$~d (FAP$\sim 10^{-4}$), $P_{\rm rot}=11.67 \pm 0.11$~d (FAP$\sim 10^{-7}$), $P_{\rm rot}=11.85 \pm 0.09$~d (FAP$\sim 10^{-8}$), and $P_{\rm rot}=11.51 \pm 0.17$~d (FAP$\sim 10^{-5}$), respectively (see the five topper panels of Fig.~\ref{fig:4}). The HIRES RV and \logrhk
series periodograms show a peak at $P_{\rm rot}=12.05 \pm 0.11$~d (FAP$\sim 2 \times 10^{-3}$) and $P_{\rm rot}=12.45 \pm 0.26$~d (FAP$\sim 7 \times 10^{-3}$) (bottom panels of Fig.~\ref{fig:4}). Analysing the complete RV  time series we obtained $P_{\rm rot}=11.85 \pm 0.09$~d(FAP$\sim 3 \times 10^{-7}$).

Averaging all the $P_{\rm rot}$ estimated from photometry and spectroscopy, we obtained  a weighted average rotation period $P_{\rm rot} = 11.9 \pm 0.3$~d.

\subsection{Coronal and chromospheric activity}
\label{sec:activity}

The chromospheric activity S-index  was measured on the spectra as described in Sect. \ref{sec:obs}.
The median value from the HARPS-N time series is $\sim 0.480$, that is \logrhk$\sim -4.48$, derived
adopting (B$-$V)$=0.888$ from \teff = 5090 K, as the observed B-V color has large uncertainty.

TOI-1430 has an X-ray detection (1RXS J200226.3+532229) within 15 arcsec
from the optical position on ROSAT Faint Source Catalog \citep{rosatfaint}. 
The corresponding X-ray luminosity, derived following \cite{hunsch1999}, is $L_{\rm X} = 2.44 \times 10^{28}$\,erg\,s$^{-1}$, and the ratio $ \log L_{\rm X}/L_{\rm bol} \simeq -4.76$.
More recently, the object was the target of XMM-Newton observations \citep{zhang2023}. The analysis of this observation by \cite{orellmiquel2023} yielded $L_{\rm X} = 1.89 \pm 0.07 \times 10^{28}$ erg s$^{-1}$ in the band 5--100\,\AA, corresponding to
$\log L_{\rm X}/L_{\rm bol}\simeq -4.86$.
Both \logrhk and  $L_{\rm X}/L_{\rm bol}$ are below the mean value for Hyades stars of similar color, but in agreement with the predictions of these emission levels based on the stellar age and rotational period (\citealt{Pizz03, 2008ApJ...687.1264M}). In particular, the X-ray luminosity is below  the $1\sigma$ lower boundary of the dispersion for the members of the Hyades cluster observed in X-rays (\citealt{Penz08a}).

\subsection{System age}
\label{sec:age}

The rotation period, lithium abundance, coronal and chromospheric emission
are all consistent with an age similar and probably slightly older than the Hyades. 
The kinematic properties are also fully compatible with such an age value but
there are no known companions or comoving objects to further refine the age determination. 
Therefore, considering the measurement uncertainties and the scatter in the indicators for coeval objects, we adopt an age of 700 $\pm$ 150 Myr.
Our adopted age is significantly older than that adopted by \cite{zhang2023} (165$\pm$30 Myr), as they obtained a rotation period which is roughly half of our determination and consider only gyrochronology for dating the star.
Our age determination is instead in good agreement with the findings by \cite{orellmiquel2023} and \cite{2024arXiv240900675F}
(600-800 Myr), based on a combination of various methods, similar to our adopted procedure\footnote{Our slightly larger uncertainty is based on a conservative choice considering the variability due to the prominent activity cycle and the possible effects on the stellar inclination.}. As noticed also by \cite{orellmiquel2023}, the expected lithium
abundance for the age by \cite{zhang2023} is highly discrepant with the observational value, ruling out such a young age.

\subsection{Stellar radius and mass}
\label{sec:massradius}

Stellar radius was obtained from Stefan-Boltzmann law, using the
adopted \teff, the observed $V$ mag from Hipparcos, and the bolometric correction from \cite{pecaut2013} \footnote{Updated values at: {\tiny \url{https://www.pas.rochester.edu/~emamajek/}}\\ {\tiny \url{EEM_dwarf_UBVIJHK_colors_Teff.txt}}}.
The stellar luminosity results in 0.360$\pm$0.026  \lsun~ and the stellar radius 0.772$\pm$0.029 \rsun.

Stellar mass was obtained using the PARAM web interface \citep{param}\footnote{\url{http://stev.oapd.inaf.it/cgi-bin/param_1.3}} which interpolated the models by \cite{bressan2012},
in a Bayesian framework to identify the most probable solution 
considering the observational errors and the lifetime of the various evolutionary phases. 
Only the age range derived by indirect methods was allowed in the retrieval, as in \citet{desidera2015}, considering the evolutionary timescales of an early K dwarf.
The stellar mass results 0.849$\pm$0.009 \msun.The small uncertainty is only due to the contribution of the "internal" error of the fitting procedure, while systematic uncertainties of the stellar models are not included; for this reason we increased the error on the prior on stellar density in next analyses. 

The mass and radius agree to be better than one sigma with \cite{orellmiquel2023}, \cite{zhang2023}, and TESS Input Catalog (TIC, \citealt{2018AJ....156..102S,2019AJ....158..138S}).

%%%%%%%%%%%%%%%%%%%%%%%%%%%%%%%%%%%%%%%%%%%%%%%%%%%%%%%%%%%%%%%%%%%%%%%%%%
\begin{table}[!htb]
  \centering
  \caption{Stellar properties of TOI-1430}
\begin{tabular}{l c c}
\hline
\hline
Parameter     &  TOI-1430 & Reference \\
\hline
\multicolumn{3}{c}{\textit{Other Target Identifiers}} \\
TIC           &     293954617       & (1)\\
2MASS         &   J20022741+5322365 & (2)\\
Gaia~DR3      & 2089570519439964416 & (3)\\
\hline
\multicolumn{3}{c}{\textit{Astrometric information}} \\
$\alpha$(J2016.0)~[deg.]  & 300.61549185196   &   (3)    \\
$\delta$(J2016.0)~[deg.]  & $+$53.37746192023 &   (3)     \\ 
$\mu_{\alpha^{\star}}$~[mas\,yr$^{-1}$] & $165.050\pm0.015$   &     (3)         \\
$\mu_{\delta}$~[mas\,yr$^{-1}$] &  $145.170 \pm 0.015$     &  (3)    \\
Parallax~[mas]  &  $24.2456 \pm 0.0121$ &    (3) \\
Distance~[pc] &   $41.21\pm 0.02$ & (4) \\
RUWE    &    0.966  &  (3) \\
%%%%%%%%%%%%%%%%%%%%%%%%%%%%%%%%%%%%%%%%%%%%%%%%%%%%%%%%%%%%%%%%%%%%%%%%%%%%%%%%%%%%%%%%%%%%%%%%%%%%%%%%%%
\hline
\multicolumn{3}{c}{\textit{Photometric information}} \\
$T$~[mag]    & $8.388  \pm      0.006$  & (1)            \\
$G$~[mag]    & $8.953  \pm      0.003$ & (3)        \\
$G_{\rm BP}$~[mag]    & $9.407  \pm      0.003$ & (3)        \\
$G_{\rm RP}$~[mag]    & $8.335  \pm      0.004$ & (3)        \\
%$B$~[mag]    & $ \pm    $   &    (??)          \\
$V$~[mag]    & $ 9.19\pm0.03 $    &   (7) \\
$J$~[mag]    & $7.646\pm 0.037$     & (2)         \\
$H$~[mag]    & $7.224\pm 0.034$ & (2)           \\
$K$~[mag]    & $7.084 \pm 0.016$             & (2)               \\
$W_1$~[mag]  & $6.995 \pm       0.050$  & (5)    \\
$W_2$~[mag]  & $7.095 \pm       0.020$  & (5)              \\
$W_3$~[mag]  & $7.075 \pm       0.016 $ & (5)      \\
$W_4$~[mag]  & $7.160 \pm       0.070$  & (5)        \\
\hline
\multicolumn{3}{c}{\it Fundamental parameters}   \\
RV [\kms]      & $ -27.33\pm 0.13 $    & (3) \\
RV [\kms]      & $ -27.246\pm 0.001 $ & (6)\\
U\ [\kms]      & $-41.728\pm0.023 $    & (8)\\	 	
V\ [\kms]      & $-21.764\pm0.375 $    & (8) \\
W\ [\kms]      & $-18.950\pm0.057 $    & (8)\\
\teff\ (spec.) [K]     &   $5075 \pm 75  $  & Sect. \ref{sec:stellar_parameters}\\
\teff\ (phot.) [K]       &   $5104 \pm 60 $  & Sect. \ref{sec:stellar_parameters}\\  
\logg\ [cgs]                   &   $ 4.55\pm0.05$ & Sect. \ref{sec:stellar_parameters} \\
\feh\ [dex]                    &  $ -0.02\pm0.03$ & Sect. \ref{sec:stellar_parameters} \\
\tih\ [dex]                    &  $0.05\pm0.06$   & Sect. \ref{sec:stellar_parameters} \\
\lstar\ [\lsun]                &  $0.360\pm$0.026 & Sect. \ref{sec:massradius} \\   
\mstar\ [$\mathrm{M_\odot}$]   &  $0.849\pm0.009$ &  Sect. \ref{sec:massradius} \\ %% 
\rstar\ [$\mathrm{R_\odot}$]   &  $0.772\pm0.029$ & Sect. \ref{sec:massradius} \\ %%   
Age [Myr]                      &  $700\pm150$  & Sect. \ref{sec:age} \\ 
$E(B-V)$ [mag]                 &  $0.0022^{+0.0141}_{-0.0022}$  & (9) \\
\vsini\ [\kms]                 & $ 1.9\pm0.6 $ & Sect. \ref{sec:vsini} \\
\prot\ [d]                     & $ 11.9 \pm 0.3 $ & Sect. \ref{sec:rotation} \\
S-index (MW)                    & $ 0.480 \pm 0.006 $  & Sect. \ref{sec:activity} \\
\logrhk\                       & $-4.480\pm 0.003 $  & Sect. \ref{sec:activity}   \\
$\log L_{X}$ [erg s$^{-1}$]    & $ (1.89\pm0.07)\times 10^{28} $  & (10) \\   
$\log L_{X}/L_{bol}$           & $ -4.86\pm0.03 $   & Sect. \ref{sec:activity} \\
$EW_{\rm Li}$ [m\AA]            & $ 0.6 \pm0.2  $ & Sect. \ref{sec:lithium}  \\
A(Li)$_{\rm NLTE}$             & $<0.1$ & Sect. \ref{sec:lithium}  \\
\hline
\end{tabular}
\tablefoot{$^{(1)}$~\tess Input Catalogue v8.2 (\citealt{2018AJ....156..102S}); $^{(2)}$~Two Micron All Sky Survey (2MASS, \citealt{2006AJ....131.1163S}); $^{(3)}$~Gaia~DR3 (\citealt{2023A&A...674A...1G}); $^{(4)}$~\citet{2021AJ....161..147B}; $^{(5)}$~Wide-field Infrared Survey Explorer (WISE, \citealt{2010AJ....140.1868W}); $^{(6)}$~This work; $^{(7)}$~Hipparcos; $^{(8)}$~\citet{smart2021}; $^{(9)}$~\citet{2021A&A...653A..98M}; $^{(10)}$~\citet{orellmiquel2023}}
  \label{tab:1}
\end{table}
%%%%%%%%%%%%%%%%%%%%%%%%%%%%%%%%%%%%%%%%%%%%%%%%%%%%%%%%%%%%%%%%%%%%%%%%%%

\begin{figure*}
  \centering
  \includegraphics[width=0.95\textwidth,bb=12 239 585 712]{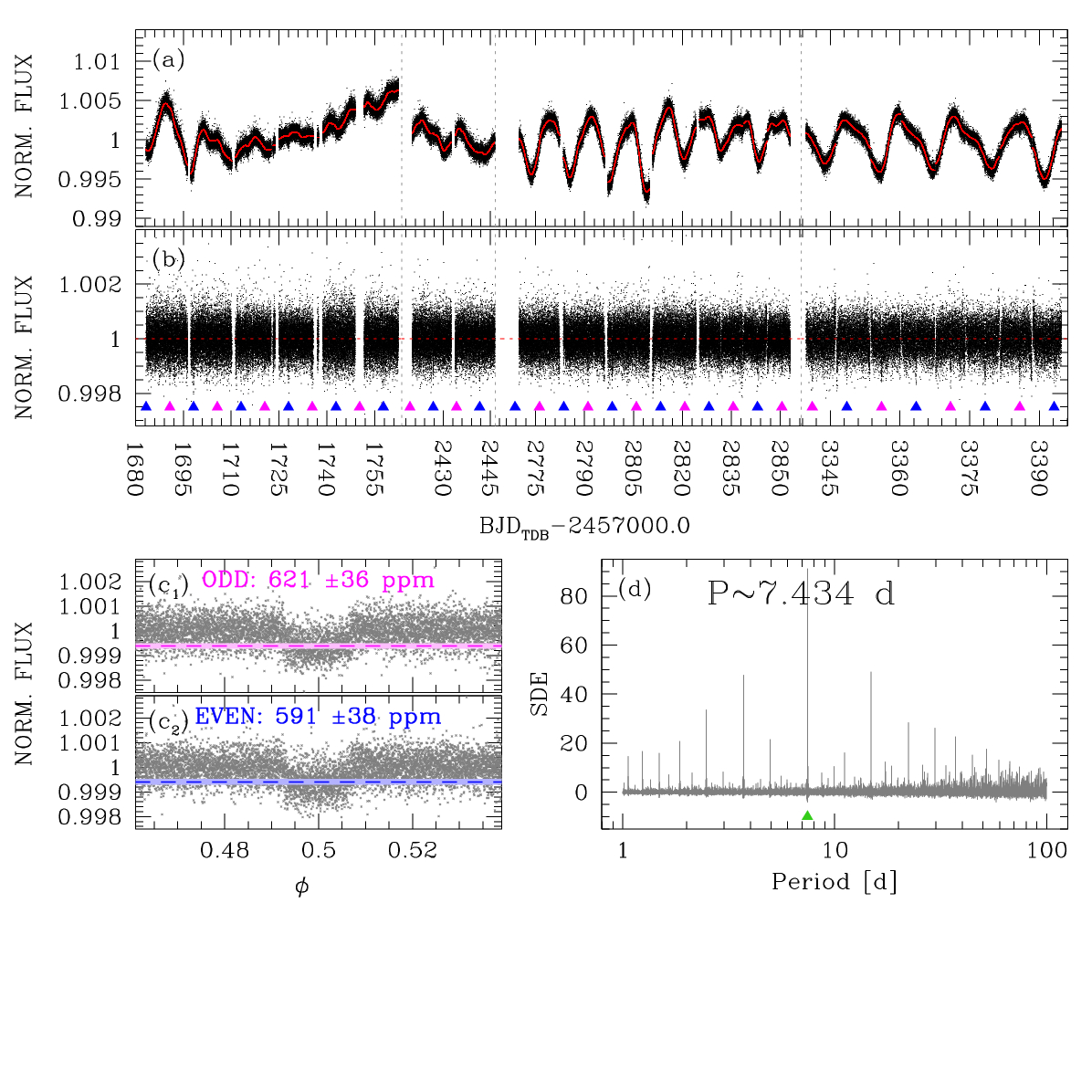}
  \caption{Detection and vetting of the candidate exoplanet TOI-1430.01. Panel (a) shows the TOI-1430's stacked light curve and the detrending model (in red). Panel (b) shows the flattened light curve and the position of the odd (magenta) and even (blue) transits. Panels (c) are a comparison between the mean depths of the odd and even transits; the dashed lines are the mean depths while the shaded colored regions represent the 3$\sigma$ confidence interval. Panel (d) is the TLS periodogram: the period of the peak is indicated with a green triangle.  \label{fig:5b}}
\end{figure*}

\begin{figure}
  \centering
  \includegraphics[width=0.5\textwidth]{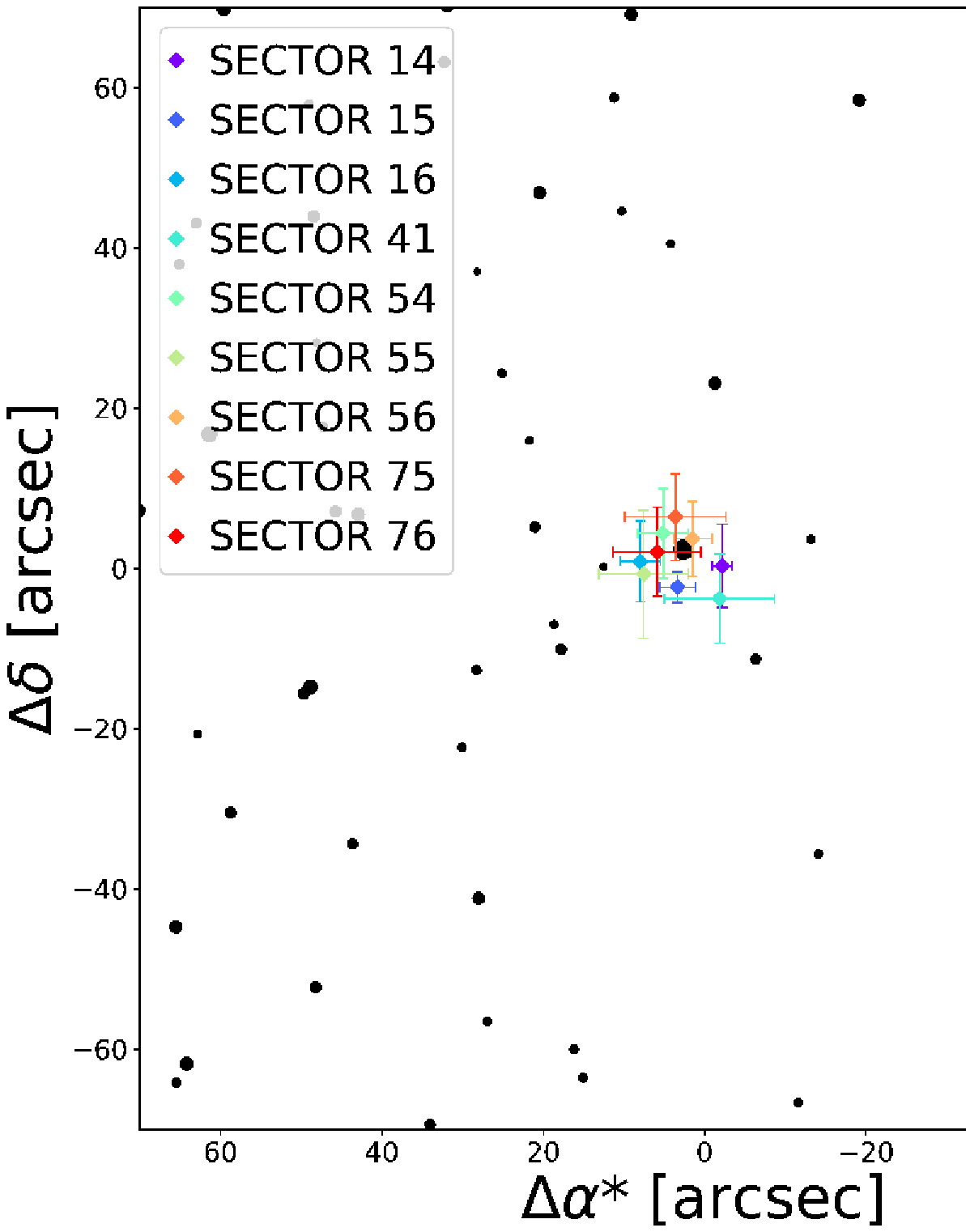}
  \caption{The in-/out-of-transit centroid analysis for TOI-1430.01. TOI-1430 is centered in (0,0). Black filled circles are the neighbor stars from Gaia~DR3; dimension of the points are proportional to the apparent luminosity of the stars.  Each colored point corresponds to the average of the centroids in each sector. Within the errors, centroids match with the position of TOI-1430. \label{fig:5a}}
\end{figure}

\section{Detection and vetting of TOI-1430~b}
\label{sec:detvet}
A transiting object in the light curve of  HD\,235088 was detected by the Science Processing Operations Center (SPOC) pipeline (\citealt{2016SPIE.9913E..3EJ}) in November 2019, and was assigned to the identifier TOI-1430.01. Through a statistical analysis,
\citet{2021AJ....161...24G} found a False Positive Probability (FPP) of $\sim 0.03$ for the transit signals, labeling this object as a likely planet. In this section, we provide further proofs of the planetary nature of this object.

First, we detrended the light curve of TOI-1430 by using the Tukey's biweight M-estimator implemented in \texttt{w\=otan} (\citealt{2019AJ....158..143H}), with a window length of 1~day. To obtain the model, we masked the likely transit events by using the ephemeris given by the SPOC pipeline. The detrending model is shown in red in panel (a) of Fig.~\ref{fig:5b}. We extracted the Transit Least Squares (TLS) periodogram (\citealt{2019A&A...623A..39H}) of the detrended light curve (panel (b) of Fig.~\ref{fig:5b}), looking for transit signals with periods between 1 and 100 days: the periodogram illustrated in panel (d) of Fig.~\ref{fig:5b} shows a peak at $P \sim 7.434$~d with a Signal Detection Efficiency (SDE) of $\sim 91.1$, and a stacked \snr$\sim 21.3$. Odd and even transits are reported with magenta and blue triangles in panel (b), respectively. We checked if the depths of the odd/even folded transits agree; the results are reported in panels (c). The significance (in standard deviations) between odd and even transit depths is $\sim 0.4$, i.e. odd and even transits have on average the same depth within $\sim 0.4 \sigma$. We also tested the in-/out-of-transit centroid associated with the signal of the transiting object as described in \citet{2023A&A...675A.158D} and \citet{2023A&A...674A.132M}. The result is shown in Fig.~\ref{fig:5a}: all the centroids, each one calculated as the average of the in-/out-of-transit centroids of each sector, agree within the errors with the position of TOI-1430. Finally, we also checked if there is any correlation between transits and (X,Y) positions of the star on the CCD and between transits' depths and different photometric apertures. The transiting object passed positively all the vetting tests and we confirmed its planetary nature (hereafter TOI-1430~b).

We masked the transit events of the planet TOI-1430~b from the detrended light curve, and we extracted three TLS periodograms to look for other transit features. In the first TLS periodogram, we sampled all the periods between 0.45~d and 1.0~d, in the second periodogram all the periods between 1.0~d and 10.0~d, and finally in the third periodogram we looked for transit signals with periods between 10.0 and 100.0~d. We did not obtain any strong peaks in the periodograms. Indeed, in the first case, we measured a peak at period P$\sim 0.85$~d with an SDE$\sim 0.09$ and \snr$\sim 1$; in the second case the peak in the periodogram corresponds to $P\sim 7.04$~d (SDE$\sim 12$, \snr$\sim 3.5$); finally in the third case we obtained a peak in the TLS periodogram at $P\sim 84.6$~d with an SDE$\sim 11$ and \snr$\sim 8.5$. Inspecting the folded light curve with the recovered periods, no evidence of transits is visible.

\section{The planetary system of TOI-1430}
\label{sec:planet}

In this section, we present the modeling of the \tess light curve and of the RV series in order to characterize TOI-1430~b. We used the publicly available code \texttt{PyORBIT}\footnote{\url{https://github.com/LucaMalavolta/PyORBIT}} (\citealt{2016ascl.soft12008M,2016A&A...588A.118M,2018AJ....155..107M}), a versatile public available software for the characterization of planetary systems that is able to model stellar activity, planetary signals, and instrumental systematics in the light curves, RV and activity index series, also adopting Gaussian processes regression with a variety of kernels. This code has been successfully adopted in many works under the GAPS Young Objects program (see, e.g., \citealt{2023arXiv231016888M, 2024A&A...682A.135C}). It firstly derives the starting values of the parameters used in the fit of the models by executing the algorithm \texttt{PyDE}\footnote{\url{https://github.com/hpparvi/PyDE}} (\citealt{1997..............S}), a global optimization code ideal to obtain the starting conditions. Secondly, these starting parameters are then used to initialize the affine invariant Markov chain Monte Carlo (MCMC) sampler \texttt{emcee} (\citealt{2013PASP..125..306F}). 

We analysed the detrended light curve shown in panel (b) Fig.~\ref{fig:5b} and the spectroscopic series shown in Fig.~\ref{fig:2} with \texttt{PyORBIT} in order to retrieve the planet's properties. We adopted as Gaussian priors the stellar parameters (stellar mass and radius, $M_{\star}$ and $R_{\star}$, respectively)  reported in Table~\ref{tab:1} as input; however, to take into account that the error on the stellar mass is underestimated, we adopted a broader prior on the stellar density.

%photometric part

We modelled the transits by using the package \texttt{batman}\footnote{\url{https://github.com/lkreidberg/batman}} (\citealt{2015PASP..127.1161K}) in order to retrieve information on the central time of the first transit ($T_{\rm 0,b}$), the orbital period ($P_{\rm b}$), the impact parameter ($b_{\rm b}$), the planetary-to-stellar-radius ratio ($R_{\rm P,b}/R_{\star}$), the duration of the transit ($T_{\rm 14,b}$, calculated as in \citealt{2010exop.book...55W}) the stellar density ($\rho_{\star}$), and the photometric jitter term ($\sigma_{\rm jit,phot}$) to be added in quadrature to the errors of the photometry to take into account any systematic or physical residual effects we did not corrected. We took into consideration the contamination of the neighbour stars that fall inside the photometric aperture, and we calculated the dilution factor ($df = 0.0073 \pm 0.0002$) that we used as Gaussian prior in the modelling. Also, by adopting the [Fe/H], $\log{g}$ and $T_{\rm eff}$ shown in Table~\ref{tab:1}, we calculated the coefficients of limb darkening (LD) by using 
the routine \texttt{PyLDTk}\footnote{\url{https://github.com/hpparvi/ldtk}}
(\citealt{Husser2013, Parviainen2015}), increasing the errors of a factor 10 in the priors in order to avoid deviations between the measured and predicted LD coefficients.
During the modeling, we adopted the LD formalism by \citet{2013MNRAS.435.2152K} and we took into account  the 2-minute cadence of the light curve (\citealt{2010MNRAS.408.1758K}). Priors for the planet transit modeling, chosen on the basis of the analysis performed in the previous sections, are reported in Table~\ref{tab:3}.

%spectroscopic part

We used \texttt{PyORBIT} to simultaneously model the stellar activity and the planetary signal in the RV time series and the stellar activity signal in the \logrhk, H$\alpha$, and BIS series.
Stellar activity was modelled through the  multidimensional Gaussian process (GP) framework. In this work, we made use of the \texttt{tinyGP}\footnote{\url{https://github.com/dfm/tinygp}}  routines for the GP regression, that works also on GPUs\footnote{In this work we used NVIDIA RTX A5500 GPU.}. 
The multidimensional GP is implemented in \texttt{PyORBIT} following the prescription of the work by \citet{2015MNRAS.452.2269R} (see \citealt{2022MNRAS.509..866B} for details). We modeled the RV, \logrhk, BIS, and H$\alpha$ spectroscopic time series. The four-dimensional GP model is described by the following equations:

\begin{align}
 \Delta{\rm RV} &= V_{\rm c} G(t) + V_{\rm r} \dot{G(t)} \\
 \log{R'_{\rm hk}} &= L1_{\rm c} G(t) \\
 {\rm BIS} &=  B_{\rm c} G(t) + B_{\rm r} \dot{G(t)} \\
 {\rm H}\alpha &= L2_{\rm c} G(t) 
\end{align}

\noindent
where $G(t)$ and $\dot{G}(t)$ are the underlying GP and its derivative, with the quasi-periodic kernel described by:

\begin{equation}
 \gamma(t_i, t_j) = \exp{ \Biggl\{ - \frac{\sin^2{ \left[\pi (t_i -t_j)/ P_{\rm rot} \right]  }  }{2 w^2} - \frac{(t_i-t_j)^2}{2 P^2_{\rm dec}}  \Biggl\} }
\end{equation}

\noindent
where $P_{\rm rot}$ is the GP period equivalent to the stellar rotation period, $w$ is the coherence scale, and $P_{\rm dec}$ is associated with the decay time scale of the active regions. The constants $V_{\rm c}$, $V_{\rm r}$, $L1_{\rm c}$, $B_{\rm c}$, $B_{\rm r}$, and $L2_{\rm c}$ are free parameters that link the individual time series to the GP and its derivative (\citealt{2022A&A...664A.163N,2023MNRAS.522.3458B}). To take into account the different data reduction and instrument properties, we treated observables coming from different observatories as independent datasets, with their own offset and jitter parameters
We adopted a Gaussian prior on the stellar rotation equal $P_{\rm rot} = 11.9 \pm 0.3$~d, on the basis of the results obtained in Sect.~\ref{sec:rotation}.  We explored the period and the semi-amplitude of the planet in  linear space, and we imposed the eccentricity equal to 0. 
The priors adopted are shown in Table~\ref{tab:3}. 

\begin{figure*}
  \centering
  \includegraphics[width=0.75\textwidth,bb=25 175 576 705]{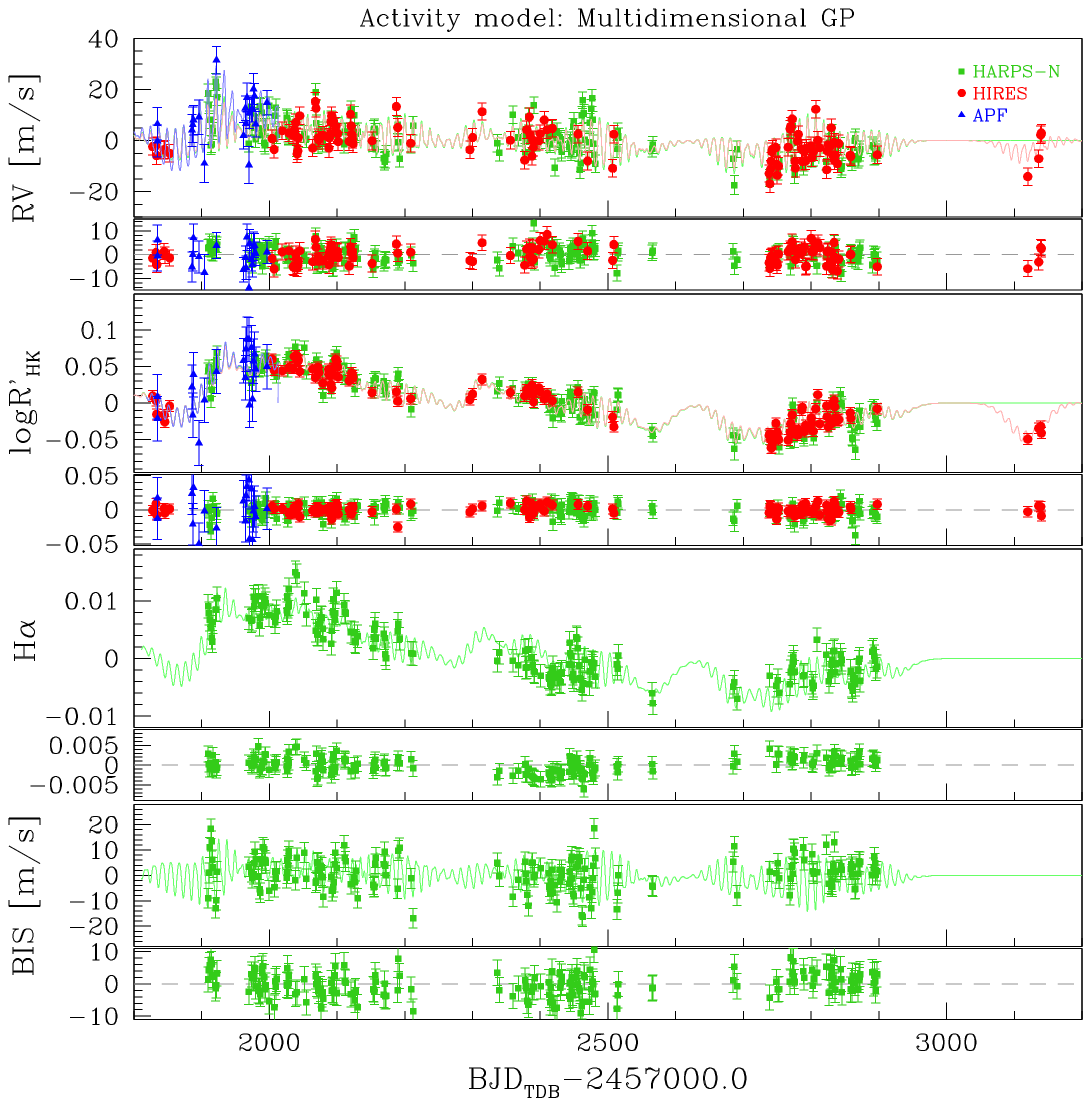} \\
   \includegraphics[width=0.35\textwidth,bb=23 367 443 677]{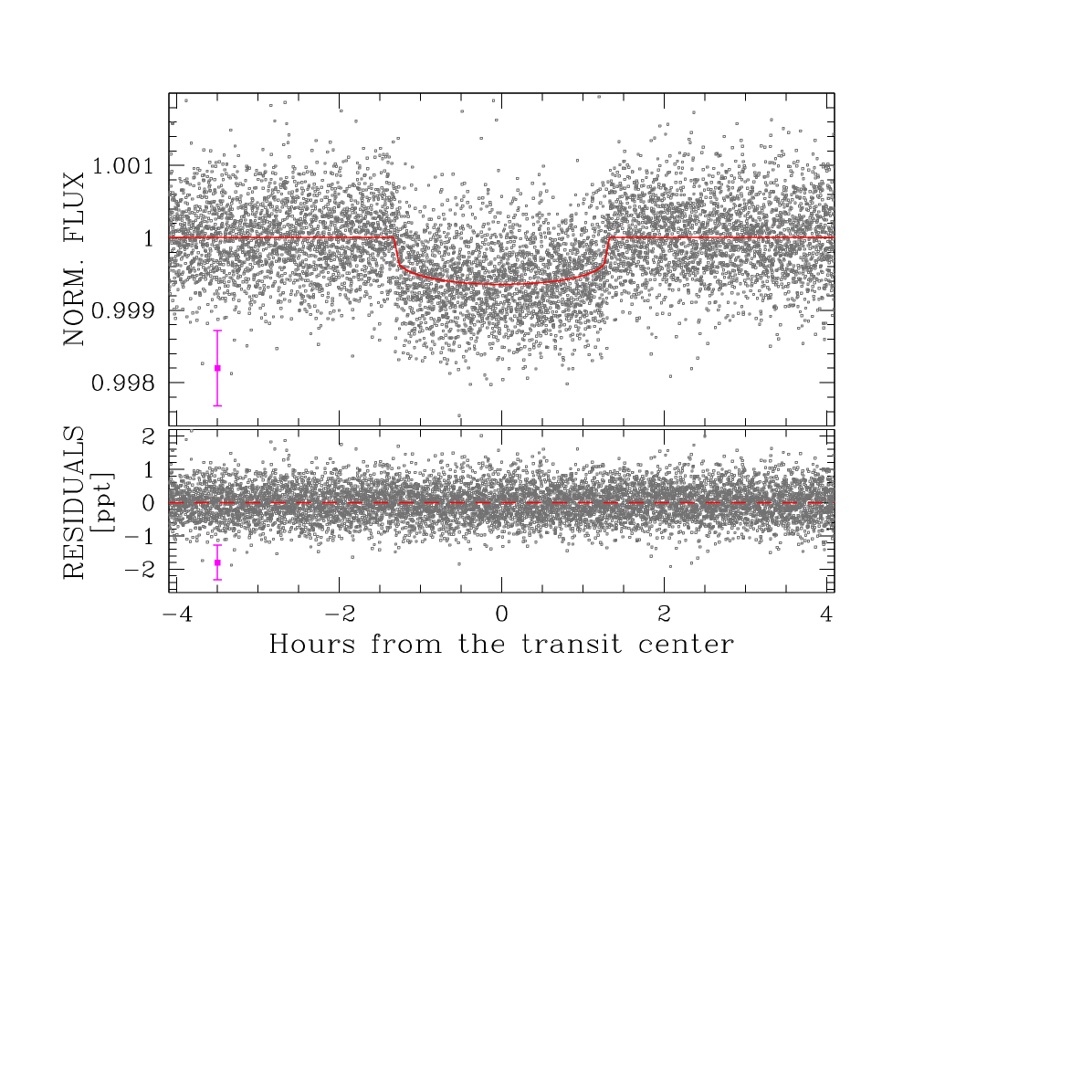} 
   \hspace{0.25cm}
  \includegraphics[width=0.35\textwidth,bb=23 367 443 677]{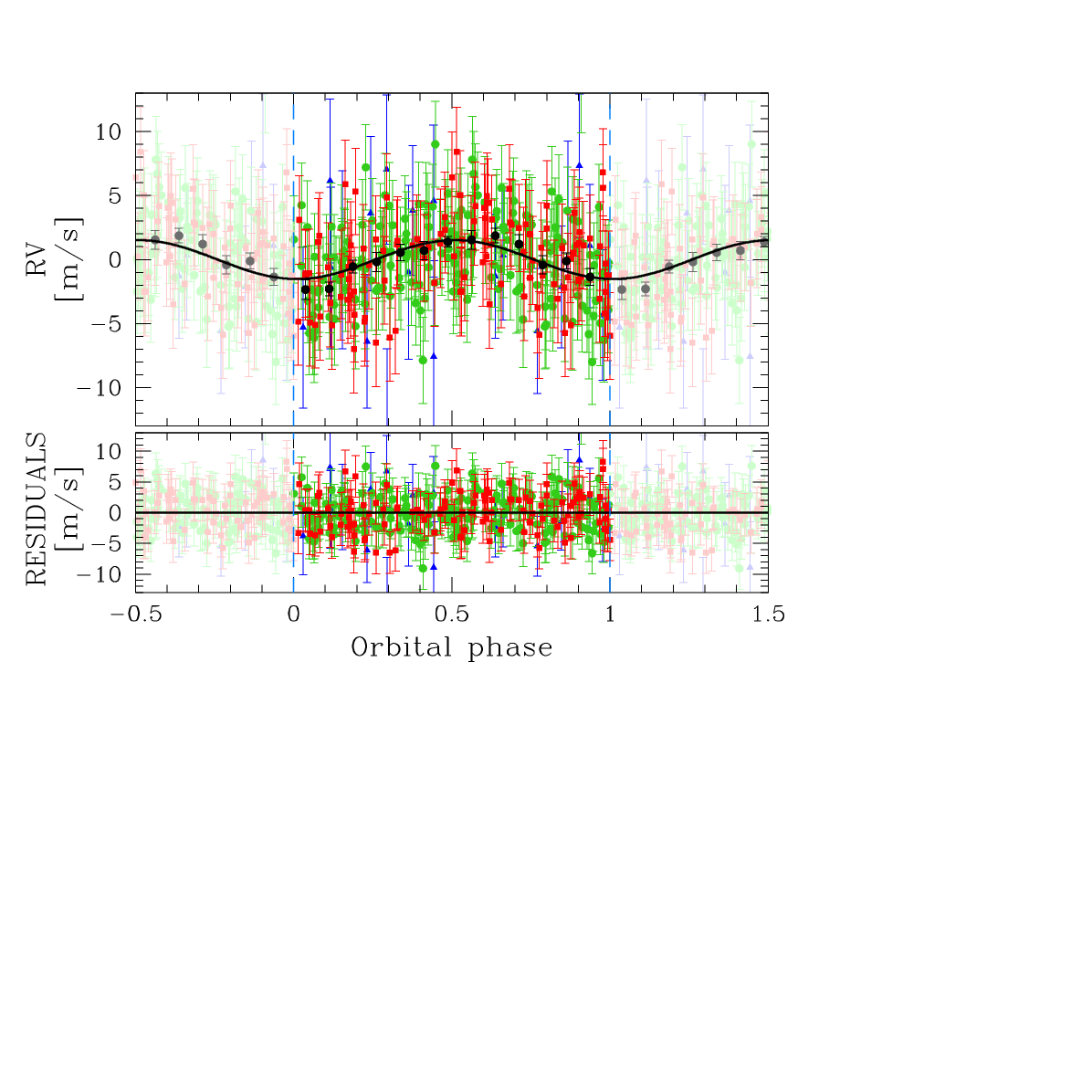}
  \caption{Overview of the modeling by using the multidimensional GP for the stellar activity. From the top, the first four panels show the RV, \logrhk, H$\alpha$, and BIS time series. In green, red and blue the HARPS-N series, HIRES, and APF time series, respectively. 
  The activity models adopted for each time series are reported with the same colors; below each panel the residuals from the activity model. Panels in the last row: on the left-hand is reported the light curve modeling of TOI-1430~b's planetary signal; the upper plot shows the folded transits of TOI-1430~b after the detrending of the light curve and, in red, the derived model of the transits. The mean photometric error (considering also the photometric jitter summed in quadrature) is shown in magenta. The lower plot shows the residuals of the light curve after the subtraction of the planetary transit model. The right-hand panels illustrate the planet detection in the RV series; the upper plot is the phase folded RV time series with the period of planet TOI-1430~b after removing the stellar activity signal with the multidimensional GP. Binned RV points are shown in black, as also the model of planet b. Lower plot shows the RV residuals after removing the planetary model.
  \label{fig:7d}}
\end{figure*}

\begin{table*}[!htb]
  \renewcommand{\arraystretch}{1.11}
  \centering
  \caption{Stellar activity with multidimensional GP and parameters of TOI-1430~b.}
\begin{tabular}{l l l l}
\hline
\hline
Parameter    &  Unit  &    Prior      &       Value   \\
\hline
\multicolumn{4}{c}{{\it TOI-1430 Stellar activity}} \\
Stellar rotational period ($P_{\rm rot}$) & days & $\mathcal{N}(11.9, 0.3)$  &   $12.17_{-0.12}^{+0.12}$ \\
Decay Timescale of activity ($P_{\rm dec}$) & days & $\mathcal{U}(13, 1000)$  &   $26.8_{-1.6}^{+1.9}$ \\
Coherence scale ($w$)  &  & ... &  $0.65_{-0.17}^{+0.17}$ \\
HARPS-N RV offset  & m\,s$^{-1}$ & ... &  $-27244.83_{-0.89}^{+0.87}$ \\
Uncorrelated HARPS-N RV jitter   & m\,s$^{-1}$ & ... & $ 3.22_{-0.23}^{+0.24}$  \\
HARPS-N $V_{\rm c}$ & m\,s$^{-1}$ & ... &  $3.51_{-0.49}^{+0.57}$ \\
HARPS-N $V_{\rm r}$ & m\,s$^{-1}$ & ... &  $31.1_{-4.1}^{+4.9}$ \\
HIRES RV offset  & m\,s$^{-1}$ & ... & $ 0.49_{-0.83}^{+0.82}$ \\
Uncorrelated HIRES RV jitter   & m\,s$^{-1}$ & ... & $ 3.21_{-0.29}^{+0.32}$  \\
HIRES $V_{\rm c}$ & m\,s$^{-1}$ & ... &  $3.12_{-0.47}^{+0.56}$ \\
HIRES $V_{\rm r}$ & m\,s$^{-1}$ & ... &  $26.3_{-4.0}^{+4.8}$ \\
APF RV offset  & m\,s$^{-1}$ & ... & $ -9.3_{-4.1}^{+3.6}$ \\
Uncorrelated APF RV jitter   & m\,s$^{-1}$ & ... & $ 3.8_{-2.0}^{+2.1}$  \\
APF $V_{\rm c}$ & m\,s$^{-1}$ & ... &  $7.0_{-2.4}^{+2.7}$ \\
APF $V_{\rm r}$ & m\,s$^{-1}$ & ... &  $40_{-14}^{+15}$ \\
HARPS-N \logrhk offset  &   & ... & $ -4.4792_{-0.0079}^{+0.0078}$ \\
Uncorrelated HARPS-N \logrhk jitter   &  & ... & $ 0.00866_{-0.00063}^{+0.00067}$ \\
HARPS-N $L1_{\rm c}$ &             & ... &  $0.0331_{-0.0038}^{+0.0046}$ \\
HIRES \logrhk offset  &   & ... & $ -4.4879_{-0.0075}^{+0.0075}$ \\
Uncorrelated HIRES \logrhk jitter   &  & ... & $ 0.00693_{-0.00066}^{+0.00071}$ \\
HIRES $L1_{\rm c}$ &             & ... &  $0.0318_{-0.0037}^{+0.0044}$ \\
APF \logrhk offset  &   & ... & $ -4.454_{-0.017}^{+0.014}$ \\
Uncorrelated APF \logrhk jitter   &  & ... & $ 0.0301_{-0.0048}^{+0.0064}$ \\
APF $L1_{\rm c}$ &             & ... &  $0.034_{-0.010}^{+0.012}$ \\
HARPS-N H$\alpha$ offset  &  & ... & $ 0.2471_{-0.0012}^{+0.0012}$ \\
Uncorrelated HARPS-N H$\alpha$ jitter   &  & ... & $ 0.00192_{-0.00012}^{+0.0013}$ \\
HARPS-N $L2_{\rm c}$ &             & ... & $0.00498_{-0.00058}^{+0.00070}$ \\
HARPS-N BIS offset  & m\,s$^{-1}$ & ... & $ 28.27_{-0.53}^{+0.50}$ \\
Uncorrelated HARPS-N BIS jitter   & m\,s$^{-1}$ & ... & $ 3.90_{-0.24}^{+0.26}$\\
HARPS-N $B_{\rm c}$ & m\,s$^{-1}$ & ... &  $1.65_{-0.37}^{+0.42}$ \\
HARPS-N $B_{\rm r}$ & m\,s$^{-1}$ & ... &  $-33.5_{-5.4}^{+4.6}$ \\
Photometric jitter ($\sigma_{\rm jitter}^{\rm LC}$) &  ppm & ... & $292.5 \pm 1.6$ \\
Stellar density ($\rho_\star$) & $\rho_\odot$ & $\mathcal{N}(1.85
, 0.25)$  &   $1.86_{-0.15}^{+0.16}$  \\
\hline
\multicolumn{4}{c}{{\it TOI-1430~b parameters}} \\
Orbital Period ($P_{\rm b}$)     & days   &  $\mathcal{N}(7.43, 0.05)$  & $  7.4341325_{-0.0000044}^{+0.0000042}$ \\
RV semi-amplitude ($K_b$) & m\,s$^{-1}$ & $\mathcal{U}(0.01, 10)$ & $1.52_{-0.30}^{+0.30}$ \\
Impact factor ($b_{\rm b}$)          &        &  ...                &  $0.435_{-0.060}^{+0.051}$ \\
Planetary-to-stellar-radius ratio ($(R_{\rm P,b}/R_{\star})$) & \% & ... & $2.360_{-0.028}^{+0.028}$ \\
Mean longitude of the ascending node ($\Omega_{\rm b}$) & deg & ... & $176.584_{-0.030}^{+0.029} $\\
Orbital eccentricity ($e_{\rm b}$)   & deg & ... & 0 (fixed)  \\
Limb darkening ($u_1$)         &   & $\mathcal{N}(0.47, 0.01)$ & $0.466_{-0.010}^{+0.010}$ \\
Limb darkening ($u_2$)         &   & $\mathcal{N}(0.14, 0.01)$ & $0.141_{-0.010}^{+0.010}$ \\
Central time of the first transit ($T_{\rm 0,b}$)      & BJD    &  $\mathcal{U}(2458705, 2458706)$  & $2458705.64614_{-0.00059}^{+0.00062}$ \\
Semi-major-axis-to-stellar-radius ratio ($(a_{\rm b}/R_{\star})$) &  & ... & $19.71_{-0.54}^{+0.55}$ \\
Orbital Semi-major axis ($a_{\rm b}$) & au & ... & $ 0.07060_{-0.00055}^{+0.00054}$ \\
Orbital inclination ($i_{\rm b}$)  & deg & ... & $88.74_{-0.19}^{+0.20}$ \\
Duration of the transit ($T_{\rm 14,b}$)  & hours & ... & $2.673_{-0.018}^{+0.017}$ \\
Planetary radius ($R_{\rm P,b}$) & ${\rm R_{\oplus}}$ & ... & $1.982_{-0.069}^{+0.072}$  \\
Planetary mass ($M_{\rm P,b}$) & $M_{\oplus}$ & ... & $4.15_{-0.83}^{+0.83}$  \\
Planetary density ($\rho_{\rm b}$) & $\rho_{\oplus}$ & ... & $0.53_{-0.12}^{+0.12}$\\
\hline
\end{tabular}
  \label{tab:3}
\end{table*}

After running the global parameter optimisation routine \texttt{PyDE}, we adopted $8 \times n_{\rm dim}$ walkers (where $n_{\rm dim}$ is the dimensionality of the model) to run the sampler with the standard ensemble method (\citealt{2010CAMCS...5...65G}) for 120\,000 steps, excluding the first 30\,000 as burn-in. To reduce the effect of the chain auto-correlation, we adopted a thinning factor of 100. We adopted the Gelman-Rubin statistics and the auto-correlation analysis to check the convergence of the chains. 

 All the results are reported in Table~\ref{tab:3}. Figure~\ref{fig:7d} shows the activity modelling for all the available time series, and the modelling of TOI-1430~b in both the photometric and RV series.

We also tested the case with non-null eccentricity: we found similar results shown in Table~\ref{tab:3} and an eccentricity $e = 0.15_{-0.11}^{+0.16}$, i.e., consistent with zero.  We run again the analysis of cases with $e=0$ and with $e \neq 0$ using the dynamic nested sampling algorithm implemented in the package \texttt{dynesty} (\citealt{2014arXiv1407.5459B, 2019S&C....29..891H, 2020MNRAS.493.3132S}) both to check the consistency of the results of the posteriors distributions of the planet parameters, and to calculate (and compare) the Bayesian evidence $\log{\mathcal{Z}}$ to evaluate the quality of the fits in the two cases. The obtained results for the stellar activity and planet parameters are consistent with those obtained with \texttt{emcee}. In the case of circular orbit, we obtained a $\log{\mathcal{Z}} \simeq 299.2$, while in the case of Keplerian orbit $\log{\mathcal{Z}} \simeq 297.2$. Even if the difference in $\log{\mathcal{Z}}$ between the models is not significant ($\Delta \log{\mathcal{Z}}\sim 2$), the case of a circular orbit is preferred.
We note that even assuming a strong tidal dissipation in the solid core of the planet (see Sect.~\ref{sec:results} for internal models) as parameterized by a modified tidal quality factor $Q^{\prime}_{\rm p} = 150$ \citep{Henningetal09}, the decay time of the eccentricity is $e/|de/dt| \sim 0.9$~Gyr that is not short in comparison with the estimated age of the system. Therefore, any primordial eccentricity should not have been reduced by tides by more than a factor of $2-3$ suggesting that the initial orbit was not remarkably eccentric.

\begin{figure*}
  \centering
  \includegraphics[width=0.9\textwidth,bb=20 541 572 715]{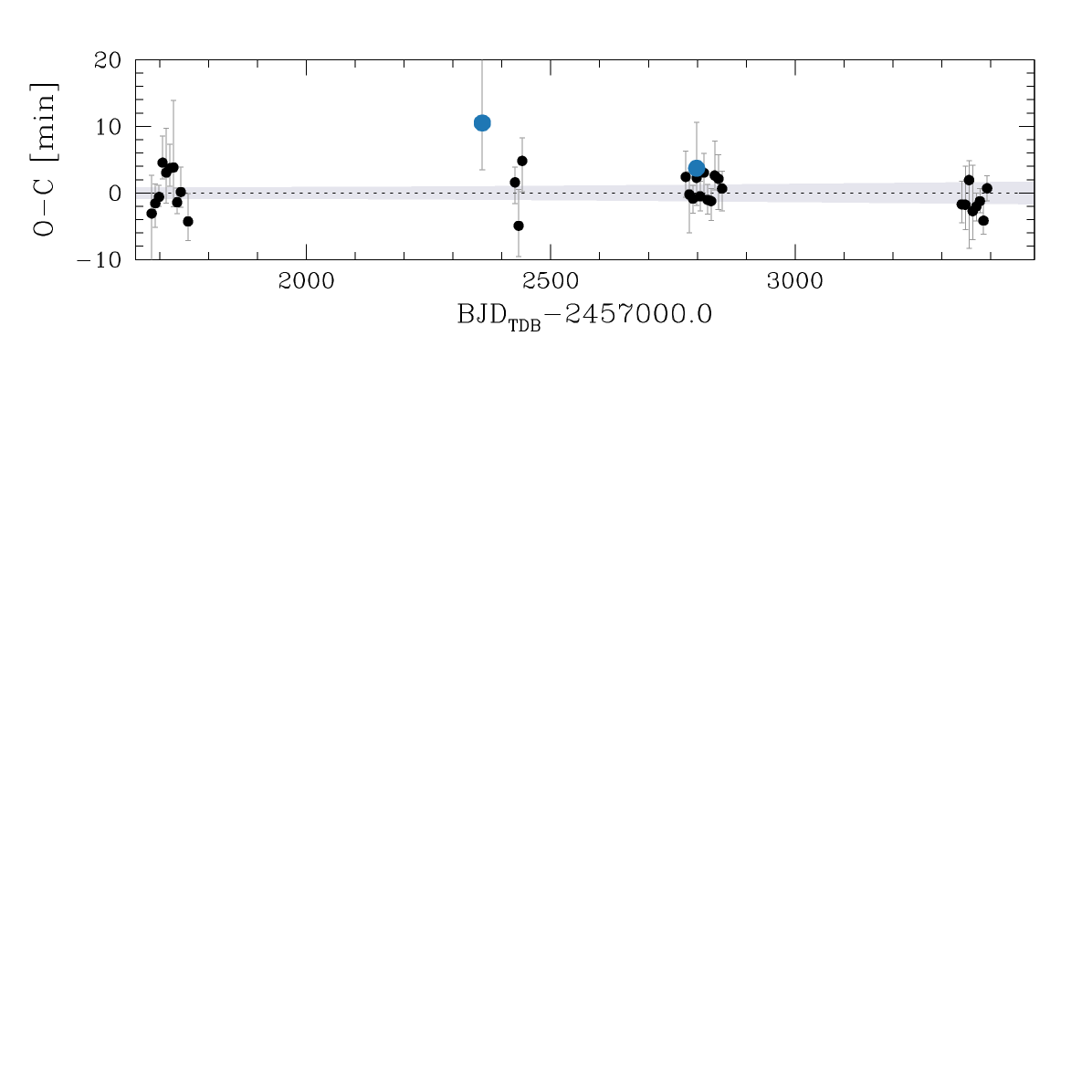} 
  \caption{Analysis of TTVs based on \tess and LCO data. The figure shows the O$-$C plot of the observed (O) and calculated (C) transit times for the linear ephemeris of TOI-1430~b.
  \label{fig:9}}
\end{figure*}

\subsection{Searching for other planets}
\label{sec:planet_c}
We investigated the presence of a second, non-transiting planet in the RV dataset. We limited the analysis to the spectroscopic series, giving a strict prior on the period of  planet b on the basis of the previous analysis (but leaving free the RV semi-amplitude).

To check the presence of a second planetary signal, we run \texttt{PyORBIT} again adopting the multidimensional GP framework for the modeling of the stellar activity, which  is less sensitive to overfitting problems of the spectroscopic series. By using the same priors of Table~\ref{tab:3}, chains converged  for $P_{\rm c} \sim 96.4$~d (that is $\sim 8\times P_{\rm rot}$) and $K_{\rm c} \sim 1.3$~m/s ($\sim 4 \sigma$ detection). We restricted this analysis to sampling orbital periods in the interval $0.4 \leq P_{\rm c} \leq 20.0$~days; the result was a signal at $P_{\rm c} \sim 12.68$~d and $K_{\rm c} \sim 1.3$ m/s ($\sim 4 \sigma$ detection). 

The disagreement between the results obtained with different models of stellar activity (see also Appendix~\ref{sec:c}) suggests that the measured signal of candidate planet c is probably due to residuals in the fit of the stellar activity, which turns out to be quite difficult to model.

For the sake of completeness, we report in Appendix~\ref{sec:c}, in Fig.~\ref{fig:C1} and Table~\ref{tab:C1} all the results related to these signals; more data and analysis are mandatory to confirm their planetary nature.

\subsection{Analysis of Transit Timing Variations}

We investigated the presence of dynamical interactions of TOI-1430~b with other potential bodies in the system. In order to perform this analysis, we calculated the transit timing variations (TTVs, \citealt{2005MNRAS.359..567A,2005Sci...307.1288H}) as the difference between the observed central times of the transits (O) and the computed central times of the transits (C) from the linear ephemeris (Table~\ref{tab:3}). We modelled every single transit by using \texttt{PyORBIT}. In this analysis, we also included the two ground-based light curves collected with LCO. The modeling of single transits is reported in Fig.~\ref{fig:9A}. Figure~\ref{fig:9} shows the O$-$C diagram.

The possible TTV amplitude, i.e. the semi-amplitude of the O$-$C series, is $A_{\rm TTV}\simeq 7.7$~minutes. To check if there is any hint of TTV we calculated the $\chi^2$ of the linear ephemeris fitting as:
\begin{equation}
\chi^2 = \sum_i^N=\frac{({\rm O}_i-{\rm C}_i)^2}{(\sigma_{+,i}^2+\sigma_{-,i}^2)}
\end{equation}
where $N$ is the number of available epochs for which we calculated the central times of the transits, and $\sigma_+$ and $\sigma_-$ are the positive and negative errors associated with the O$-$C. We obtained $\chi^2 \sim 9.4$ and a reduced  $\chi^2$, $\chi^2_{\rm red}(32) \simeq 0.29$, where 32 are the degrees of freedom. The linear ephemeris fits well  the data and there is no hint of TTV.
In Table~\ref{tab:D1} the $T_0$ values for the different epochs are reported.

\section{Results}
\label{sec:results}

\subsection{System inclination}
\label{sec:systeminclination}
The combination of the measured stellar rotation period, projected rotational velocity (Sect.~\ref{sec:vsini}, Table~\ref{tab:1}), and stellar radius (Table~\ref{tab:1}) yield a stellar inclination of $35^{+15}_{-12}$ deg. The equatorial velocity implied by stellar rotation period and stellar radius is 3.28$\pm$0.17 \kms. 
The possibility that the true rotation period is two times the observed one (double-dip behavior) can be dismissed considering the levels of chromospheric and coronal activity, which are fully consistent with the adopted rotation period, derived from both spectroscopic and photometric time series.  
Therefore, there are indications that the star rotation is not aligned with the planetary orbit, although this needs to be confirmed through measurement of the Rossiter-McLaughlin effect (\citealt{2000A&A...359L..13Q}, expected maximum amplitude $\sim 1$m\,s$^{-1}$). 
Tides induced by the planet in the star are not sufficiently strong to modify the obliquity of the orbit because of the small mass of the planet. Therefore, the current obliquity should be close to the primordial obliquity of the system.

\subsection{Mass-radius diagram}

\begin{figure*}
    \sidecaption
  \includegraphics[width=12.95cm,bb=25 172 509 645]{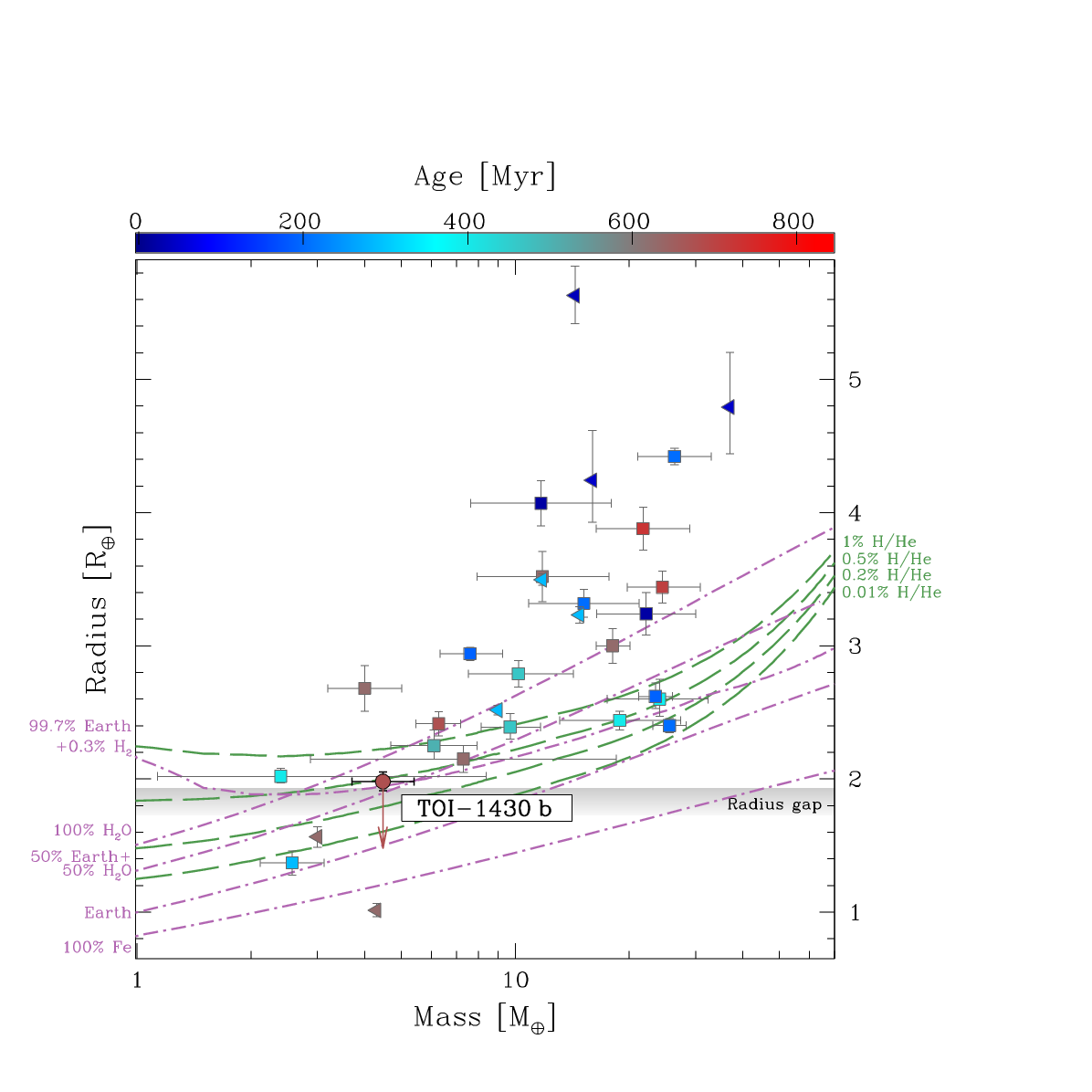} 
  \caption{ The mass-radius diagram for young ($<1$~Gyr) planets with precise age estimate (relative errors $< 50 \%$). TOI-1430~b is located at the edge of the Fulton gap; the arrow indicates the final position in the diagram in the next 200~Myr, when it will  totally lose its atmosphere. The \citet{2019PNAS..116.9723Z}'s tracks are reported in magenta (dot-dashed lines) in the case of planets composed of 100~\% Fe, 32.5~\% Fe+67.5~\% MgSiO$_3$ (Earth-like), 50~\% Earth+50~\% H$_2$O, 99.7~\% Earth+0.3~\% H$_2$; in green are reported the \citet{LopFor14}'s tracks for exoplanets with H/He envelopes between 0.01 ~\% and 1~\%. Exoplanets with accurate age estimation are color-coded on the basis of the colorbar on the top of the figure; square points are the exoplanets with accurate mass measurement, and triangles are the exoplanets with a mass upper limit measurement. 
  \label{fig:11}}
\end{figure*}

In this work, we have measured the mass and the radius of the exoplanet TOI-1430~b with an high-level of precision ($\sim 19~\%$ and $\sim 4~\%$, respectively).
In Fig.~\ref{fig:11} we report the results in the mass-radius diagram, where all other young planets ($<1$~Gyr) with accurate (based on different techniques) and precise (relative errors $< 50~\%$) age measurements  are also reported. 
Combining the values reported in Tables~\ref{tab:2} and \ref{tab:3}, we obtained a planetary density of $\sim 0.53 \rho_{\oplus} \sim 3~$g\,cm$^{-3}$. The low density of this exoplanet is due to its extended atmosphere; the detection of an escaping atmosphere is reported by \citet{zhang2023} and \citet{orellmiquel2023}: both the works reported the detection of helium excess absorption ($\sim -0.6$~--~$-0.9$\,\%) due to the atmosphere of TOI-1430~b. It is among the few small planets with a robust detection of the atmosphere. 

As shown in Fig.~\ref{fig:11}, TOI-1430~b is located at the upper limit of the "radius gap" (\citealt{Fulton17,2018AJ....156..264F}), that region in the mass-radius diagram sparsely populated by exoplanets that divides the rocky Earths/super-Earths from the low-density mini-Neptunes. This gap could be an effect of the atmospheric evolution of mini-Neptunes characterized by a rocky core and an extended H/He atmosphere that inflates their radius. This atmosphere is then stripped away through mechanisms like, e.g., photoevaporation (\citealt{2013ApJ...775..105O,2017ApJ...847...29O}) or core-powered mass loss (\citealt{2018MNRAS.476..759G}), resulting in the exposure of the rocky core, i.e the rocky planets we observe today below the radius gap. 

By comparing the derived planet parameters with the tracks by \citet{2019PNAS..116.9723Z}\footnote{\url{https://lweb.cfa.harvard.edu/~lzeng/planetmodels.html}}, it results that TOI-1430~b is compatible with an Earth-like rocky core (32.5~\% Fe+67.5~\% MgSiO3) + 0.3~\% H$_2$ atmosphere/envelope, or with a water world (50~\% Earth-like rocky core $+$ 50~\% H$_2$O). Also, the comparison of the mass and radius of TOI-1430~b with the 700~Myr old tracks calculated by \citet{LopFor14} is in agreement with a planet with a 0.2--0.5~\% H/He envelope.
The atmospheric evolution model described in the next section, based on the assumption the planet is a rocky world with an extended atmosphere, predicts that in $\sim 200$~Myr this planet will lose its envelope showing its $\sim 1.5$~R$_{\oplus}$ rocky core, as illustrated by the arrow in Fig.~\ref{fig:11}.

\subsection{TOI-1430~b atmospheric evolution}

Given its young age (700 $\pm$ 150 Myr), TOI-1430~b represents a prime target for photoevaporation studies. 

To assess the current atmospheric mass loss rate, the past history and the fate of the TOI-1430~b atmosphere subjected to photoevaporation, we followed a modeling approach proposed by \cite{Locci19}, which has been  utilized and updated in several previous works (e.g. \citealt{benatti2021,maggio2022}) and most recently described in detail by \citet{Mantovan24}. 
In brief, we coupled the new ATES photoionization hydrodynamics code \citep{Caldiroli+2021, Caldiroli+2022,2023AJ....165..200S}
with the planetary core-envelope models by \cite{Fortney2007} and \cite{LopFor14}, the MESA Stellar Tracks (MIST; \citealt{choi+2016}), and the XUV luminosity time evolution by using different prescriptions \citep{Penz08a,SF22,Jo2021}. 

Simulating a grid of exoplanets with different levels of X-ray and Extreme Ultraviolet (XUV) irradiation and different gravitational potential energy, \citet{Caldiroli+2022} found an analytical approximation for the evaporation efficiency, $\eta_{\rm eff}$. Using this evaporating efficiency as a function of the gravitational potential energy of the planet and the XUV irradiation in the classical energy-limited formula \citep{Erkaev+2007}, it is possible to derive an approximation of the atmospheric mass loss rate evaluated with hydrodynamic simulations:
\begin{equation}
    \Dot{M} = \eta_{\rm eff} (F_{\rm XUV}, \Psi) \frac{3 F_{\rm XUV}}{4 G K \rho_{\rm p}}~,
    \label{eq:mdot}
\end{equation}
where $F_{\rm XUV}$ is the XUV flux at the (average) orbital distance, $\Psi$ the gravitational potential energy of the planet,
$\rho_{\rm p}$ is the mean planetary mass density and the factor K accounts
for the host star tidal forces \citep{Erkaev+2007}. %At each time step 
We used this analytical approximation to speed up the modeling of the past and future evolution of the planet. 

\begin{figure}
  \centering
\includegraphics[width=0.475\textwidth]{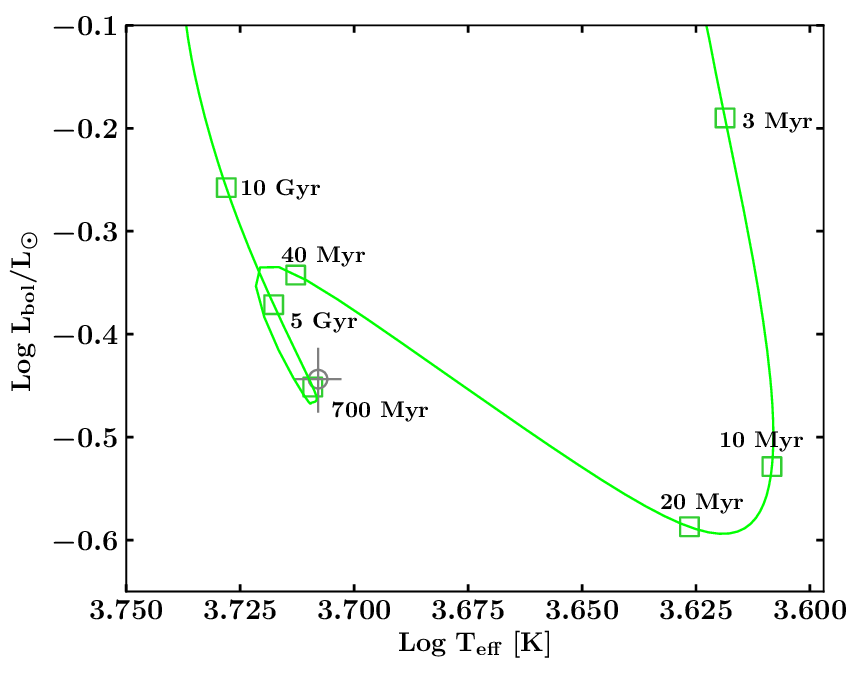}
  \caption{Evolutionary track of TOI\,1430  in the effective temperature-bolometric luminosity plane. A grey dot marks the current location of the star on the track.
  \label{fig:evoltrack}}
\end{figure}

We considered the stellar bolometric and XUV luminosity evolution. We obtained the stellar evolutionary track (the theoretical temperature-luminosity diagram shown in Fig. \ref{fig:evoltrack}) through the web-based interpolator\footnote{\url{https://waps.cfa.harvard.edu/MIST/interp_tracks.html}} of the MESA Isochrones and Stellar Tracks (MIST, \citealt{choi+2016}).

We assumed two different descriptions to estimate the evolution of stellar X-ray and EUV emission. The first method is the X-ray luminosity versus age analytical relation for G-type stars derived by \cite{Penz08a}, anchored to the current value of the X-ray luminosity (Sect.\ \ref{sec:activity}). In this case, to estimate the stellar irradiation in the EUV band,  we used the scaling law between EUV (100--920\,\AA) and X-ray (5--100\,\AA) luminosities, derived by \cite{SF22}:
\begin{equation}
\log L_{\rm EUV} = (0.793 \pm 0.058) \log L_{\rm x} + (6.53 \pm 1.61)~,
\label{eq:euv}
\end{equation}

\begin{figure}
  \centering
\includegraphics[width=0.475\textwidth]{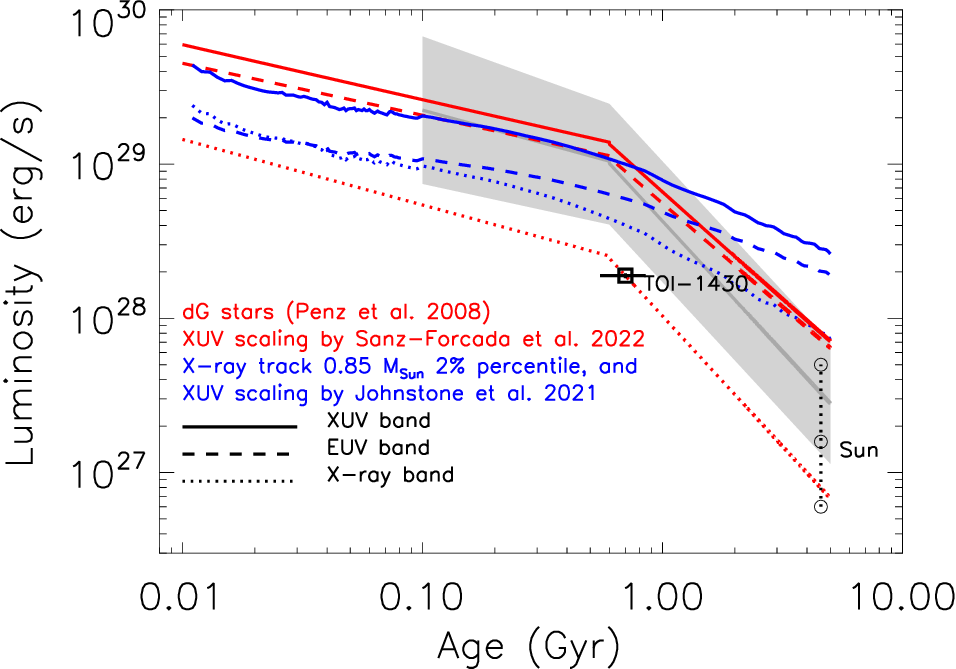}
  \caption{Time evolution of X-ray (5--100\,\AA), EUV (100--920\,\AA), and total XUV luminosity of TOI\,1430, according to \cite{Penz08a} and the X-ray/EUV scaling by 
\citet{SF22} (red lines) and according to \cite{Jo2021} (blue lines). Uncertainties on the age and X-ray luminosity of TOI\,1430 are also indicated. The gray area is the original locus for dG stars in \cite{Penz08a}.
  \label{fig:evolx}}
\end{figure}

\noindent
In Fig.\ref{fig:evolx}, we show with red lines the predicted evolution of the X-ray and EUV stellar luminosities obtained with this method.

The second method that we adopted to model the evolution of TOI-1430 X-ray and EUV emission is  the semi-empirical modeling by \cite{Jo2021}, where the X-ray to bolometric luminosity ratio, $L_{\rm X}/L_{\rm bol}$ follows a broken power law with the Rossby number \citep{Pizz03,Wright+2011}, i.e., the ratio of the rotation period to the convective turnover time. In this case, the initial rotation rate of the star at early ages ($\Omega_0$ at $t = 1$\,Myr) determines the evolutionary track of $L_x/L_{bol}$ in time, for any star with a given mass. For TOI-1430 we selected the track of a star with $M_* = 0.85$\,\msun which predicts the X-ray luminosity nearest to the measured value, at an age in the range 550--850\,Myr. Then, we computed the EUV luminosity following again \cite{Jo2021}, who proposed an empirical mass-independent power-law scaling between the surface EUV and X-ray fluxes calibrated on a sample of stars observed with the Extreme Ultraviolet Explorer satellite and solar spectra derived from the TIMED/SEE mission. The X-EUV luminosity tracks derived with this second method are also shown in Fig.\ref{fig:evolx} with blue lines.

It must be noted that the actual (measured) $L_{\rm X}$ value (Sect.~\ref{sec:activity}) remains   lower than the emission level expected   in both the descriptions by \cite{Penz08a} and \cite{Jo2021}. According to the latter, the expected X-ray luminosity for the stars with the lowest activity level should be about $4 \times 10^{28}$\,erg s$^{-1}$ (Fig.~\ref{fig:evolx}), corresponding to a Rossby number $R_{\rm o} = 0.62^{+0.03}_{-0.02}$ \citep{Wright+2018}, that is more than a factor 2 higher than the observed $L_{\rm X}$ value. In other words, TOI-1430 appears to have a high-energy emission level relatively lower than expected for stars with similar mass and age, such as the members of the Hyades open cluster.

Assuming the planetary parameters in Table~\ref{tab:3},  
the mass loss rate estimated by running the ATES hydrodynamic simulation gives a value between 3.1--3.5$\times$10$^{10}$\,g\,s$^{-1}$ depending on the XUV prescriptions (\citealt{Jo2021} and \citealt{SF22} respectively). This value is consistent with the estimation performed by \cite{orellmiquel2023} (1.5--5 $\times$10$^{10}$~g\,s$^{-1}$), who used the 1D hydrodynamic model by \cite{lampon2020}. 

To evaluate the past and the future evolution of the planetary atmosphere subjected to photoevaporation we determined the planetary core mass, the atmospheric mass fraction and the radius at the current age by adopting the core-envelope model described in \cite{LopFor14}. This core-envelope model has four unknowns: the mass and radius of the core, $M_{\rm core}$ and $R_{\rm core}$, the radius of the envelope, $R_{\rm env}$, and the atmospheric mass fraction, $f_{\rm atm}$. Four relations and two observational constraints link these quantities: the measured planet radius and total mass, $R_{\rm p}$ and $M_{\rm p}$; the \cite{LopFor14} relation which links $R_{\rm env}$, $f_{\rm atm}$, and $M_{\rm p}$; and a relation between $R_{\rm core}$, $R_{\rm p}$, and $M_{\rm p}$ at the given age and star--planet distance. For the latter, we adopted the internal structure models by \cite{Fortney2007}, where cores can be composed of different ice-rock or rock-iron mixtures.

We note that the relation by \cite{LopFor14} was developed for atmospheres dominated by H--He, and also takes into account the cooling and contraction of the envelope as a consequence of its thermal evolution \citep{LopForMil12}.
In the \cite{LopFor14} relation, the radius time dependence is linked to the metallicity. For atmospheres with solar opacity, the radius of the envelope decreases with age as $\propto t^{-0.11}$, while for an enhanced opacity, the variation is $\propto t^{-0.18}$. Allowing for the time variation of bolometric luminosity and surface temperature equilibrium introduces a further time dependence on the envelope size.

Assuming an Earth-like core composed of 67\% rocks and 33\% iron, and an envelope with solar opacity, we found for TOI-1430-b a solution with $M_{\rm core,b} = 4.13 $\,\mearth, $R_{\rm core,b} = 1.45$\,\rearth, $R_{\rm env,b} = 0.53$\,\rearth, and $f_{\rm atm} = 0.42\%$, corresponding to the measured planetary mass and optical radius. With an enhanced metallicity envelope we found another solution with $M_{\rm core,b} = 4.14$\,\mearth, $R_{\rm core,b} = 1.45$\,\rearth,  $R_{\rm env,b} = 0.53$\,\rearth,  and $f_{\rm atm} = 0.33\%$.

\begin{figure*}
\centering 
\includegraphics[width=0.45\textwidth]{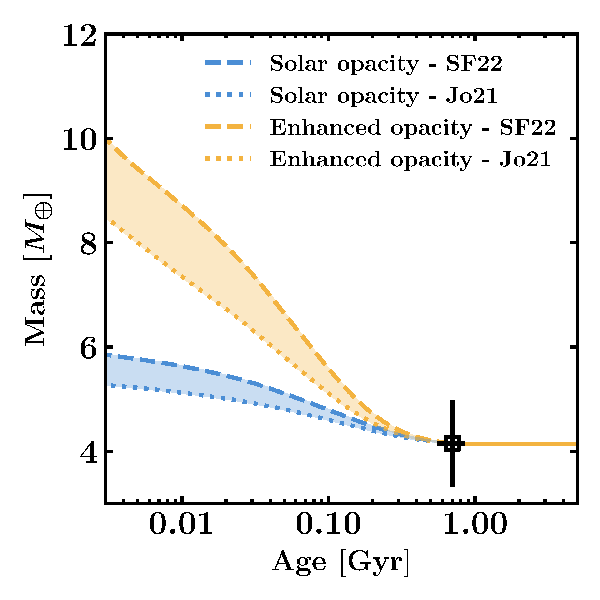}
\includegraphics[width=0.45\textwidth]{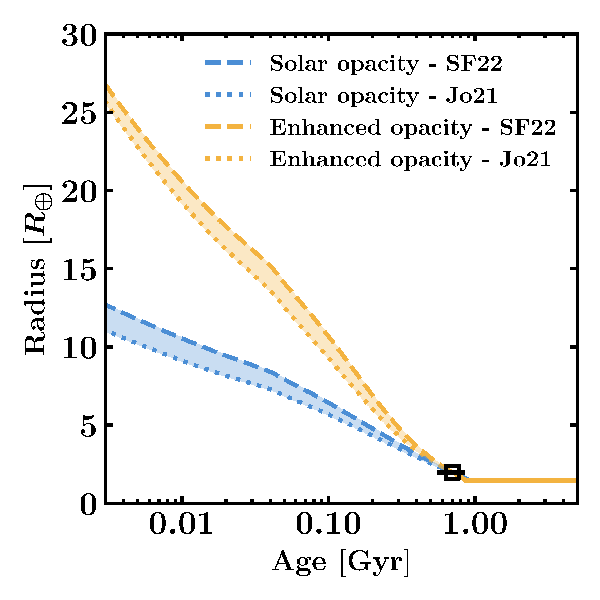} \\
\includegraphics[width=0.45\textwidth]{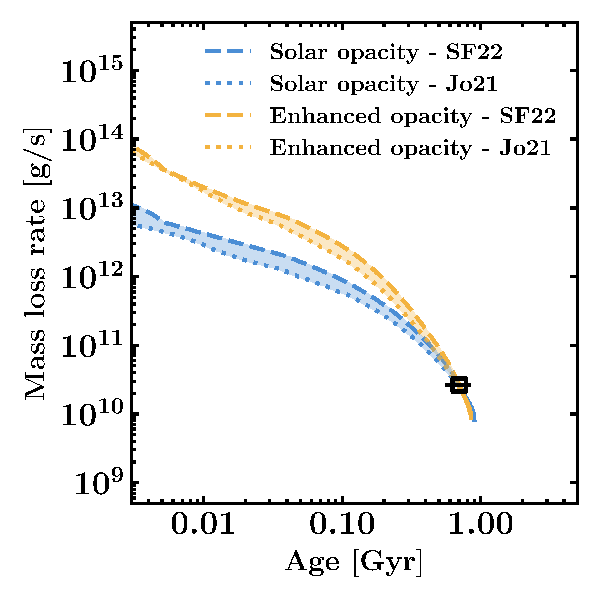}
\includegraphics[width=0.45\textwidth]{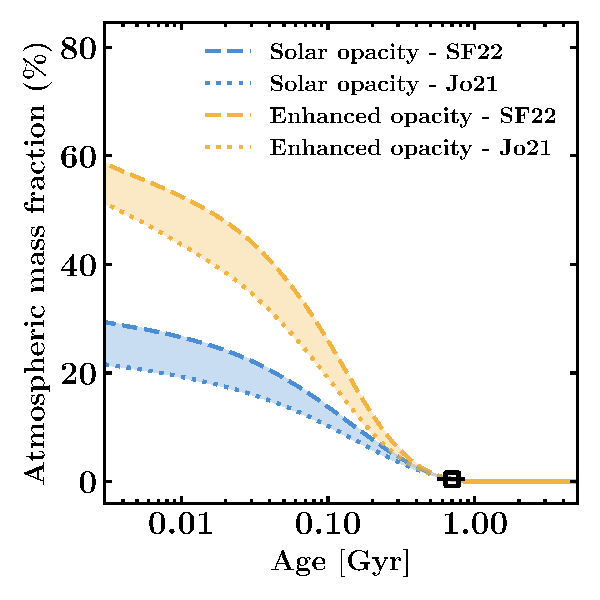}
\caption{The evolution of the planetary mass (upper-left panel), planetary radius (upper-right panel), atmospheric mass loss rate (bottom-left panel) and atmospheric mass fraction (bottom-right panel) according to our models with different assumptions. Black squares mark the current location of the star in each plot. \label{fig:evolution}
}
\end{figure*}

We then proceeded with the time evolution of the photoevaporation. For each time step of the simulation, we computed the mass-loss rate and updated $f_{\rm atm}$ and the planetary mass, obtaining a new value of $R_{\rm env}$ with the relation by \cite{LopFor14}. The latter quantity added to the core radius (assumed constant) provides the updated planetary radius. 

According to the aforementioned scenario and assuming a core that does not change in size or mass, we followed the planetary evolution back in time. We stopped our simulations in the past at 3\,Myr, the end time of planetary formation models, that is, when the circumstellar disc has already disappeared and each planet is in its final, stable orbit. For  future evolution, we let the system evolve from the current age (700\,Myr) until 5\,Gyr.
The results of the simulations dependent on different assumptions are shown in Figure \ref{fig:evolution}.

Currently, the planet is close the radius gap (\citealt{Fulton17}) with a radius of 1.982 \rearth. As shown in Figure \ref{fig:evolution}, our model shows that the planet experienced significant evolution in the past, from a birth as a (super-)puffy planet to the current sub-Neptune properties. 

In order to retain a part of this envelope at the current age (700 Myr, an envelope mass fraction of $\sim$0.3-0.4\%) our model suggests that the planet was surrounded by a very large envelope at the beginning of its evolution. More specifically, the model with a solar-opacity envelope suggests an envelope mass fraction at 3 Myr of 29\% or 21\%, assuming the \cite{SF22} or \cite{Jo2021} XUV evolution, respectively. 
The corresponding planetary radius was 12.7~\rearth\, or 11.0~\rearth, with a planetary mass of 5.9~\mearth\, or 5.3~\mearth. The model with an enhanced-opacity envelope suggests an even larger envelope. Assuming the \cite{SF22} XUV evolution, this model predicts an envelope mass fraction at 3 Myr of 59\%, with a radius of 26.8~\rearth\, and a mass of 10.0~\mearth, while assuming \cite{Jo2021} the envelope mass fraction was 51\%, with a planetary radius of 25.7~\rearth\, and a planetary mass of 8.5~\mearth. At 3 Myr the exoplanetary mass loss rate was 
in the range 5.8--11.1$\times$10$^{12}$\,g\,s$^{-1}$ for the solar-opacity model, and 6.5--8.2$\times$10$^{13}$\,g\,s$^{-1}$ for the enanched-opacity envelope model. 

Under the effect of photoevaporation, the planet lost much of its envelope in the first 300\,Myr under the effect of high-energy stellar irradiation. At 300\,Myr the envelope mass fraction was between 3.6\% and 6.1\% depending on the different assumptions that we adopted, with a planetary radius and a planetary mass between 3.5-4.9~\rearth\, and 4.3-4.4~\mearth, respectively. 

Regarding the future of the planet, our model predicts that it will completely lose its envelope in the next 150--200\,Myr, and it will become an envelope-stripped super-Earth-size planet with a radius equal to the core radius (1.48\,\rearth). 
Our simulations show that the planet undergoes a noticeable radius evolution. In fact, the planet started its evolution as a Jovian or puffy-Jovian (depending on the model). Between around 100 and 600 Myr, it passed through the first peak of the bimodal distribution, reaching its current position at the edge of the Fulton gap. Within the next hundreds of Myr, the planet will completely lose its atmosphere, crossing the Fulton gap and ending its evolution as a Chthonian planet in the second peak of the distribution.

\section{Conclusions}

In this work, we measured the density of the young planet TOI-1430~b with high precision thanks to the combination of exquisite data-sets coming from {\it TESS}, HARPS-N, HIRES, and APF instruments.

First, we derived the stellar parameters adopting information from Gaia and analysing complementary spectroscopic and photometric data. In particular, on the basis of the stellar rotation period ($P_{\rm rot}\sim 12$~d), the kinematics, the lithium abundance, and the coronal and chromospheric activity, we found that TOI-1430 is $700\pm 150$~Myr old.

We combined the \tess light curves of nine sectors with spectroscopic series collected with HARPS-N@TNG, HIRES@Keck and APF in a $\gtrsim$3-year campaign to simultaneously model the activity and  the signal of the planet. To model the stellar activity in the spectroscopic series we adopted two different approaches: (i) GP regression with a quasi-periodic kernel, and (ii) multidimensional GP framework. Planet parameters are reported in Table \ref{tab:3}. We measured a stellar rotational period $P_{\rm rot}=12.2 \pm 0.1$~d and a planetary mass $M_{\rm P, b} = 4.2 \pm 0.8~M_{\oplus}$.
The planetary radius we measured is $R_{\rm P,b}=1.98 \pm 0.07~R_{\oplus}$. From these values we inferred a density $\rho_{\rm b} \sim 0.53 \pm 0.12~\rho_{\oplus} = 2.9 \pm 0.7$~g\,cm$^{-3}$, i.e. TOI-1430~b is a low-density mini-Neptune with an extended atmosphere located at the upper edge of the radius gap in the mass-radius diagram (see Fig.~\ref{fig:11}).

\citet{zhang2023} and \citet{orellmiquel2023} detected an evaporating He atmosphere around TOI-1430~b. By adopting the planet parameters derived in this work, we traced the history of the atmosphere of this planet, from 3 Myr after its birth to the present age and also for the next 5 Gyr. We found that the planet's radius has significantly evolved during its lifetime: starting from a Jovian/puffy-Jovian ($\sim$10--25 $R_{\oplus}$ and $\sim$5--10 $M_{\oplus}$, on the basis of the model adopted), the planet lost most of its envelope in the first 100--300~Myr. With a radius of $\sim 2~R_{\oplus}$, actually, the planet is entering the Fulton gap and continues to lose atmosphere. In the next 100--200~Myr, TOI-1430~b will lose its entire envelope, becoming an Earth-size rocky planet of $\sim 1.5~R_{\oplus}$.

\begin{acknowledgements}

The authors thank the anonymous referee for helping to improve the quality of the paper.
This work has been supported by the PRIN-INAF 2019 "Planetary systems at young ages (PLATEA)" and ASI-INAF agreement n.2018-16-HH.0. V.N., G.P., L.M. acknowledge financial support from the Bando Ricerca Fondamentale INAF 2023, Data Analysis Grant: "Characterization of transiting exoplanets by exploiting the unique synergy between TASTE and TESS". A.M. acknowledges partial support by the project HOT-ATMOS (PRIN INAF 2019), and by the PRIN/MUR EXO-CASH.  
R.S. acknowledges the support of the ARIEL ASI/INAF agreement no. 2021-5-HH.0 and the support of grant no. 2022J7ZFRA – Exo-planetary Cloudy Atmospheres and Stellar High energy (Exo-CASH) funded by MUR – PRIN 2022.
L.M. acknowledges financial contribution from PRIN MUR 2022 project 2022J4H55R.
We acknowledge the Italian center for Astronomical Archives (IA2, https://www.ia2.inaf.it), part of the Italian National Institute for Astrophysics (INAF), for providing technical assistance, services and supporting activities of the GAPS collaboration. 
This work includes data collected with the TESS mission, obtained from
the MAST data archive at the Space Telescope Science Institute
(STScI). Funding for the TESS mission is provided by the NASA Explorer
Program. STScI is operated by the Association of Universities for
Research in Astronomy, Inc., under NASA contract NAS 5–26555. We
acknowledge the use of public TESS data from pipelines at the TESS
Science Office and at the TESS Science Processing Operations Center.
Resources supporting this work were provided by the NASA High-End
Computing (HEC) Program through the NASA Advanced Supercomputing (NAS)
Division at Ames Research Center for the production of the SPOC data
products. Funding for the TESS mission is provided by NASA's Science
Mission Directorate.
This work has made use of data from the European Space Agency (ESA) mission Gaia (\url{https://www.cosmos.esa.int/gaia}), processed by the Gaia Data Processing and Analysis Consortium (DPAC, \url{https://www.cosmos.esa.int/web/gaia/dpac/consortium}). Funding for the DPAC has been provided by national institutions, in particular the institutions participating in the Gaia Multilateral Agreement. This work makes use of observations collected at the Asiago Schmidt 67/92 cm telescope (Asiago, Italy) of the INAF – Osservatorio Astronomico di Padova. This paper made use of data from the KELT survey (\citealt{2007PASP..119..923P}).

\end{acknowledgements}

\bibliographystyle{aa}
\bibliography{biblio}

%%%%%%%%%%%%%%%%%%%%%%%%%%%%%%%%%%%%% APPENDIX #######################################################
\begin{appendix}

%%%%%%%
\section{Light curve correction}
\label{sec:light_curve_corr}
\begin{figure*}
  \centering
  \includegraphics[width=0.9\textwidth, bb=21 275 566 675]{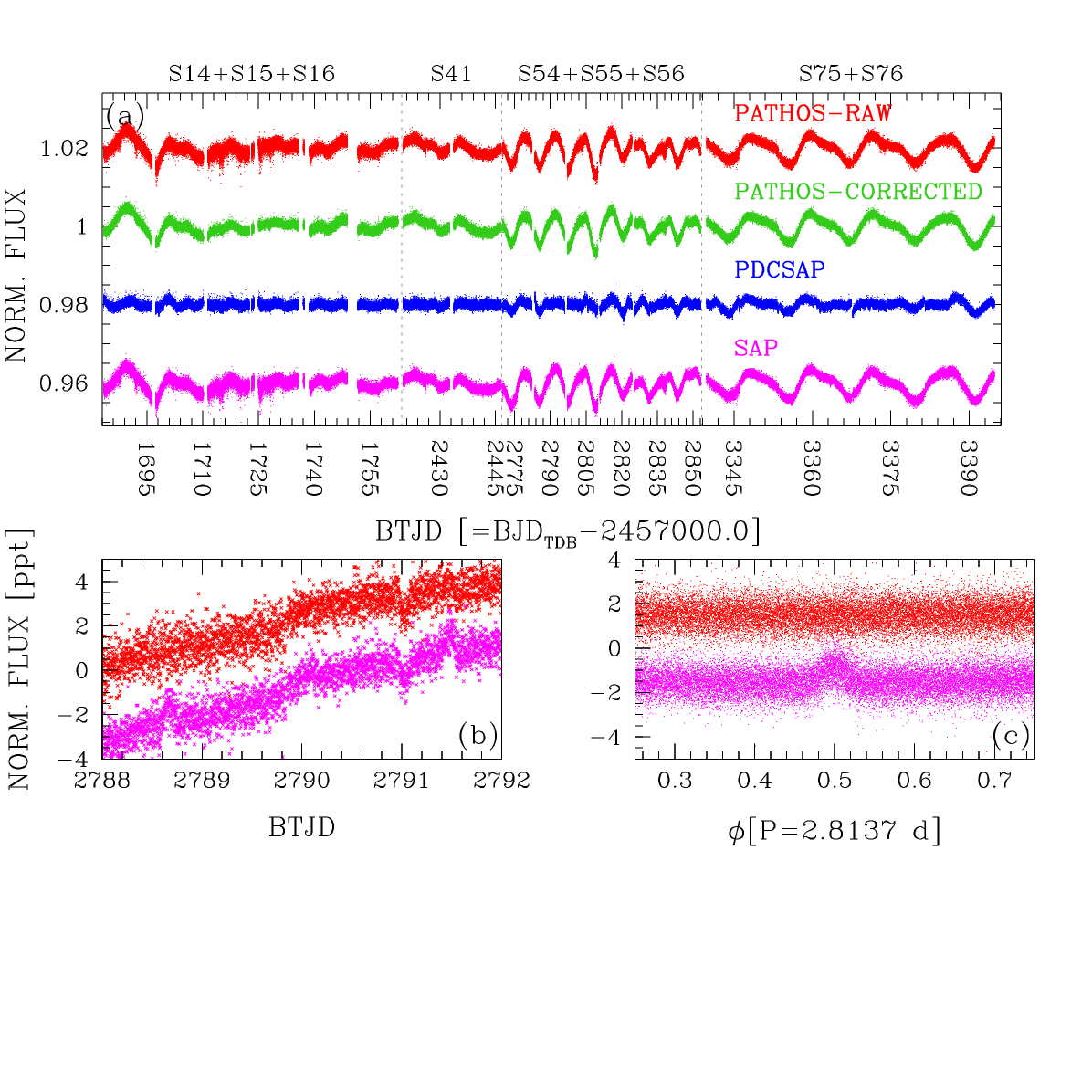} \\
  \caption{Comparison between the PATHOS and official \tess light curves of TOI-1430. Panel (a) shows the SAP and the PDCSAP light curves (in magenta and blue, respectively) compared to the PATHOS raw and corrected light curves (in red and green, respectively). Grey dashed vertical lines separate the different years in which TOI-1430 was observed by \tess. Panel (b) shows a zoom-in of the raw PATHOS (red) and SAP (magenta) light curves at BTJD$=2790 \pm 2$~d: spurious signals are detectable at BTJD$\simeq 2788.7$ and BTJD$\simeq 2791.5$  Panel (c) shows the detrended raw PATHOS and SAP light curves phased with a period P$= 2.8137$~d, and demonstrates that the systematic signal at 2.8137~d is not present in the PATHOS light curve. \label{fig:A1}}
\end{figure*}

In Fig.~\ref{fig:A1} we compare the light curves of TOI-1430 extracted with the PATHOS and the official pipelines.
The need to extract the light curves with a different pipeline arises from two different reasons: (i) from SAP to PDCSAP light curves any original information on the stellar variability is compromised, as demonstrated in panel (a); (ii) the presence of a positive signal in the SAP light curves (P$\sim 2.81$~d) due to a close-by eclipsing binary and bad estimation of the local background (panels (b) and (c)). 

In the PATHOS pipeline, the photometry of the stars is extracted after the subtraction of neighbor contaminants that are modeled by using an input catalog (Gaia DR3) and the TESS PSF models. The local background is measured by using the pixels inside an annulus with inner radius 6 pixel and outer radius 20 pixel and centered on the target star. To avoid the contamination of potential variable stars (like the eclipsing binary TIC~293954660) in the calculation of the local background, the PATHOS routine masks the pixels corresponding to the location of the neighbor stars (on the basis of the Gaia DR3 catalog) and where the contribution of neighbors' PSF is $>0.1$~\%.

% Figure A1

\section{Lithium measurament }

\begin{figure}
  \centering
  \includegraphics[width=0.39\textwidth]{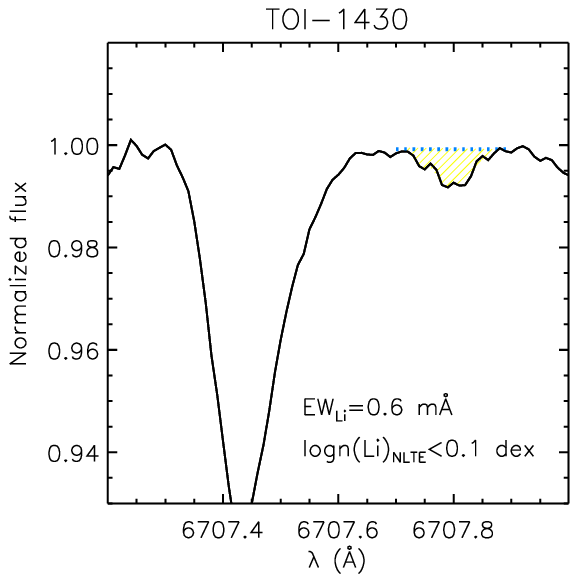} 
  \caption{Coadded spectrum of TOI-1430 around the Li line (solid line). Shaded region represents the line integration for the lithium EW measurement.
  \label{fig:lithium}}
\end{figure}

In Fig.~\ref{fig:lithium} we show the co-added spectrum of the target around the region of the lithium line at $\sim$6707.8 m\AA\,(solid line). In the same figure, we mark with shaded region the area where the lithium equivalent width measurement was performed.

\section{Planet TOI-1430~b: results }
\begin{figure*}
  \centering
  \includegraphics[width=0.99\textwidth]{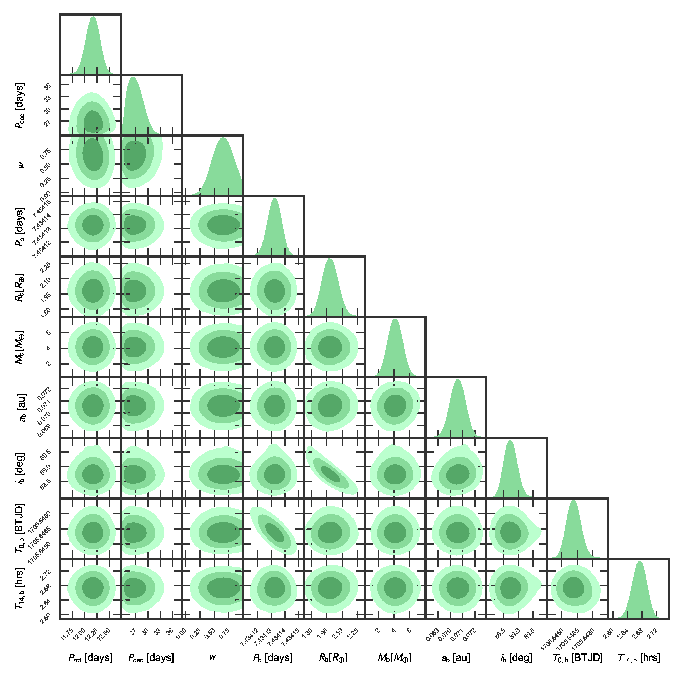} 
  \caption{Corner plot of the posteriors for the main derived parameters of TOI-1430~b and of the stellar activity (multidimensional GP analysis).
  \label{fig:corner}}
\end{figure*}

Figure~\ref{fig:corner} shows the corner plot of the posteriors for the main derived parameters of the stellar activity (from the multidimensional GP analysis of Sect.~\ref{sec:planet}) and of the planet TOI-1430~b.

\section{Joint Photometric and Spectroscopic analysis of TOI-1430~b through quasiperiodic-GP}

\begin{table*}[!htb]
  \renewcommand{\arraystretch}{1.25}
  \centering
  \caption{Stellar activity with quasiperiodic GP and orbital parameters of TOI-1430~b}
\begin{tabular}{l l l l}
\hline
\hline
Parameter    &  Unit  &    Prior      &       Value   \\
\hline
\multicolumn{4}{c}{{\it TOI-1430 Stellar activity}} \\
Stellar rotational period ($P_{\rm rot}$) & days & $\mathcal{N}(11.9, 0.3)$  &   $12.058_{-0.082}^{+0.084}$ \\
Decay Timescale of activity ($P_{\rm dec}$) & days & $\mathcal{U}(13, 1000)$  &   $41.4_{-7.8}^{+6.6}$ \\
Coherence scale ($w$)  &  & ... &  $0.842_{-0.070}^{+0.085}$ \\
Amplitude of the HARPS-N RV signal & m\,s$^{-1}$ & $\mathcal{U}(0.01, 100)$  &   $8.6_{-1.2}^{+1.5}$  \\
HARPS-N RV offset  & m\,s$^{-1}$ & ... &  $-27244.9_{-2.4}^{+2.4}$ \\
Uncorrelated HARPS-N RV jitter   & m\,s$^{-1}$ & ... & $ 2.86_{-0.24}^{+0.25}$  \\
Amplitude of the HIRES RV signal & m\,s$^{-1}$ & $\mathcal{U}(0.01, 100)$  &   $7.8_{-1.2}^{+1.5}$  \\
HIRES RV offset  & m\,s$^{-1}$ & ... & $ -0.7_{-2.2}^{+2.2}$ \\
Uncorrelated HIRES RV jitter   & m\,s$^{-1}$ & ... & $ 2.64_{-0.42}^{+0.47}$  \\
Amplitude of the APF RV signal & m\,s$^{-1}$ & $\mathcal{U}(0.01, 100)$  &   $7.8_{-5.7}^{+9.6}$  \\
APF RV offset  & m\,s$^{-1}$ & ... & $ -0.3_{-5.7}^{+4.9}$ \\
Uncorrelated APF RV jitter   & m\,s$^{-1}$ & ... & $ 7.2_{-3.1}^{+2.6}$  \\
Amplitude of the HARPS-N \logrhk signal &  & ...  & $  0.0254_{-0.0026}^{+0.0031}$ \\
HARPS-N \logrhk offset  &   & ... & $ -4.4768_{-0.0071}^{+0.0070}$ \\
Uncorrelated HARPS-N \logrhk jitter   &  & ... & $ 0.00776_{-0.00079}^{+0.00077}$ \\
Amplitude of the HIRES \logrhk signal &  & ...  & $  0.0242_{-0.0026}^{+0.0031}$ \\
HIRES \logrhk offset  &   & ... & $ -4.44890_{-0.0068}^{+0.0067}$ \\
Uncorrelated HIRES \logrhk jitter   &  & ... & $ 0.00443_{-0.00087}^{+0.00083}$ \\
Amplitude of the APF \logrhk signal &  & ...  & $  0.030_{-0.012}^{+0.017}$ \\
APF \logrhk offset  &   & ... & $ -4.427_{-0.020}^{+0.019}$ \\
Uncorrelated APF \logrhk jitter   &  & ... & $ 0.0321_{-0.0058}^{+0.0078}$ \\
Amplitude of the HARPS-N H$\alpha$ signal &  & ...  & $  0.00367_{-0.00038}^{+0.00045}$ \\
HARPS-N H$\alpha$ offset  &  & ... & $ 0.2478_{-0.0010}^{+0.0010}$ \\
Uncorrelated HARPS-N H$\alpha$ jitter   &  & ... & $ 0.00143_{-0.00013}^{+0.0012}$ \\
Amplitude of the HARPS-N BIS signal & m\,s$^{-1}$ & ...  & $  9.6_{-1.4}^{+1.7}$ \\
HARPS-N BIS offset  & m\,s$^{-1}$ & ... & $ 28.0_{-2.7}^{+2.7}$ \\
Uncorrelated HARPS-N BIS jitter   & m\,s$^{-1}$ & ... & $ 3.13_{-0.21}^{+0.23}$\\
Stellar density ($\rho_\star$) & $\rho_\odot$ & $\mathcal{N}(1.85
, 0.25)$  &   $1.86_{-0.15}^{+0.16}$  \\
\hline
\multicolumn{4}{c}{{\it TOI-1430~b parameters}} \\
Orbital Period ($P_{\rm b}$)     & days   &  $\mathcal{N}(7.43, 0.05)$  & $  7.4341325_{-0.0000044}^{+0.0000042}$ \\
RV semi-amplitude ($K_{\rm b}$) & m\,s$^{-1}$ & $\mathcal{U}(0.01, 10)$ & $1.63_{-0.31}^{+0.31}$ \\
Orbital eccentricity ($e_{\rm b}$)   & deg & ... & 0 (fixed)  \\
Orbital Semi-major axis ($a_{\rm b}$) & au & ... & $ 0.07060_{-0.00055}^{+0.00054}$ \\
Planetary mass ($M_{\rm P,b}$) & $M_{\oplus}$ & ... & $4.47_{-0.84}^{+0.84}$  \\
Planetary density ($\rho_{\rm b}$) & $\rho_{\oplus}$ & ... & $0.57_{-0.13}^{+0.13}$\\
\hline
\end{tabular}
  \label{tab:2}
\end{table*}

\begin{figure*}
  \centering
  \includegraphics[width=0.75\textwidth,bb=25 175 576 705]{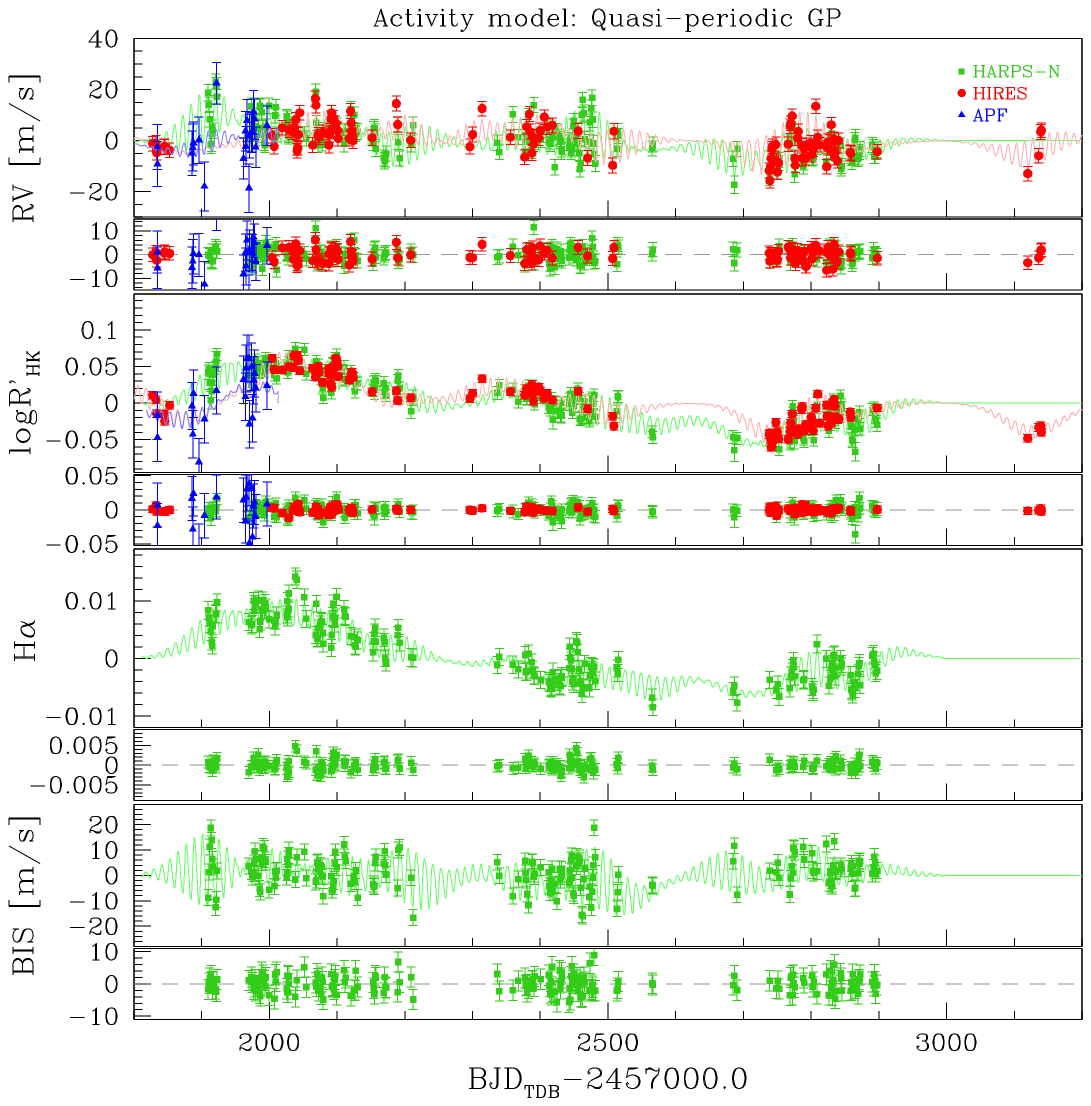} \\
   \includegraphics[width=0.35\textwidth,bb=23 367 443 677]{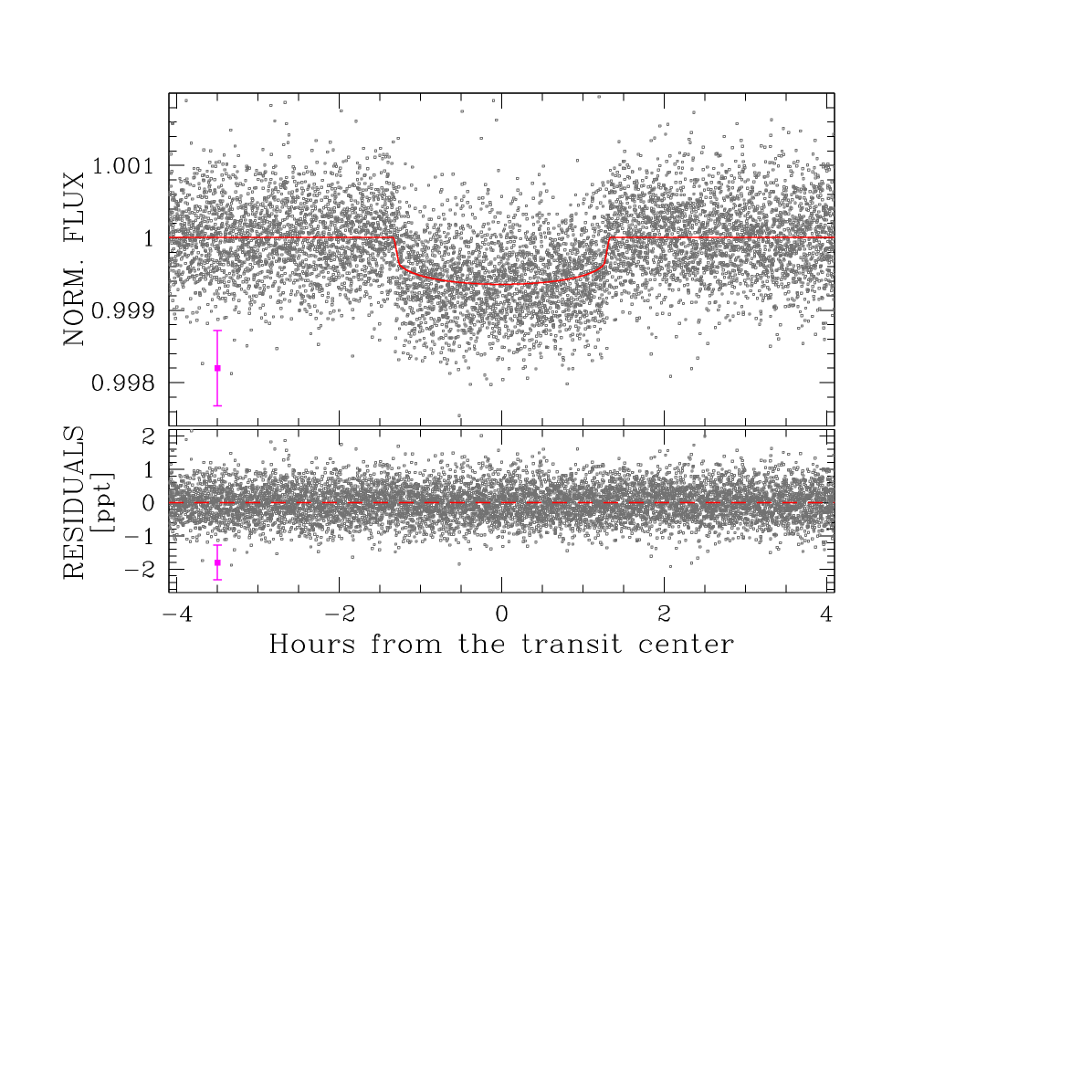} 
   \hspace{0.25cm}
  \includegraphics[width=0.35\textwidth,bb=23 367 443 677]{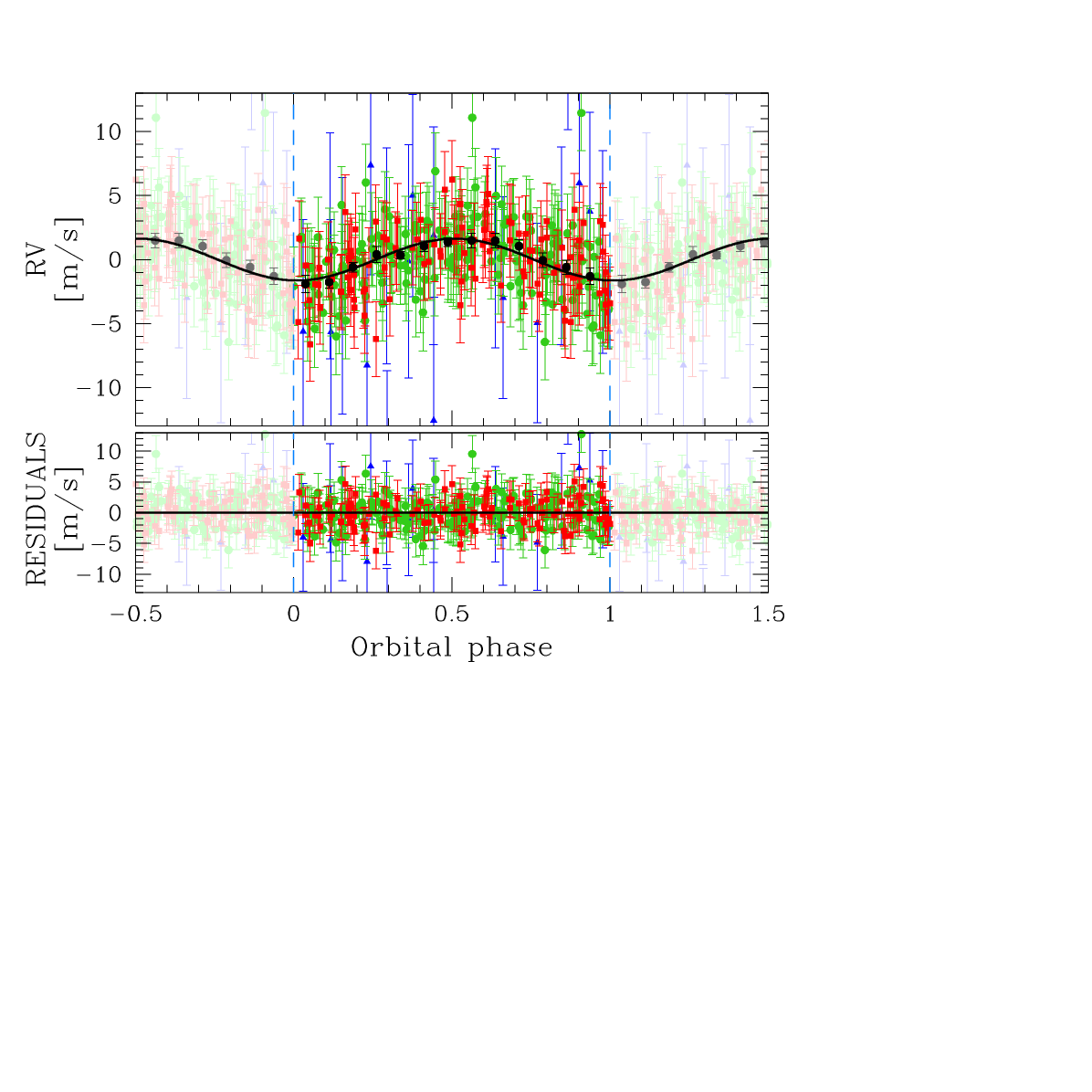}
%  \includegraphics[width=0.8\textwidth,bb=25 228 576 705]{figure/fig7a.ps} \\
%   \includegraphics[width=0.4\textwidth,bb=23 367 443 677]{figure/fig7b.ps} 
%   \hspace{0.25cm}
%  \includegraphics[width=0.4\textwidth,bb=23 367 443 677]{figure/fig7c.ps}
  \caption{As Fig.~\ref{fig:7d}, by adopting the quasi-periodic GP regression for the stellar activity.
  \label{fig:7a}}
\end{figure*}

We carried out a modelling similar to that described in the Section~\ref{sec:planet}, with the exception of using the GP regression, adopting a quasi-periodic kernel as defined by  \citet{2015ApJ...808..127G}:
 \begin{equation}
 \gamma(t_i, t_j) = H^2_{\rm amp} \exp{ \Biggl\{ - \frac{\sin^2{ \left[\pi (t_i -t_j)/ P_{\rm rot} \right]  }  }{2 w^2} - \frac{(t_i-t_j)^2}{2 P^2_{\rm dec}}  \Biggl\} }
\end{equation}
where $H_{\rm amp}$ is the amplitude of the signal, and the other parameters are defined in Sect.~\ref{sec:planet} In this analysis, we treated each dataset with an independent GP regression where only the three main hyper-parameters of the quasi-periodic kernel are shared between the datasets, namely (i) the stellar rotation $P_{\rm rot}$, (ii) the characteristics decay time scale $P_{\rm dec}$, (iii) the coherence scale $w$. In other words, each dataset is characterized by its own independent covariance matrix, with the activity indicators serving as training datasets for the hyper-parameters. We adopted the same priors used in Sect.~\ref{sec:planet} and reported, with the results of the model fitting, in Table~\ref{tab:2}. Figure~\ref{fig:7a} shows the results of the fitting.
 The most important difference with the multi-GP modeling is that the activity models for different datasets are different; indeed the multidimensional GP regression minimizes the risks of overfitting  the activity variations as the underlying GP must be compatible with all the provided radial velocity and activity indicator datasets, even when coming from independent instruments, thus providing more accurate results both inherent in stellar activity and for planetary parameters. 
Despite the less robust treatment of stellar activity by the quasi-periodic GP, we achieved a comparable precision in the planet RV semi-amplitude compared to the multi-dimensional GP analysis, with results in agreement well within $<1 \sigma$, although the median value is slightly higheer (1.6~m\,s$^{-1}$  versus 1.5 m\,s$^{-1}$).

Also in this case we tested the non-null eccentricity case, obtaining $e = 0.130^{+0.150}_{-0.091}$ and similar results of Table~\ref{tab:2}. Running the dynamic nested sampling algorithm, in the case of circular orbit, we obtained a $\log{\mathcal{Z}} \simeq 286.6$, while in the case of Keplerian orbit $\log{\mathcal{Z}} \simeq 285.0$; the circular model is preferred also in this case.

\section{The candidate planet c}
\label{sec:c}

\begin{figure*}
  \centering
    \includegraphics[width=0.33\textwidth,bb=21 417 431 696]{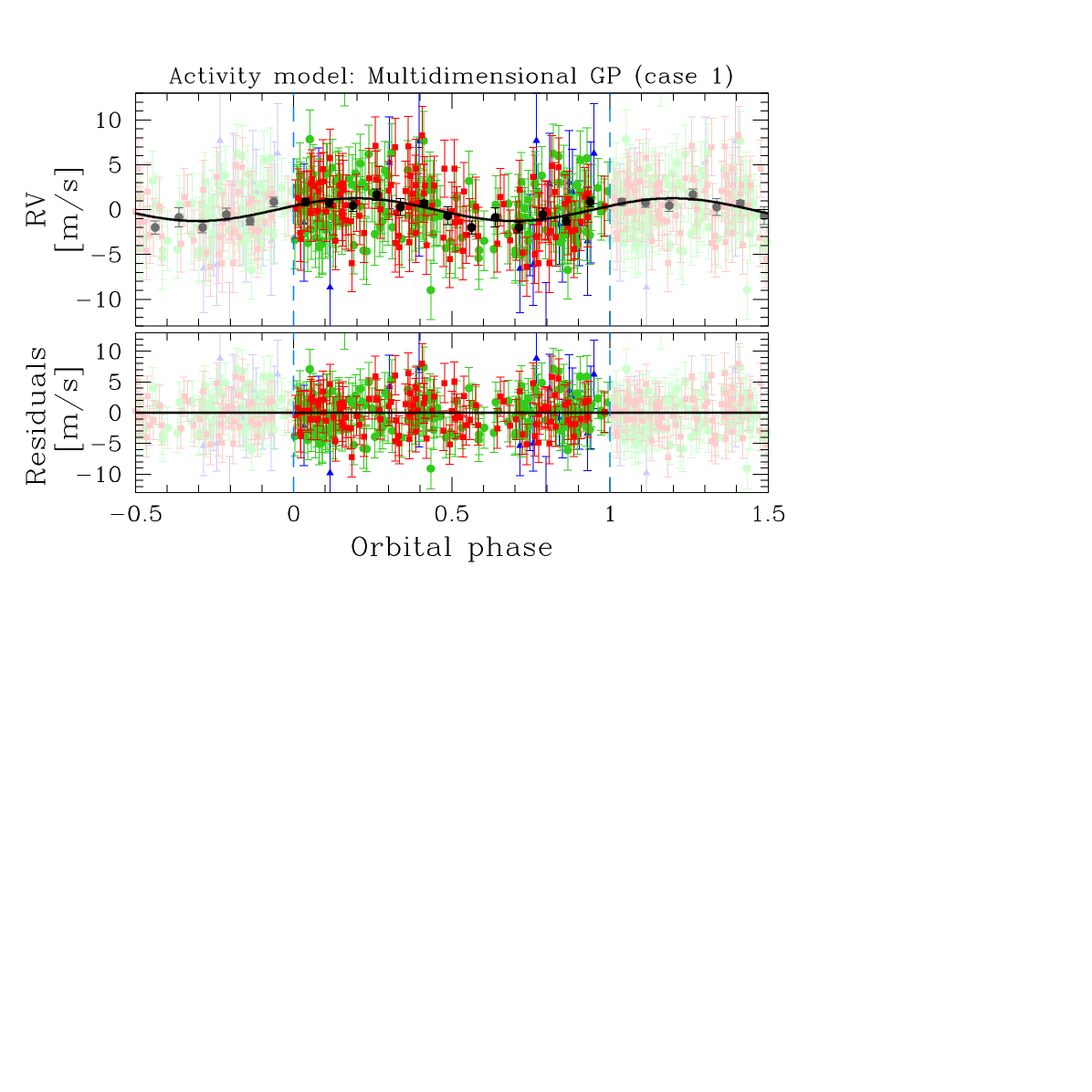} 
    \includegraphics[width=0.33\textwidth,bb=21 417 431 696]{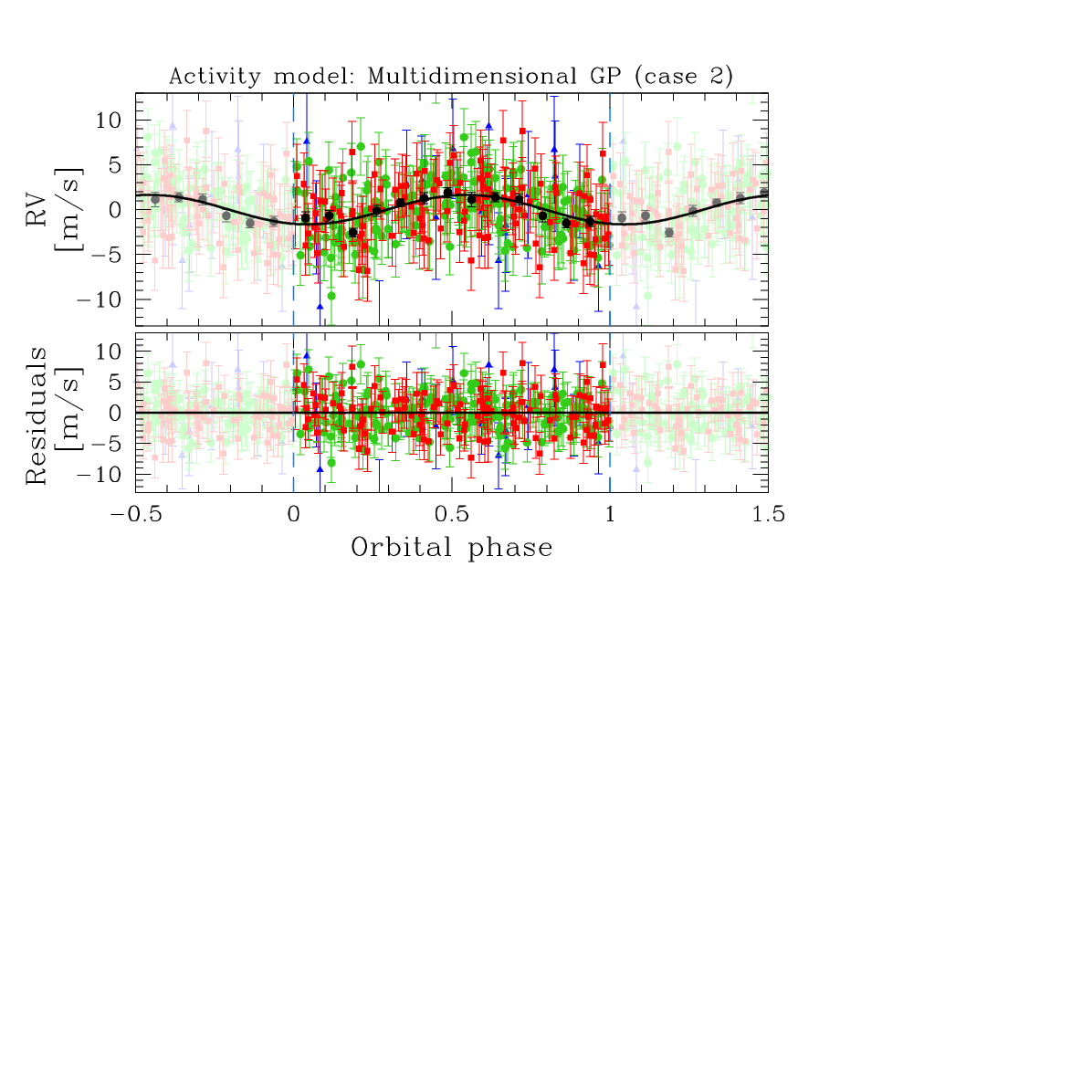}  
   \includegraphics[width=0.33\textwidth,bb=21 417 431 696]{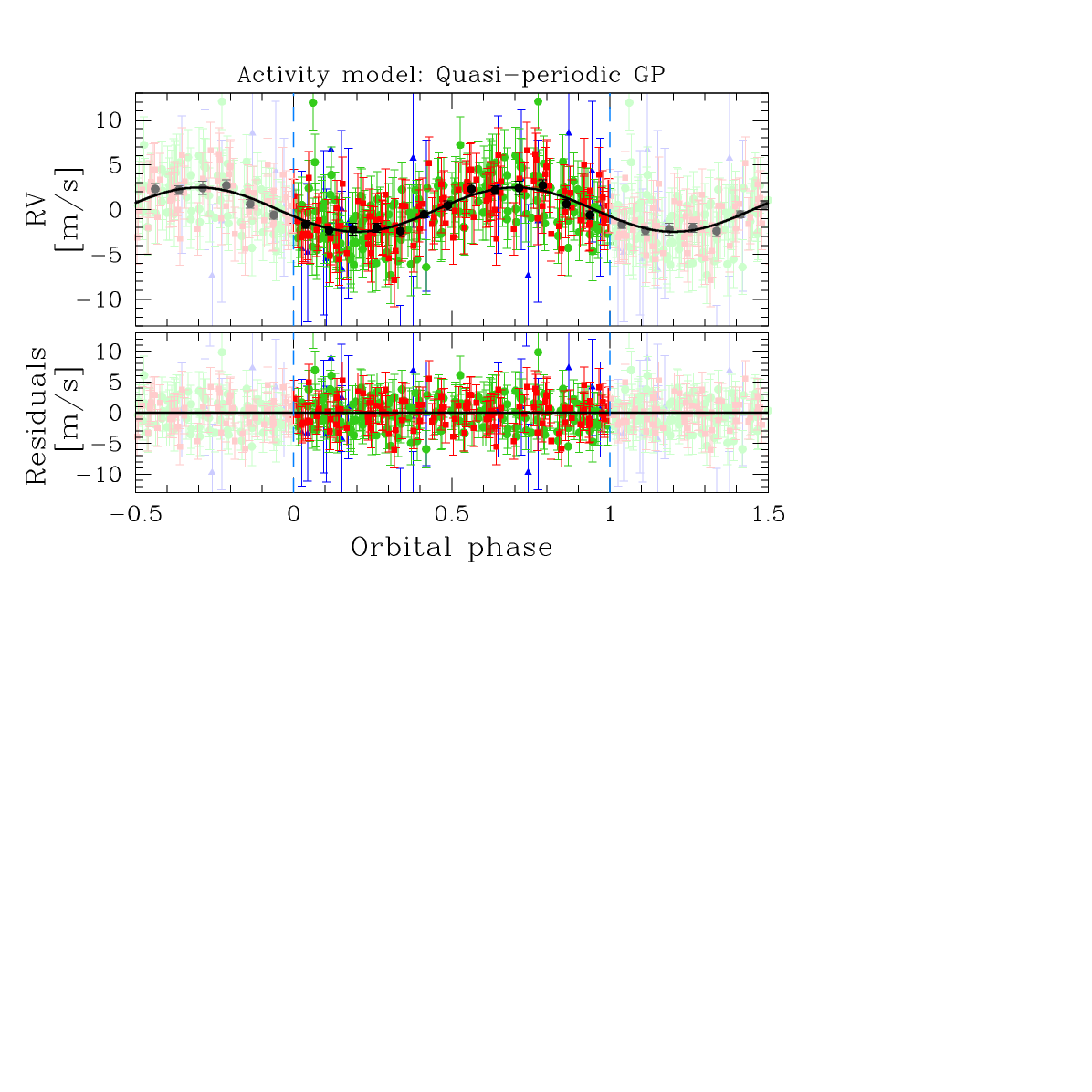} 
  \caption{Candidate planet c signal obtained with different stellar activity models. From the left to the right:  activity modelled with multidimensional GP and orbital period boundaries between 0.4~d and 100.0~d ($P_{\rm c} \sim 96.4$~d); activity modelled with multidimensional GP and orbital period boundaries between 0.4~d and 20.0~d ($P_{\rm c} \sim 12.7$~d); activity modelled with quasi-periodic GP ($P_{\rm c} \sim 13.3$~d).\label{fig:C1}}
\end{figure*}

In addition to the analysis done in section~\ref{sec:planet_c}, we searched for the signal of a second planet modeling the activity with the quasi-periodic GP regression and the same priors reported in Table~\ref{tab:2}. We run \texttt{PyORBIT} to search for planetary signals in the range of periods $0.4 \leq P_{\rm c} \leq 100.0$~days and RV semi-amplitude $0.01 \leq K_{\rm c} \leq 10.00$ m/s. Chains converged after 100\,000 steps, resulting in a signal with period $P_{\rm c} = 13.27 \pm 0.02$~d and RV semi-amplitude $K_{\rm c}=2.5 \pm 0.5$ m/s (corresponding to a mass $M_{\rm P, c} \sin{i_{\rm c}} = 8.2 \pm 1.7 M_{\oplus}$); the resulting parameters for planet TOI-1430~b are the same reported in Table~\ref{tab:2}. Different models give different results (Table~\ref{tab:C1}, Fig.~\ref{fig:C1}), leading to the conclusion that the signal of candidate TOI-1430~c is actually due to residual of stellar activity.

\begin{table*}[!htb]
  \renewcommand{\arraystretch}{1.25}
  \centering
  \caption{Orbital parameters of the candidate TOI-1430~c from RV analysis with different activity models} 
\begin{tabular}{l l l l}
\hline
\hline
Parameter    &  Unit  &    Prior      &       Value   \\
\hline
\multicolumn{4}{c}{{\it Candidate TOI-1430~c -- Activity modelled through multidimensional GP (case 1)}} \\
Orbital Period ($P_{\rm c}$)     & days   &  $\mathcal{U}(0.4, 100.0)$  & $  96.4_{-1.1}^{+1.1}$ \\
Semi-major-axis-to-stellar-radius ratio ($(a_{\rm c}/R_{\star})$) &  & ... & $108.5_{-4.1}^{+4.4}$ \\
Orbital Semi-major axis ($a_{\rm c}$) & au & ... & $ 0.3897_{-0.0043}^{+0.0043}$ \\
RV semi-amplitude ($K_{\rm c}$) & m\,s$^{-1}$ & $\mathcal{U}(0.01, 10)$ & $1.27_{-0.33}^{+0.32}$ \\
Orbital eccentricity ($e_{\rm c}$)   & deg & ... & 0 (fixed)  \\
Mean longitude of the ascending node ($\Omega_{\rm c}$) & deg & ... & $289_{-31}^{+30} $\\
Planetary mass ($M_{\rm P,c} \sin{i_{\rm c}}$) & $M_{\oplus}$ & ... & $8.2_{-2.1}^{+2.1}$  \\
\hline
\multicolumn{4}{c}{{\it Candidate TOI-1430~c -- Activity modelled through multidimensional GP (case 2)}} \\
Orbital Period ($P_{\rm c}$)     & days   &  $\mathcal{U}(0.4 ,20.0)$  & $  12.685_{-0.015}^{+0.015}$ \\
Semi-major-axis-to-stellar-radius ratio ($(a_{\rm c}/R_{\star})$) &  & ... & $28.1_{-1.0}^{+1.1}$ \\
Orbital Semi-major axis ($a_{\rm c}$) & au & ... & $ 0.10079_{-0.0081}^{+0.0080}$ \\
RV semi-amplitude ($K_{\rm c}$) & m\,s$^{-1}$ & $\mathcal{U}(0.01, 10)$ & $1.65_{-0.40}^{+0.39}$ \\
Orbital eccentricity ($e_{\rm c}$)   & deg & ... & 0 (fixed)  \\
Mean longitude of the ascending node ($\Omega_{\rm c}$) & deg & ... & $164_{-31}^{+32} $\\
Planetary mass ($M_{\rm P,c} \sin{i_{\rm c}}$) & $M_{\oplus}$ & ... & $5.4_{-1.3}^{+1.3}$  \\
\hline
\multicolumn{4}{c}{{\it Candidate TOI-1430~c -- Activity modelled through quasiperiodic GP}} \\
Orbital Period ($P_{\rm c}$)     & days   &  $\mathcal{U}(0.4, 100.0)$  & $  13.267_{-0.016}^{+0.016}$ \\
Semi-major-axis-to-stellar-radius ratio ($(a_{\rm c}/R_{\star})$) &  & ... & $28.9_{-1.1}^{+1.1}$ \\
Orbital Semi-major axis ($a_{\rm c}$) & au & ... & $ 0.10385_{-0.00083}^{+0.00081}$ \\
RV semi-amplitude ($K_{\rm c}$) & m\,s$^{-1}$ & $\mathcal{U}(0.01, 10)$ & $2.47_{-0.50}^{+0.50}$ \\
Orbital eccentricity ($e_{\rm c}$)   & deg & ... & 0 (fixed)  \\
Mean longitude of the ascending node ($\Omega_{\rm c}$) & deg & ... & $108_{-23}^{+23} $\\
Planetary mass ($M_{\rm P,c} \sin{i_{\rm c}}$) & $M_{\oplus}$ & ... & $8.2_{-1.7}^{+1.7}$  \\
\hline
\end{tabular}
  \label{tab:C1}
\end{table*}

\section{TTV Analysis}

\begin{figure*}
  \centering
  \includegraphics[width=0.3\textwidth,bb=71 145 285 715]{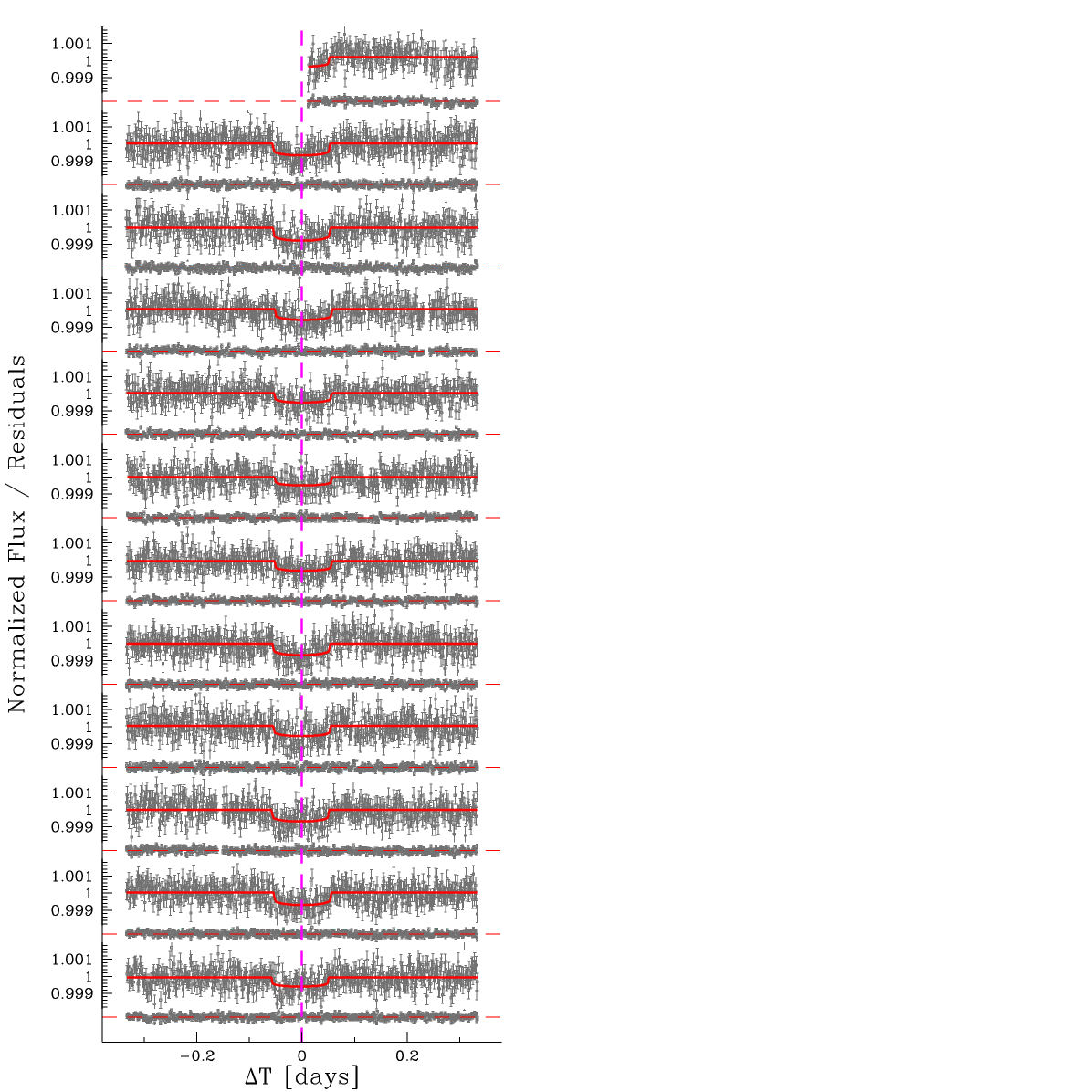} 
  \includegraphics[width=0.3\textwidth,bb=71 145 285 715]{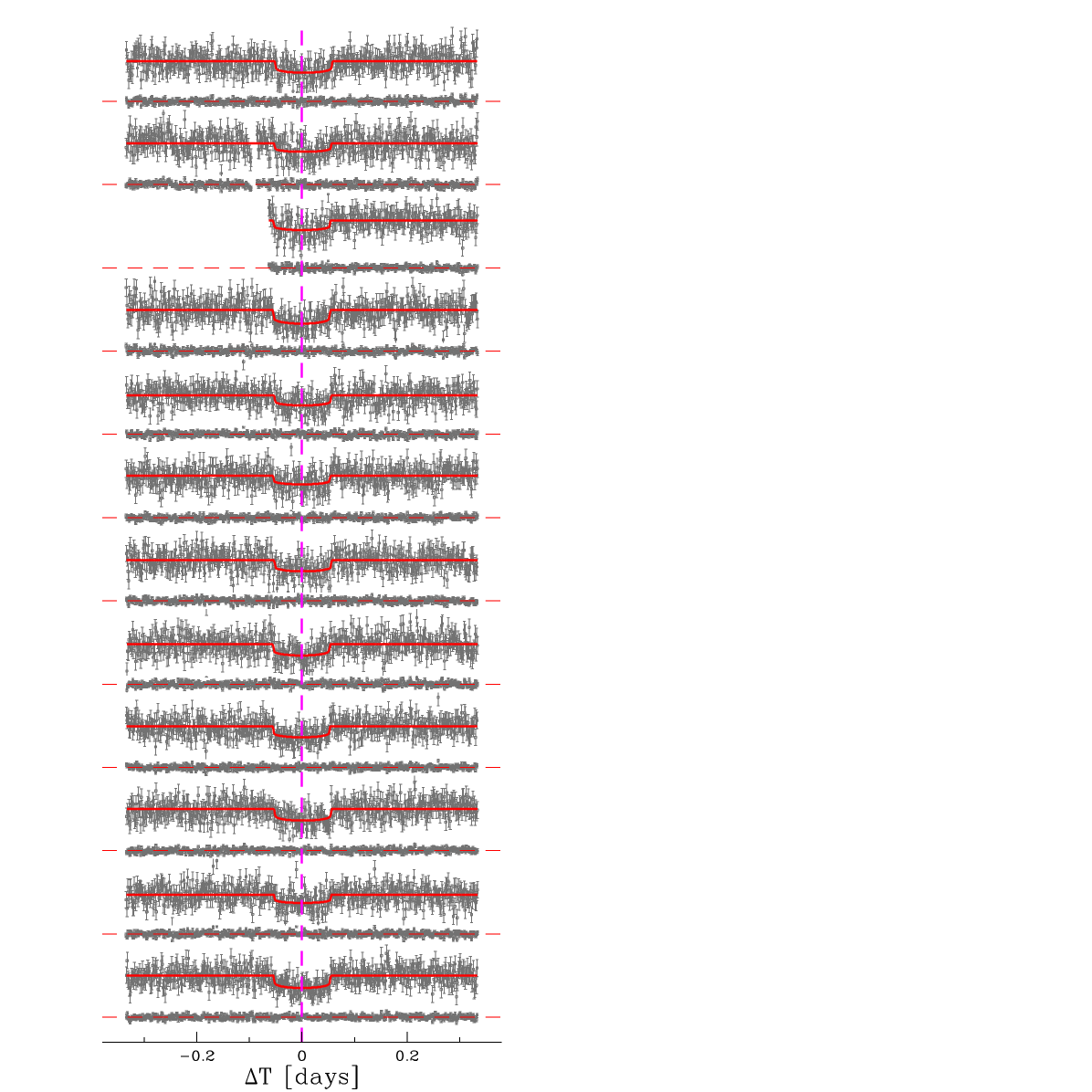} 
  \includegraphics[width=0.3\textwidth,bb=71 145 285 715]{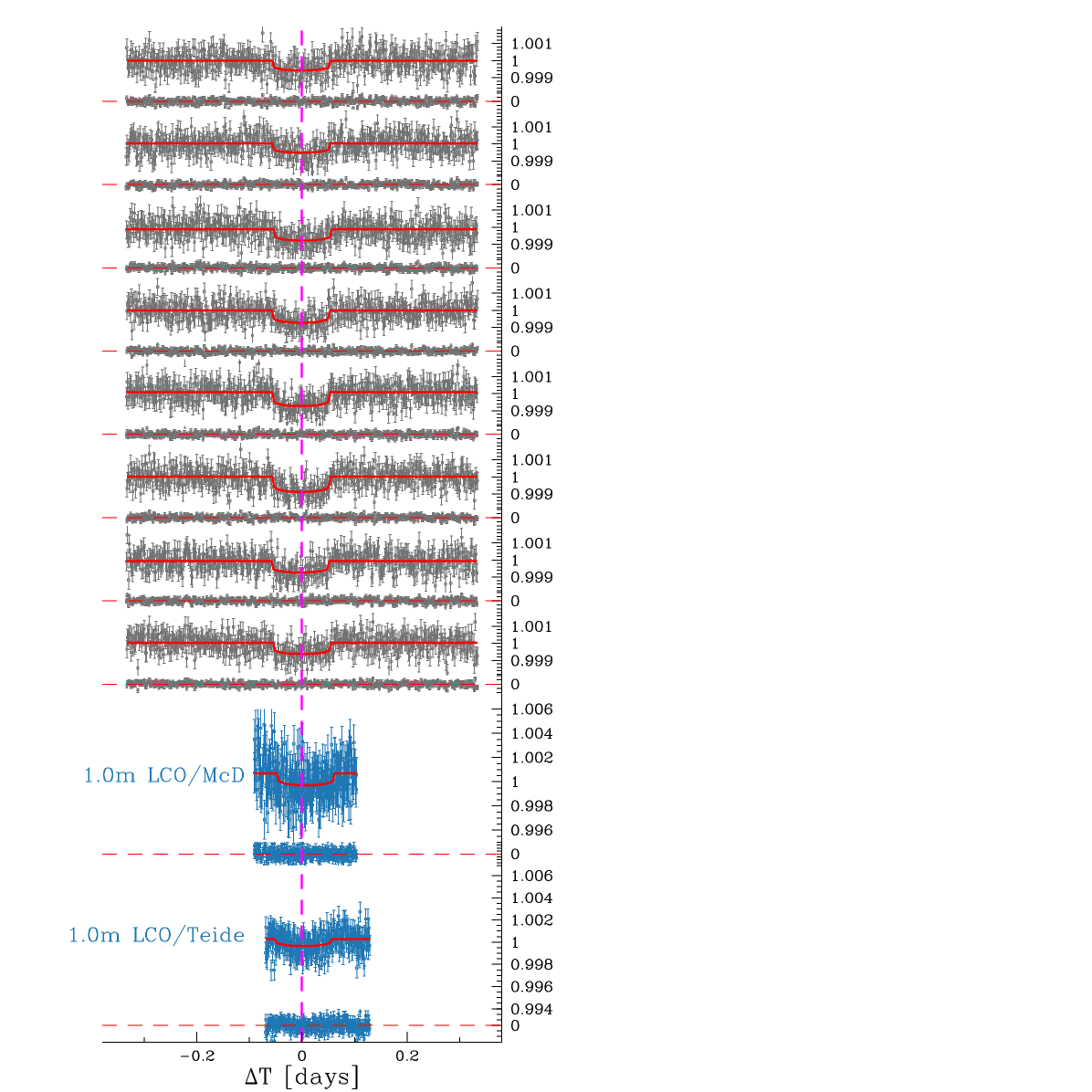} 
  \caption{Analysis of TTVs based on \tess (grey points) and LCO (azure points) data. Panels shows  every single transit centred on the expected $T_0$ (magenta dashed line) and the fitted model (in red). Below each transit, the residuals are reported. 
  \label{fig:9A}}
\end{figure*}

In Table~\ref{tab:D1} the Central times and Planetary-to-stellar-radius ratio from the transits observed with \tess and LCO facilities are reported. Figure~\ref{fig:9A} shows the modeling of single transits observed by TESS and LCO. 

\begin{table*}[!htb]
  \renewcommand{\arraystretch}{1.25}
  \centering
  \caption{Central times and Planetary-to-stellar-radius ratio from the transits observed with \tess and LCO facilities} 
  \resizebox{\textwidth}{!}{
\begin{tabular}{l c c l c c l c c}
\hline
\hline
Telescope    &  $T_0$ [BJD$_{\rm TDB}-2457000.0$] &    $R_{\rm P,b}/R_\star$  & Telescope    &  $T_0$ [BJD$_{\rm TDB}-2457000.0$] &    $R_{\rm P,b}/R_\star$ & Telescope    &  $T_0$ [BJD$_{\rm TDB}-2457000.0$] &    $R_{\rm P,b}/R_\star$  \\
\hline 
TESS   &      $1683.3416_{-0.0130}^{+0.0040}$ & $0.0225_{-0.0032}^{+0.0029}$ &  TESS   &      $2441.6286_{-0.0032}^{+0.0024}$ & $0.0245_{-0.0013}^{+0.0012}$ & TESS   &      $2850.5030_{-0.0023}^{+0.0018}$ & $0.0258_{-0.0013}^{+0.0012}$ \\        
TESS   &      $1690.7768_{-0.0025}^{+0.0020}$ & $0.0251_{-0.0023}^{+0.0021}$ &  TESS   &      $2776.1629_{-0.0021}^{+0.0027}$ & $0.0212_{-0.0015}^{+0.0014}$ & TESS   &      $3341.1541_{-0.0019}^{+0.0024}$ & $0.0224_{-0.0015}^{+0.0014}$ \\        
TESS   &      $1698.2116_{-0.0010}^{+0.0012}$ & $0.0263_{-0.0027}^{+0.0028}$ &  TESS   &      $2783.5952_{-0.0040}^{+0.0024}$ & $0.0224_{-0.0013}^{+0.0013}$ & TESS   &      $3348.5882_{-0.0026}^{+0.0040}$ & $0.0221_{-0.0012}^{+0.0012}$ \\        
TESS   &      $1705.6493_{-0.0017}^{+0.0028}$ & $0.0241_{-0.0013}^{+0.0013}$ &  TESS   &      $2791.0289_{-0.0015}^{+0.0014}$ & $0.0268_{-0.0016}^{+0.0015}$ & TESS   &      $3356.0249_{-0.0071}^{+0.0020}$ & $0.0243_{-0.0015}^{+0.0014}$ \\        
TESS   &      $1713.0824_{-0.0032}^{+0.0046}$ & $0.0225_{-0.0012}^{+0.0012}$ &  TESS   &      $2798.4652_{-0.0019}^{+0.0014}$ & $0.0230_{-0.0015}^{+0.0014}$ & TESS   &      $3363.4558_{-0.0030}^{+0.0048}$ & $0.0255_{-0.0015}^{+0.0014}$ \\        
TESS   &      $1720.5170_{-0.0019}^{+0.0025}$ & $0.0210_{-0.0014}^{+0.0013}$ &  TESS   &      $2805.8974_{-0.0015}^{+0.0016}$ & $0.0211_{-0.0012}^{+0.0012}$ & TESS   &      $3370.8904_{-0.0015}^{+0.0014}$ & $0.0272_{-0.0014}^{+0.0013}$ \\        
TESS   &      $1727.9512_{-0.0041}^{+0.0070}$ & $0.0228_{-0.0014}^{+0.0013}$ &  TESS   &      $2813.3340_{-0.0024}^{+0.0020}$ & $0.0241_{-0.0013}^{+0.0013}$ & TESS   &      $3378.3251_{-0.0012}^{+0.0013}$ & $0.0283_{-0.0013}^{+0.0012}$ \\        
TESS   &      $1735.3817_{-0.0012}^{+0.0015}$ & $0.0247_{-0.0016}^{+0.0015}$ &  TESS   &      $2820.7653_{-0.0015}^{+0.0016}$ & $0.0245_{-0.0015}^{+0.0014}$ & TESS   &      $3385.7572_{-0.0014}^{+0.0020}$ & $0.0247_{-0.0013}^{+0.0012}$ \\        
TESS   &      $1742.8169_{-0.0016}^{+0.0026}$ & $0.0232_{-0.0014}^{+0.0013}$ &  TESS   &      $2828.1993_{-0.0020}^{+0.0013}$ & $0.0241_{-0.0013}^{+0.0012}$ & TESS   &      $3393.1947_{-0.0013}^{+0.0013}$ & $0.0243_{-0.0011}^{+0.0011}$ \\        
TESS   &      $1757.6821_{-0.0020}^{+0.0029}$ & $0.0249_{-0.0014}^{+0.0013}$ &  TESS   &      $2835.6361_{-0.0015}^{+0.0036}$ & $0.0249_{-0.0013}^{+0.0013}$ & LCO/1.0m~McDonald   & $2359.8571_{-0.0049}^{+0.0096}$ & $0.0296_{-0.0014}^{+0.0013}$ \\
TESS   &      $2426.7581_{-0.0022}^{+0.0016}$ & $0.0256_{-0.0014}^{+0.0014}$ &  TESS   &      $2843.0699_{-0.0032}^{+0.0025}$ & $0.0210_{-0.0013}^{+0.0012}$ & LCO/1.0m~Teide      & $2798.4662_{-0.0039}^{+0.0048}$ & $0.0233_{-0.0013}^{+0.0013}$ \\
TESS   &      $2434.1877_{-0.0032}^{+0.0038}$ & $0.0222_{-0.0013}^{+0.0013}$ &         &                                                             &        &                 \\
\hline
\end{tabular}}
  \label{tab:D1}
\end{table*}

\section{Spectroscopic series}
Tables~\ref{tab:E1}, ~\ref{tab:E2}, and ~\ref{tab:E3} include the HARPS-N, HIRES, and APF spectroscopic time series used in this work.

\begin{table*}[!htb]
  \renewcommand{\arraystretch}{0.99}
  \centering
  \caption{HARPS-N@TNG spectroscopic series} 
  \resizebox{0.99\textwidth}{!}{
  \begin{tabular}{l c c c c c c c || l c c c c c c c}
\hline
\hline
BJD$_{\rm TDB}-2457000.0$   &   RV  &   $\sigma_{\rm RV}$ & log$R'_{\rm HK}$ &  $\sigma_{{\rm log} R'_{\rm HK}}$ &H$\alpha$ &  $\sigma_{{\rm H \alpha}}$ & BIS & BJD$_{\rm TDB}-$2450000   &   RV  &   $\sigma_{\rm RV}$ & log$R'_{\rm HK}$ &  $\sigma_{{\rm log} R'_{\rm HK}}$ &H$\alpha$ &  $\sigma_{{\rm H \alpha}}$ & BIS \\
               & (m~s$^{-1}$) & (m~s$^{-1}$) &        &      &    &     &  (m~s$^{-1}$)  &                  & (m~s$^{-1}$) & (m~s$^{-1}$) &        &      &    &     &  (m~s$^{-1}$) \\
\hline
1909.771903611   & $-$27226.2 & 1.5 & $-$4.4349 & 0.0063  & 0.25619  & 0.00065  & 19.3  & 2413.651394081 & $-$27245.3  & 0.9 & $-$4.4825  & 0.0033  & 0.24362  & 0.00038  & 23.8 \\
1910.774268941   & $-$27230.9 & 1.0 & $-$4.4300 & 0.0034  & 0.25495  & 0.00045  & 29.3  & 2414.677514241 & $-$27246.2  & 0.9 & $-$4.4861  & 0.0031  & 0.24280  & 0.00039  & 21.2 \\
1911.778914971   & $-$27236.6 & 1.2 & $-$4.4492 & 0.0051  & 0.25389  & 0.00055  & 39.3  & 2415.709768291 & $-$27242.1  & 0.8 & $-$4.4812  & 0.0026  & 0.24358  & 0.00030  & 26.0 \\
1913.754504871   & $-$27246.6 & 1.8 & $-$4.4729 & 0.0091  & 0.25258  & 0.00068  & 46.7  & 2416.619079731 & $-$27240.1  & 0.6 & $-$4.4705  & 0.0018  & 0.24262  & 0.00026  & 22.6 \\
1914.776849511   & $-$27246.2 & 0.8 & $-$4.4609 & 0.0027  & 0.25077  & 0.00036  & 41.8  & 2417.612952531 & $-$27240.2  & 0.7 & $-$4.4779  & 0.0025  & 0.24295  & 0.00033  & 30.6 \\
1915.776150601   & $-$27239.8 & 0.7 & $-$4.4488 & 0.0022  & 0.25009  & 0.00029  & 34.4  & 2418.611176811 & $-$27243.7  & 1.8 & $-$4.4997  & 0.0095  & 0.24550  & 0.00071  & 21.4 \\
1916.774885121   & $-$27241.1 & 0.9 & $-$4.4377 & 0.0032  & 0.25345  & 0.00041  & 31.4  & 2421.604168551 & $-$27255.4  & 1.0 & $-$4.4760  & 0.0037  & 0.24429  & 0.00042  & 24.5 \\
1920.762930621   & $-$27221.8 & 0.9 & $-$4.4286 & 0.0032  & 0.25558  & 0.00039  & 15.4  & 2424.688743011 & $-$27241.4  & 1.2 & $-$4.4883  & 0.0057  & 0.24379  & 0.00047  & 17.8 \\
1921.740151511   & $-$27223.8 & 2.1 & $-$4.4199 & 0.0109  & 0.25562  & 0.00079  & 18.5  & 2428.549931931 & $-$27243.5  & 1.1 & $-$4.4873  & 0.0045  & 0.24320  & 0.00044  & 30.5 \\
1922.766736901   & $-$27227.8 & 0.8 & $-$4.4103 & 0.0025  & 0.25759  & 0.00034  & 29.8  & 2429.673499641 & $-$27245.0  & 1.0 & $-$4.4940  & 0.0038  & 0.24446  & 0.00043  & 26.1 \\
1969.723335231   & $-$27246.0 & 0.7 & $-$4.4385 & 0.0021  & 0.25416  & 0.00037  & 31.7  & 2430.616138591 & $-$27244.2  & 1.0 & $-$4.4881  & 0.0039  & 0.24305  & 0.00042  & 27.5 \\
1973.695124831   & $-$27231.5 & 1.4 & $-$4.4514 & 0.0059  & 0.25365  & 0.00053  & 29.5  & 2431.526253551 & $-$27249.2  & 1.5 & $-$4.4938  & 0.0071  & 0.24476  & 0.00061  & 28.6 \\
1974.678489081   & $-$27239.4 & 0.9 & $-$4.4394 & 0.0029  & 0.25387  & 0.00042  & 27.3  & 2431.631058011 & $-$27248.4  & 0.8 & $-$4.4766  & 0.0030  & 0.24450  & 0.00035  & 30.4 \\
1976.676665861   & $-$27235.2 & 0.8 & $-$4.4218 & 0.0027  & 0.25786  & 0.00037  & 28.8  & 2442.652055351 & $-$27247.5  & 1.3 & $-$4.4777  & 0.0057  & 0.24614  & 0.00050  & 25.1 \\
1977.694383561   & $-$27235.9 & 0.7 & $-$4.4245 & 0.0022  & 0.25618  & 0.00034  & 33.0  & 2443.594814961 & $-$27245.8  & 1.0 & $-$4.4718  & 0.0037  & 0.24988  & 0.00046  & 29.7 \\
1978.700827911   & $-$27239.4 & 0.8 & $-$4.4181 & 0.0024  & 0.25728  & 0.00040  & 37.6  & 2444.540525481 & $-$27244.7  & 0.9 & $-$4.4638  & 0.0033  & 0.24808  & 0.00043  & 26.6 \\
1979.709784071   & $-$27244.2 & 0.7 & $-$4.4175 & 0.0020  & 0.25726  & 0.00031  & 34.6  & 2445.568342301 & $-$27249.3  & 0.8 & $-$4.4711  & 0.0026  & 0.24756  & 0.00037  & 34.2 \\
1983.697900081   & $-$27239.9 & 1.9 & $-$4.4304 & 0.0086  & 0.25797  & 0.00069  & 27.8  & 2446.594908531 & $-$27248.4  & 0.8 & $-$4.4705  & 0.0027  & 0.24608  & 0.00035  & 35.7 \\
1985.634317931   & $-$27231.3 & 0.8 & $-$4.4348 & 0.0026  & 0.25296  & 0.00033  & 22.8  & 2447.590872231 & $-$27247.8  & 0.8 & $-$4.4897  & 0.0030  & 0.24385  & 0.00036  & 27.1 \\
1986.665081121   & $-$27231.6 & 0.8 & $-$4.4301 & 0.0025  & 0.25425  & 0.00033  & 21.5  & 2448.565249681 & $-$27239.4  & 0.6 & $-$4.4757  & 0.0018  & 0.24447  & 0.00027  & 21.0 \\
1988.693909891   & $-$27233.9 & 0.8 & $-$4.4271 & 0.0027  & 0.25626  & 0.00036  & 33.0  & 2449.590147511 & $-$27237.2  & 0.7 & $-$4.4807  & 0.0025  & 0.24338  & 0.00030  & 20.0 \\
1989.706837041   & $-$27242.7 & 0.8 & $-$4.4276 & 0.0026  & 0.25637  & 0.00038  & 34.7  & 2453.495508481 & $-$27243.7  & 0.7 & $-$4.4704  & 0.0023  & 0.25085  & 0.00044  & 31.6 \\
1990.696742641   & $-$27240.1 & 0.8 & $-$4.4223 & 0.0028  & 0.25792  & 0.00044  & 39.4  & 2454.480889611 & $-$27245.7  & 0.6 & $-$4.4657  & 0.0021  & 0.25059  & 0.00036  & 35.4 \\
1992.701485301   & $-$27245.7 & 1.0 & $-$4.4329 & 0.0034  & 0.25623  & 0.00045  & 38.6  & 2455.564533911 & $-$27246.2  & 0.7 & $-$4.4575  & 0.0021  & 0.24892  & 0.00029  & 32.7 \\
1993.684214791   & $-$27246.7 & 0.7 & $-$4.4199 & 0.0021  & 0.25746  & 0.00033  & 34.2  & 2457.534784051 & $-$27249.8  & 0.7 & $-$4.4703  & 0.0022  & 0.24597  & 0.00031  & 35.0 \\
1994.689734401   & $-$27239.5 & 0.9 & $-$4.4155 & 0.0028  & 0.25673  & 0.00034  & 33.2  & 2458.610127891 & $-$27256.2  & 1.1 & $-$4.4986  & 0.0046  & 0.24374  & 0.00052  & 33.8 \\
1999.615280101   & $-$27231.2 & 1.0 & $-$4.4232 & 0.0031  & 0.25467  & 0.00040  & 22.3  & 2459.507960211 & $-$27253.6  & 0.8 & $-$4.4984  & 0.0032  & 0.24347  & 0.00043  & 23.2 \\
2007.672813081   & $-$27244.0 & 0.6 & $-$4.4384 & 0.0019  & 0.25442  & 0.00034  & 23.7  & 2461.547152341 & $-$27234.8  & 0.8 & $-$4.4975  & 0.0029  & 0.24172  & 0.00033  & 12.6 \\
2008.683551781   & $-$27235.1 & 0.7 & $-$4.4313 & 0.0022  & 0.25342  & 0.00030  & 29.7  & 2462.514704991 & $-$27229.0  & 0.8 & $-$4.4960  & 0.0027  & 0.24286  & 0.00033  & 12.0 \\
2009.687368831   & $-$27231.9 & 0.8 & $-$4.4224 & 0.0025  & 0.25446  & 0.00030  & 27.2  & 2464.515271211 & $-$27234.2  & 1.3 & $-$4.4826  & 0.0058  & 0.24273  & 0.00054  & 26.5 \\
2010.704693101   & $-$27242.6 & 0.7 & $-$4.4428 & 0.0021  & 0.25534  & 0.00036  & 29.9  & 2465.538319511 & $-$27244.4  & 0.8 & $-$4.4743  & 0.0028  & 0.24691  & 0.00040  & 31.0 \\
2025.634870751   & $-$27235.8 & 0.7 & $-$4.4260 & 0.0023  & 0.25551  & 0.00036  & 27.3  & 2472.496831081 & $-$27245.4  & 0.6 & $-$4.5012  & 0.0021  & 0.24266  & 0.00030  & 19.7 \\
2026.637885491   & $-$27236.3 & 0.8 & $-$4.4285 & 0.0025  & 0.25499  & 0.00042  & 29.1  & 2473.473590691 & $-$27237.8  & 1.7 & $-$4.4958  & 0.0082  & 0.24495  & 0.00064  & 23.6 \\
2026.728050091   & $-$27237.3 & 0.8 & $-$4.4348 & 0.0028  & 0.25779  & 0.00046  & 34.5  & 2475.480140101 & $-$27232.3  & 1.4 & $-$4.4888  & 0.0063  & 0.24647  & 0.00055  & 15.3 \\
2027.709871781   & $-$27242.1 & 0.9 & $-$4.4346 & 0.0030  & 0.25614  & 0.00051  & 32.2  & 2476.481758921 & $-$27228.2  & 0.9 & $-$4.4699  & 0.0033  & 0.24589  & 0.00036  & 21.6 \\
2028.646460451   & $-$27243.2 & 0.8 & $-$4.4217 & 0.0023  & 0.25651  & 0.00033  & 35.8  & 2477.476363431 & $-$27236.7  & 1.0 & $-$4.4722  & 0.0036  & 0.24640  & 0.00036  & 27.2 \\
2028.707550111   & $-$27242.8 & 0.7 & $-$4.4127 & 0.0020  & 0.25914  & 0.00030  & 38.0  & 2478.472437311 & $-$27246.0  & 1.1 & $-$4.4754  & 0.0043  & 0.24556  & 0.00043  & 32.1 \\
2038.612913011   & $-$27238.6 & 0.8 & $-$4.4022 & 0.0026  & 0.26207  & 0.00038  & 26.4  & 2479.464320911 & $-$27250.3  & 1.0 & $-$4.4659  & 0.0035  & 0.24633  & 0.00035  & 46.8 \\
2040.627498351   & $-$27247.5 & 1.3 & $-$4.4098 & 0.0046  & 0.26155  & 0.00051  & 35.2  & 2481.474903301 & $-$27253.9  & 1.1 & $-$4.4953  & 0.0046  & 0.24400  & 0.00044  & 35.1 \\
2051.686634181   & $-$27242.3 & 0.8 & $-$4.4036 & 0.0025  & 0.25848  & 0.00033  & 37.3  & 2513.361547711 & $-$27248.1  & 1.0 & $-$4.4883  & 0.0044  & 0.24613  & 0.00044  & 14.9 \\
2054.696144781   & $-$27244.1 & 0.9 & $-$4.4369 & 0.0029  & 0.25474  & 0.00036  & 27.0  & 2514.410475121 & $-$27241.2  & 0.7 & $-$4.4676  & 0.0025  & 0.24533  & 0.00027  & 21.4 \\
2068.472124481   & $-$27225.8 & 1.0 & $-$4.4328 & 0.0033  & 0.25194  & 0.00034  & 25.5  & 2515.339209041 & $-$27243.1  & 0.6 & $-$4.4686  & 0.0021  & 0.24760  & 0.00028  & 28.3 \\
2069.513350341   & $-$27238.0 & 0.5 & $-$4.4199 & 0.0015  & 0.25736  & 0.00028  & 26.2  & 2565.300429561 & $-$27248.0  & 0.6 & $-$4.5161  & 0.0025  & 0.24108  & 0.00028  & 24.3 \\
2070.643766261   & $-$27244.9 & 0.9 & $-$4.4303 & 0.0030  & 0.25137  & 0.00034  & 33.0  & 2566.307700821 & $-$27246.5  & 0.7 & $-$4.5237  & 0.0030  & 0.23940  & 0.00033  & 23.9 \\
2071.667675931   & $-$27246.5 & 1.0 & $-$4.4312 & 0.0034  & 0.25129  & 0.00039  & 33.4  & 2684.735542791 & $-$27252.0  & 0.9 & $-$4.5233  & 0.0042  & 0.24222  & 0.00043  & 33.6 \\
2072.658610231   & $-$27241.9 & 0.9 & $-$4.4251 & 0.0030  & 0.25309  & 0.00033  & 30.6  & 2686.698800471 & $-$27262.3  & 1.9 & $-$4.5415  & 0.0132  & 0.24305  & 0.00071  & 39.7 \\
2075.556664051   & $-$27243.0 & 0.8 & $-$4.4322 & 0.0025  & 0.25379  & 0.00035  & 32.0  & 2687.657020611 & $-$27252.8  &10.3 & $-$4.6857  & 0.1459  & 0.25976  & 0.00210  & 45.0 \\
2076.637249551   & $-$27243.6 & 1.0 & $-$4.4428 & 0.0033  & 0.25432  & 0.00045  & 28.2  & 2690.728216121 & $-$27248.2  & 0.7 & $-$4.5251  & 0.0030  & 0.24017  & 0.00029  & 20.4 \\
2077.648169671   & $-$27244.7 & 0.9 & $-$4.4379 & 0.0027  & 0.25421  & 0.00033  & 31.4  & 2738.690721451 & $-$27259.5  & 0.9 & $-$4.5265  & 0.0036  & 0.24415  & 0.00040  & 29.8 \\
2078.605269511   & $-$27246.1 & 0.5 & $-$4.4562 & 0.0016  & 0.25227  & 0.00030  & 27.5  & 2748.706048891 & $-$27249.8  & 1.0 & $-$4.5216  & 0.0044  & 0.24239  & 0.00039  & 29.2 \\
2079.595586571   & $-$27240.8 & 0.7 & $-$4.4695 & 0.0024  & 0.25045  & 0.00037  & 19.8  & 2750.722279071 & $-$27256.9  & 1.0 & $-$4.5188  & 0.0045  & 0.24338  & 0.00040  & 33.1 \\
2091.533468191   & $-$27236.6 & 1.1 & $-$4.4628 & 0.0040  & 0.24970  & 0.00044  & 27.7  & 2752.718287791 & $-$27252.6  & 0.7 & $-$4.5400  & 0.0030  & 0.24120  & 0.00036  & 27.1 \\
2092.491648451   & $-$27239.0 & 1.0 & $-$4.4417 & 0.0036  & 0.25192  & 0.00041  & 22.2  & 2768.660377451 & $-$27243.5  & 1.5 & $-$4.5295  & 0.0084  & 0.24269  & 0.00061  & 20.5 \\
2093.513091861   & $-$27239.9 & 1.4 & $-$4.4241 & 0.0054  & 0.25515  & 0.00055  & 22.2  & 2769.634996091 & $-$27245.7  & 2.0 & $-$4.5237  & 0.0125  & 0.24480  & 0.00074  & 26.9 \\
2094.460171571   & $-$27240.6 & 0.7 & $-$4.4236 & 0.0022  & 0.25739  & 0.00037  & 23.7  & 2771.666062901 & $-$27240.7  & 1.5 & $-$4.5009  & 0.0078  & 0.24553  & 0.00059  & 31.7 \\
2095.463708841   & $-$27239.5 & 0.6 & $-$4.4245 & 0.0016  & 0.25474  & 0.00033  & 25.6  & 2772.664407801 & $-$27239.2  & 0.6 & $-$4.4815  & 0.0021  & 0.24623  & 0.00026  & 29.0 \\
2096.487824141   & $-$27239.4 & 0.7 & $-$4.4208 & 0.0021  & 0.25390  & 0.00032  & 32.7  & 2773.629915901 & $-$27245.7  & 0.7 & $-$4.4907  & 0.0025  & 0.24744  & 0.00034  & 38.8 \\
2097.491001471   & $-$27239.1 & 0.7 & $-$4.4158 & 0.0020  & 0.25497  & 0.00031  & 37.3  & 2774.609521161 & $-$27251.5  & 0.6 & $-$4.5017  & 0.0021  & 0.24658  & 0.00032  & 38.5 \\
2099.553106361   & $-$27239.0 & 0.8 & $-$4.4112 & 0.0026  & 0.25857  & 0.00038  & 31.2  & 2775.612309401 & $-$27258.1  & 1.4 & $-$4.5286  & 0.0076  & 0.24482  & 0.00063  & 37.2 \\
2110.553320351   & $-$27246.2 & 1.4 & $-$4.4281 & 0.0056  & 0.25640  & 0.00059  & 40.2  & 2788.586848201 & $-$27254.9  & 1.1 & $-$4.5248  & 0.0047  & 0.24406  & 0.00048  & 39.4 \\
2111.525413931   & $-$27241.4 & 0.7 & $-$4.4451 & 0.0021  & 0.25523  & 0.00040  & 35.3  & 2789.591944861 & $-$27253.3  & 1.1 & $-$4.5282  & 0.0049  & 0.24426  & 0.00051  & 34.3 \\
2112.572887891   & $-$27241.3 & 0.8 & $-$4.4441 & 0.0025  & 0.25511  & 0.00042  & 33.8  & 2801.520875441 & $-$27251.9  & 0.9 & $-$4.5124  & 0.0037  & 0.24294  & 0.00035  & 36.7 \\
2120.583719121   & $-$27232.7 & 1.0 & $-$4.4343 & 0.0037  & 0.25174  & 0.00035  & 24.8  & 2802.545479351 & $-$27251.5  & 0.7 & $-$4.5168  & 0.0024  & 0.24200  & 0.00030  & 29.9 \\
2125.441460841   & $-$27234.8 & 0.9 & $-$4.4434 & 0.0028  & 0.25153  & 0.00033  & 27.3  & 2803.567721571 & $-$27251.5  & 0.8 & $-$4.5274  & 0.0032  & 0.24241  & 0.00037  & 29.0 \\
2126.443697021   & $-$27242.1 & 1.3 & $-$4.4603 & 0.0050  & 0.25018  & 0.00051  & 25.6  & 2808.646921611 & $-$27247.2  & 1.6 & $-$4.4792  & 0.0086  & 0.25036  & 0.00069  & 29.6 \\
2127.448085481   & $-$27242.2 & 0.8 & $-$4.4501 & 0.0024  & 0.24993  & 0.00029  & 30.0  & 2822.492876511 & $-$27253.0  & 0.8 & $-$4.4964  & 0.0030  & 0.24780  & 0.00043  & 40.4 \\
2130.433123751   & $-$27241.4 & 1.1 & $-$4.4500 & 0.0041  & 0.25100  & 0.00043  & 19.3  & 2826.357931311 & $-$27245.6  & 1.2 & $-$4.5147  & 0.0053  & 0.24316  & 0.00045  & 25.0 \\
2153.339238581   & $-$27245.9 & 0.7 & $-$4.4599 & 0.0022  & 0.24892  & 0.00032  & 24.6  & 2828.486241411 & $-$27249.4  & 1.1 & $-$4.5053  & 0.0046  & 0.24424  & 0.00045  & 29.6 \\
2154.325261341   & $-$27240.5 & 1.3 & $-$4.4489 & 0.0053  & 0.25082  & 0.00054  & 23.6  & 2829.449522331 & $-$27242.7  & 0.7 & $-$4.4955  & 0.0027  & 0.24471  & 0.00033  & 23.8 \\
2155.311607911   & $-$27240.2 & 2.2 & $-$4.4587 & 0.0108  & 0.25217  & 0.00079  & 23.9  & 2830.466413451 & $-$27245.5  & 1.3 & $-$4.4946  & 0.0059  & 0.24625  & 0.00053  & 25.9 \\
2156.344121791   & $-$27241.4 & 1.0 & $-$4.4452 & 0.0035  & 0.25316  & 0.00048  & 32.9  & 2830.561586381 & $-$27242.0  & 1.2 & $-$4.4961  & 0.0060  & 0.24603  & 0.00050  & 24.7 \\
2157.306336721   & $-$27248.2 & 0.7 & $-$4.4470 & 0.0022  & 0.25077  & 0.00035  & 35.3  & 2831.443859871 & $-$27246.9  & 0.8 & $-$4.4808  & 0.0030  & 0.24691  & 0.00035  & 23.8 \\
2169.298026221   & $-$27251.7 & 1.3 & $-$4.4496 & 0.0053  & 0.25143  & 0.00054  & 32.7  & 2832.503052761 & $-$27248.9  & 0.7 & $-$4.4728  & 0.0023  & 0.24836  & 0.00030  & 31.8 \\
2170.297159161   & $-$27253.0 & 1.2 & $-$4.4616 & 0.0043  & 0.25068  & 0.00043  & 37.5  & 2834.480060701 & $-$27247.6  & 1.6 & $-$4.4970  & 0.0074  & 0.24849  & 0.00059  & 41.4 \\
2171.301419151   & $-$27252.3 & 0.7 & $-$4.4571 & 0.0022  & 0.24788  & 0.00028  & 32.1  & 2835.491578681 & $-$27248.4  & 1.2 & $-$4.5024  & 0.0054  & 0.24478  & 0.00045  & 30.0 \\
2172.297160831   & $-$27250.0 & 0.7 & $-$4.4664 & 0.0020  & 0.24717  & 0.00028  & 28.3  & 2836.545248431 & $-$27245.4  & 0.8 & $-$4.4943  & 0.0028  & 0.24619  & 0.00032  & 29.5 \\
2189.362228101   & $-$27238.7 & 1.1 & $-$4.4463 & 0.0043  & 0.25189  & 0.00049  & 23.0  & 2843.463762141 & $-$27246.9  & 0.7 & $-$4.4671  & 0.0025  & 0.24781  & 0.00030  & 30.5 \\
2190.326304531   & $-$27245.5 & 0.8 & $-$4.4394 & 0.0027  & 0.25318  & 0.00035  & 38.1  & 2844.438655401 & $-$27255.9  & 0.8 & $-$4.4798  & 0.0031  & 0.24644  & 0.00035  & 31.8 \\
2192.300810341   & $-$27251.9 & 0.8 & $-$4.4562 & 0.0026  & 0.25059  & 0.00033  & 39.1  & 2845.439543131 & $-$27254.9  & 0.8 & $-$4.4830  & 0.0033  & 0.24694  & 0.00038  & 30.5 \\
2209.309890941   & $-$27246.2 & 1.4 & $-$4.4879 & 0.0063  & 0.24807  & 0.00063  & 27.1  & 2858.400437271 & $-$27251.0  & 0.9 & $-$4.5031  & 0.0034  & 0.24498  & 0.00040  & 30.8 \\
2212.303855161   & $-$27239.0 & 1.6 & $-$4.4696 & 0.0073  & 0.24796  & 0.00060  & 11.4  & 2859.425646371 & $-$27251.7  & 1.0 & $-$4.5091  & 0.0041  & 0.24243  & 0.00043  & 32.7 \\
2336.716973911   & $-$27245.8 & 1.1 & $-$4.4629 & 0.0039  & 0.24673  & 0.00042  & 33.2  & 2860.418459891 & $-$27253.2  & 0.9 & $-$4.5268  & 0.0041  & 0.24213  & 0.00046  & 28.7 \\
2339.701511501   & $-$27249.5 & 0.6 & $-$4.4525 & 0.0019  & 0.24812  & 0.00026  & 27.9  & 2861.415566871 & $-$27251.4  & 0.8 & $-$4.5289  & 0.0032  & 0.24159  & 0.00039  & 31.6 \\
2359.671725181   & $-$27234.7 & 1.3 & $-$4.4588 & 0.0054  & 0.24676  & 0.00048  & 19.9  & 2864.439014131 & $-$27244.8  & 0.8 & $-$4.5136  & 0.0034  & 0.24374  & 0.00039  & 22.8 \\
2367.673367981   & $-$27243.6 & 0.7 & $-$4.4752 & 0.0025  & 0.24596  & 0.00035  & 26.6  & 2865.444302091 & $-$27243.9  & 1.7 & $-$4.5433  & 0.0102  & 0.24269  & 0.00066  & 23.1 \\
2377.577219291   & $-$27249.3 & 0.6 & $-$4.4593 & 0.0017  & 0.24846  & 0.00028  & 33.7  & 2869.425821311 & $-$27252.1  & 1.0 & $-$4.4993  & 0.0045  & 0.24520  & 0.00046  & 28.9 \\
2378.592945881   & $-$27245.3 & 0.7 & $-$4.4664 & 0.0022  & 0.24551  & 0.00029  & 28.9  & 2870.325589271 & $-$27251.4  & 1.0 & $-$4.4963  & 0.0042  & 0.24653  & 0.00045  & 33.8 \\
2379.570488221   & $-$27242.8 & 0.7 & $-$4.4705 & 0.0021  & 0.24557  & 0.00030  & 26.2  & 2871.387042391 & $-$27249.4  & 0.9 & $-$4.4896  & 0.0037  & 0.24706  & 0.00041  & 32.7 \\
2380.530807611   & $-$27242.7 & 0.9 & $-$4.4708 & 0.0032  & 0.24564  & 0.00036  & 20.7  & 2872.360279921 & $-$27248.2  & 0.8 & $-$4.5119  & 0.0033  & 0.24324  & 0.00039  & 33.6 \\
2382.713363621   & $-$27235.4 & 0.8 & $-$4.4531 & 0.0024  & 0.24870  & 0.00030  & 16.3  & 2890.306801661 & $-$27246.6  & 0.6 & $-$4.4857  & 0.0018  & 0.24833  & 0.00032  & 28.6 \\
2387.690481311   & $-$27249.0 & 0.6 & $-$4.4588 & 0.0020  & 0.24718  & 0.00036  & 33.3  & 2892.305880341 & $-$27247.5  & 0.7 & $-$4.4836  & 0.0023  & 0.24867  & 0.00034  & 31.2 \\
2388.714307721   & $-$27244.7 & 0.6 & $-$4.4534 & 0.0019  & 0.24594  & 0.00027  & 33.2  & 2893.339527091 & $-$27251.0  & 0.7 & $-$4.4833  & 0.0026  & 0.24749  & 0.00034  & 35.4 \\
2390.702883931   & $-$27231.0 & 0.7 & $-$4.4692 & 0.0021  & 0.24492  & 0.00029  & 27.6  & 2894.319807371 & $-$27250.9  & 0.6 & $-$4.4963  & 0.0021  & 0.24680  & 0.00033  & 33.0 \\
2391.708873351   & $-$27245.6 & 0.6 & $-$4.4782 & 0.0019  & 0.24476  & 0.00027  & 28.6  & 2895.305897481 & $-$27249.4  & 0.8 & $-$4.5019  & 0.0029  & 0.24532  & 0.00043  & 28.5 \\
2400.659335481   & $-$27244.6 & 0.7 & $-$4.4694 & 0.0022  & 0.24413  & 0.00030  & 31.2  & 2897.315483451 & $-$27248.5  & 1.2 & $-$4.5065  & 0.0055  & 0.24573  & 0.00051  & 29.9 \\
2412.676269891   & $-$27242.6 & 1.1 & $-$4.4822 & 0.0046  & 0.24390  & 0.00047  & 29.2  &                &             &     &            &         &          &          &      \\
\hline                                                                         
\end{tabular}                                                                  

  \label{tab:E1}
  }
\end{table*}
\begin{table*}[!htb]
  \renewcommand{\arraystretch}{0.99}
  \centering
  \caption{HIRES@Keck~I spectroscopic series} 
  \resizebox{0.99\textwidth}{!}{
  \begin{tabular}{l c c c c || l c c c c}
\hline
\hline
BJD$_{\rm TDB}-2457000.0$   &   RV  &   $\sigma_{\rm RV}$ & log$R'_{\rm HK}$ &  $\sigma_{{\rm log} R'_{\rm HK}}$   & BJD$_{\rm TDB}-2457000.0$  &  RV  &   $\sigma_{\rm RV}$ & log$R'_{\rm HK}$ &  $\sigma_{{\rm log} R'_{\rm HK}}$   \\
               & (m~s$^{-1}$) & (m~s$^{-1}$) &        &      & &  (m~s$^{-1}$) & (m~s$^{-1}$) &        &      \\
\hline
1827.685434  &   $-$1.86 & 1.07  & $-$4.4783  & 0.0024 & 2398.092858  &    0.39 & 1.41  & $-$4.4672  & 0.0023 \\
1832.690571  &   $-$1.63 & 1.27  & $-$4.4846  & 0.0024 & 2400.024250  &    3.21 & 1.20  & $-$4.4727  & 0.0024 \\
1833.692228  &   $-$5.37 & 1.21  & $-$4.5035  & 0.0025 & 2406.080881  &    8.55 & 1.02  & $-$4.4827  & 0.0024 \\
1844.700628  &   $-$3.22 & 1.19  & $-$4.5040  & 0.0025 & 2410.075833  &    4.77 & 1.39  & $-$4.4756  & 0.0024 \\
1845.695992  &   $-$3.01 & 1.29  & $-$4.5138  & 0.0026 & 2417.945933  &    5.29 & 1.23  & $-$4.4848  & 0.0024 \\
1852.698773  &   $-$4.52 & 1.13  & $-$4.4921  & 0.0025 & 2455.919418  &    3.05 & 1.22  & $-$4.4730  & 0.0024 \\
2004.068819  &    1.26 & 1.14  & $-$4.4278  & 0.0021   & 2469.889300  &   $-$7.49 & 1.40  & $-$4.4971  & 0.0025 \\
2007.059160  &   $-$3.00 & 1.09  & $-$4.4430  & 0.0022 & 2506.828387  &  $-$10.38 & 1.29  & $-$4.5070  & 0.0025 \\
2019.048114  &    4.34 & 1.17  & $-$4.4439  & 0.0022   & 2508.838944  &    2.97 & 1.36  & $-$4.5205  & 0.0026 \\
2029.032907  &    3.48 & 1.12  & $-$4.4394  & 0.0022   & 2737.945987  &  $-$12.40 & 1.17  & $-$4.5321  & 0.0027 \\
2035.042662  &    2.66 & 1.07  & $-$4.4405  & 0.0022   & 2739.117009  &  $-$16.43 & 1.07  & $-$4.5302  & 0.0027 \\
2036.929885  &    4.54 & 1.00  & $-$4.4243  & 0.0021   & 2740.096890  &  $-$10.72 & 1.21  & $-$4.5382  & 0.0027 \\
2039.071431  &    7.54 & 1.10  & $-$4.4236  & 0.0021   & 2741.057523  &   $-$7.62 & 1.20  & $-$4.5494  & 0.0028 \\
2040.081965  &    1.95 & 1.08  & $-$4.4234  & 0.0021   & 2742.073701  &   $-$3.86 & 1.20  & $-$4.5488  & 0.0028 \\
2040.917436  &   $-$4.72 & 1.12  & $-$4.4236  & 0.0021 & 2742.913323  &   $-$2.87 & 1.17  & $-$4.5435  & 0.0028 \\
2041.914577  &   $-$3.49 & 1.05  & $-$4.4310  & 0.0021 & 2748.106598  &   $-$2.25 & 1.19  & $-$4.5156  & 0.0026 \\
2043.078551  &    1.64 & 1.05  & $-$4.4292  & 0.0021   & 2749.071431  &   $-$8.87 & 1.20  & $-$4.5338  & 0.0027 \\
2044.865558  &   10.19 & 1.22  & $-$4.4448  & 0.0022   & 2749.889789  &  $-$13.12 & 1.10  & $-$4.5284  & 0.0027 \\
2064.026575  &   $-$2.43 & 1.08  & $-$4.4415  & 0.0022 & 2751.984574  &   $-$9.52 & 1.12  & $-$4.5379  & 0.0027 \\
2067.996032  &   15.77 & 1.53  & $-$4.4534  & 0.0022   & 2766.065823  &   $-$2.13 & 1.14  & $-$4.5386  & 0.0027 \\
2068.932292  &   13.13 & 1.27  & $-$4.4420  & 0.0022   & 2769.029658  &    4.95 & 1.11  & $-$4.5278  & 0.0027 \\
2072.069879  &    1.55 & 1.16  & $-$4.4390  & 0.0022   & 2770.047506  &    6.05 & 1.13  & $-$4.5165  & 0.0026 \\
2073.040975  &    0.69 & 1.09  & $-$4.4467  & 0.0022   & 2771.970469  &    8.94 & 1.03  & $-$4.5040  & 0.0025 \\
2079.038339  &    2.51 & 1.32  & $-$4.4608  & 0.0023   & 2776.068894  &   $-$7.61 & 1.21  & $-$4.5286  & 0.0027 \\
2087.933810  &    1.59 & 1.37  & $-$4.4419  & 0.0022   & 2776.868691  &  $-$10.32 & 1.14  & $-$4.5264  & 0.0027 \\
2088.003994  &   $-$2.45 & 1.79  & $-$4.4480  & 0.0022 & 2780.057784  &    0.26 & 1.00  & $-$4.5319  & 0.0027 \\
2088.999167  &    1.70 & 1.28  & $-$4.4464  & 0.0022   & 2781.016252  &    2.89 & 1.08  & $-$4.5242  & 0.0026 \\
2090.029466  &    2.95 & 1.42  & $-$4.4506  & 0.0022   & 2785.989815  &   $-$2.71 & 1.04  & $-$4.4979  & 0.0025 \\
2091.054580  &    7.91 & 1.31  & $-$4.4626  & 0.0023   & 2787.026314  &   $-$7.67 & 1.14  & $-$4.4938  & 0.0025 \\
2092.026083  &   10.17 & 1.40  & $-$4.4674  & 0.0023   & 2789.882817  &   $-$3.75 & 1.10  & $-$4.5271  & 0.0027 \\
2093.022272  &    8.18 & 1.25  & $-$4.4481  & 0.0022   & 2791.898576  &   $-$7.48 & 1.07  & $-$4.5186  & 0.0026 \\
2094.819325  &    8.10 & 1.28  & $-$4.4402  & 0.0022   & 2792.928521  &   $-$2.47 & 1.11  & $-$4.5230  & 0.0026 \\
2097.969541  &    6.06 & 1.35  & $-$4.4280  & 0.0021   & 2800.087459  &   $-$1.50 & 1.15  & $-$4.5081  & 0.0026 \\
2099.823047  &    3.57 & 1.45  & $-$4.4310  & 0.0021   & 2800.829769  &   $-$6.21 & 1.11  & $-$4.5158  & 0.0026 \\
2100.720042  &    3.18 & 1.41  & $-$4.4387  & 0.0022   & 2803.048863  &   $-$3.95 & 1.23  & $-$4.5266  & 0.0027 \\
2101.720405  &    0.13 & 1.44  & $-$4.4526  & 0.0022   & 2806.852536  &   12.74 & 1.04  & $-$4.4958  & 0.0025 \\
2117.978869  &    0.15 & 1.32  & $-$4.4576  & 0.0023   & 2809.874370  &    0.56 & 1.12  & $-$4.4767  & 0.0024 \\
2119.865087  &   10.86 & 1.35  & $-$4.4542  & 0.0023   & 2812.773524  &   $-$2.31 & 1.14  & $-$4.5165  & 0.0026 \\
2120.961444  &    6.16 & 1.35  & $-$4.4542  & 0.0023   & 2822.943211  &  $-$10.91 & 1.11  & $-$4.4919  & 0.0025 \\
2121.715990  &    3.01 & 1.34  & $-$4.4466  & 0.0022   & 2823.895604  &   $-$3.54 & 1.11  & $-$4.5068  & 0.0025 \\
2122.952852  &    1.14 & 1.46  & $-$4.4506  & 0.0022   & 2824.996317  &   $-$1.10 & 1.21  & $-$4.5205  & 0.0026 \\
2123.713243  &   $-$0.77 & 1.31  & $-$4.4531  & 0.0022 & 2828.758960  &   $-$1.42 & 1.18  & $-$4.5166  & 0.0026 \\
2151.780594  &    0.40 & 1.42  & $-$4.4738  & 0.0024   & 2829.760272  &    5.51 & 1.27  & $-$4.5160  & 0.0026 \\
2187.788126  &   13.83 & 1.25  & $-$4.4724  & 0.0023   & 2831.924735  &   $-$5.17 & 1.24  & $-$4.4885  & 0.0024 \\
2189.799169  &    5.61 & 1.41  & $-$4.4856  & 0.0024   & 2833.902658  &   $-$0.24 & 1.25  & $-$4.4841  & 0.0024 \\
2208.692942  &   $-$0.57 & 1.18  & $-$4.4820  & 0.0024 & 2834.864443  &   $-$7.41 & 1.13  & $-$4.4933  & 0.0025 \\
2296.150339  &   $-$3.08 & 1.27  & $-$4.4832  & 0.0024 & 2835.843045  &   $-$5.29 & 1.20  & $-$4.5080  & 0.0025 \\
2300.149619  &    1.71 & 1.22  & $-$4.4757  & 0.0024   & 2838.839529  &   $-$8.54 & 1.24  & $-$4.5120  & 0.0026 \\
2314.144947  &   11.83 & 0.96  & $-$4.4557  & 0.0023   & 2840.733325  &   $-$0.68 & 1.15  & $-$4.5106  & 0.0026 \\
2355.977621  &    0.60 & 1.20  & $-$4.4735  & 0.0024   & 2857.905436  &   $-$5.29 & 1.21  & $-$4.5013  & 0.0025 \\
2377.126881  &   $-$7.15 & 1.11  & $-$4.4783  & 0.0024 & 2858.790522  &   $-$5.74 & 1.22  & $-$4.5086  & 0.0026 \\
2380.065539  &    4.77 & 1.00  & $-$4.4696  & 0.0023   & 2897.795444  &   $-$5.06 & 1.18  & $-$4.4958  & 0.0025 \\
2384.073278  &    9.73 & 1.29  & $-$4.4751  & 0.0024   & 3119.918974  &  $-$13.62 & 1.08  & $-$4.5373  & 0.0027 \\
2385.060044  &    2.80 & 1.29  & $-$4.4697  & 0.0023   & 3136.062563  &   $-$6.59 & 1.12  & $-$4.5225  & 0.0026 \\
2387.858959  &   $-$5.59 & 1.23  & $-$4.4633  & 0.0023 & 3139.030178  &    2.72 & 1.18  & $-$4.5200  & 0.0026 \\
2389.060128  &    0.71 & 1.35  & $-$4.4761  & 0.0024   & 3140.084275  &    3.43 & 1.09  & $-$4.5291  & 0.0027 \\
2390.061199  &   $-$2.05 & 1.20  & $-$4.4865  & 0.0024 &                 &         &       &          &        \\
\hline                                                                         
\end{tabular}

  \label{tab:E2}
  }
\end{table*}
\begin{table*}[!htb]
  \renewcommand{\arraystretch}{0.99}
  \centering
  \caption{APF spectroscopic series} 
  \resizebox{0.99\textwidth}{!}{
  \begin{tabular}{l c c c c || l c c c c}
\hline
\hline
BJD$_{\rm TDB}-2457000.0$   &   RV  &   $\sigma_{\rm RV}$ & log$R'_{\rm HK}$ &  $\sigma_{{\rm log} R'_{\rm HK}}$   & BJD$_{\rm TDB}-2457000.0$  &  RV  &   $\sigma_{\rm RV}$ & log$R'_{\rm HK}$ &  $\sigma_{{\rm log} R'_{\rm HK}}$   \\
               & (m~s$^{-1}$) & (m~s$^{-1}$) &        &      & &  (m~s$^{-1}$) & (m~s$^{-1}$) &        &      \\
\hline
      1834.674350    &     $-$2.79 & 5.15 & $-$4.4750 & 0.0047   &     1965.923857    &     $-$2.61 & 3.06 & $-$4.3801 & 0.0038 \\
      1834.689268    &     $-$9.51 & 5.04 & $-$4.4449 & 0.0044   &     1966.913030    &      7.55 & 4.12 & $-$4.3665 & 0.0037 \\
      1886.079199    &     $-$5.32 & 5.02 & $-$4.4316 & 0.0043   &     1969.822088    &    $-$18.94 & 6.20 & $-$4.3667 & 0.0037 \\
      1886.999954    &     $-$3.06 & 5.83 & $-$4.4704 & 0.0047   &     1970.916404    &      1.43 & 4.57 & $-$4.4564 & 0.0045 \\
      1888.042942    &     $-$1.36 & 4.44 & $-$4.4147 & 0.0041   &     1973.922285    &      2.77 & 2.88 & $-$4.3780 & 0.0038 \\
      1895.992414    &     $-$0.16 & 5.62 & $-$4.5083 & 0.0051   &     1974.896088    &      3.06 & 3.39 & $-$4.4484 & 0.0044 \\
      1904.017076    &    $-$18.27 & 6.48 & $-$4.4495 & 0.0045   &     1976.876174    &     10.85 & 4.69 & $-$4.3965 & 0.0039 \\
      1922.037585    &     22.19 & 3.88 & $-$4.4104 & 0.0041     &   1977.861553    &      8.07 & 3.34 & $-$4.3870 & 0.0039 \\
      1961.921692    &     $-$7.38 & 3.57 & $-$4.3956 & 0.0039   &     1979.983485    &     $-$2.95 & 3.33 & $-$4.4072 & 0.0040 \\
      1964.937358    &      3.32 & 3.02 & $-$4.4191 & 0.0042     &   1996.890942    &      5.56 & 2.95 & $-$4.4039 & 0.0040 \\
\hline                                                                         
\end{tabular}                                                                  

  \label{tab:E3}
  }
\end{table*}

\end{appendix}

\end{document}